\documentclass[12pt]{iopart}
\usepackage{iopams} 
\usepackage{epsf}
\def\Ar{\AA$^{-1}$}
\def\Tg{$T_{\rm g}$} 
\def\Tm{$T_{\rm m}$} 

\def\SQw{S(Q,\omega)}

\def\arx{$\alpha$--relaxation}
\def\brx{$\beta$--relaxation}
\def\t_s{$t_{\sigma}$}

\def\tm{\langle \tau \rangle}
\def\Sqw{$S(Q,\omega)$}
\def\vdW{van der Waals}

\def\Mob{M\"o{\ss}bauer}
\def\bib{\bibitem}

\begin{document}
\review[Neutron Scattering Studies of the 
Model Glass Former Ortho-terphenyl]
{Neutron Scattering Studies 
of the Model Glass Former Ortho-terphenyl}
\author{Albert~T\"olle}
\address{
Department of Biophysics, Biocenter University of Basel, 
CH-4056 Basel, Switzerland}
\ead{Albert.Toelle@unibas.ch}
\date{\today}
\begin{abstract}
The van der Waals liquid orthoterphenyl has long been used as a model
system in the study of the glass transition.
Motivated by mode--coupling theory,
extensive experiments have been undertaken to monitor the
onset of structural relaxation on microscopic time and length scales.
Using in particular quasielastic neutron scattering,
the decay of density and tagged--particle correlations 
has been measured as a function of temperature, pressure and wave number.
A consistent picture is developped
in which the mode--coupling singularity appears as a change of transport
mechanism in the moderately viscous liquid,
at temperatures far above the conventional (caloric) glass transition.\\
\end{abstract}
\pacs{
64.70.Pf, 
62.50.+p, 
61.25.Em, 
61.12.-q  
}
\submitto{\RPP}
\maketitle
\section{Introduction} \label{Introduction}
A glass may simply be defined as a non--crystalline solid
and is described by the predominance of local order.
Translational symmetry, {\it i.\ e.} long range order, is lacking.
The term {\it frozen liquid} stresses the structural
similarity to a liquid.

By {\it glass transition} we define the process of solidification
of a supercooled liquid to a non-crystalline solid, more exactly
we may call it a {\it liquid--to--glass transition}. 
When crystallization during cooling 
can be avoided the structural \arx\ time 
$\tau$ increases dramatically.
In a normal liquid the relaxation time is of the order of 1\,psec,
it increases continuously within a narrow temperature range over 
many orders of magnitude to attain values in the range of seconds, 
hours or even days.
Then the supercooled liquid is conventionally called a glass.

Many theoretical concepts have been proposed to explain the glass transition
among these free--volume and thermodynamic theories.
In the free--volume theories, introduced by Cohen and Turnbull \cite{CoTu59}
and substantiated by Cohen and Grest \cite{CoGr79}
the slowing down of the dynamics is related to the decrease of free volume 
and excess entropy.
Thermodynamic theories first formulated by Gibbs and Di--Marzio 
\cite{GiDi58} and Adams and Gibbs \cite{AdGi65}
assume an underlying second--order phase transition
in the limit $\tau \to \infty$ to account for the Kauzmann paradox.

A new approach to the problem was opened up after the mode--coupling
approximation has been introduced first by Leutheusser \cite{Leu84} and
independently by Bengtzelius, G\"otze and Sj\"ogren \cite{BeGS84}.
It provides an appproximation scheme which allows for the calculation of 
the evolution of time correlations from a microscipic equation of motion
and basically deals with the two well known phenomena: cage and backflow
effect.
In order to understand the structural relaxation 
it is proposed to investigate the short--time limit 
$\tau \simeq 1$\,psec where relaxational motion begins to evolve from
elementary vibrations of atoms or molecules.

Even in its most reduced schematic form mode--coupling theory (MCT) is able to 
describe essential features of structural relaxation and provides new 
predictions which can be tested experimentally.
Numeric solutions of mode--coupling equations describe the dynamics
of computer liquids or colloids with an accuracy of better than 10\% 
{\it without} any adjustable parameters 
\cite{MeUn93a,MeUn93b,MeUn94}.
The structure of the theory suggests that the asymptotic results hold also
for more complicated ``real'' glass fromers.
However, only experiments can decide to which extent ``real'' systems
reach the asymptotic behaviour predicted by theory.
An increasing number of measurements on many {\it different} materials
has been devoted to this question and 
give evidence for the relevance of the mode--coupling theory \cite{Got99}. 

In the present review, we survey results 
for {\it one} often studied model system, the molecular liquid 
{\it orthoterphenyl} (OTP) (section~\ref{Orthoterphenyl}).
The glass former orthoterphenyl can be considered to be a very good 
candidate to test the validity of such a theory as the molecules are
relatively symmetric and rigid.
The model character is underlined by the recurrent use for experimental 
studies and tests of various theories \cite{VoGG97}.

Neutron scattering is particularly well suited to test microscopic
theories as it measures directly the density and tagged--particle 
correlations on time scales of psec $\ldots$ nsec and on length scales 
of intermolecular distances and therefore provides an ideal link between 
theory and experiment.
Thus, neutron scattering experiments have been performed in a large
variety of different glass forming systems
like ionic glasses \cite{MeKF87,KnMF88,KaCC96,WuOG00}, 
polymers \cite{RiFF88,FrFR90,ArBW96} 
and even metallic glasses \cite{MeWP96,MeWP98,MeBS99}.
A comparison can be found in \cite{PeWu95}. 

To establish the context, we first describe some glass 
transition signatures (section~\ref{GT})
and some examples of the evolution of structural relaxation on time scales 
$\tau \gg 1$\,nsec accessible by conventional techniques.
After some remarks on the experimental method neutron scattering 
(section~\ref{Experimental_Methods}) a short introduction into 
the mode--coupling theory 
of the glass transtion (section~\ref{Mode_Coupling_Theory}) is given.
We then concentrate on the fast dynamics on the psec$\ldots$nsec time scale.
Tagged--particle and collective motion in liquid, glassy and crystalline 
(section~\ref{Vibrational_dynamics}) 
orthoterphenyl have been investigated as a function of temperature 
(section~\ref{Temperature_Dependent_Experiments}), pressure 
(section~\ref{Experiments_under_Pressure}) and scattering vector.
In this range time correlation functions evolve from vibrations 
(section~\ref{Vibrational_dynamics})
at very short times to structural relaxation through an intermediate
regime for which the mode--coupling theory makes the most prominent
predictions.

The neutron scattering experiments on OTP are probably
the most intensive study on a
molecular glass former up to now.
Here, we not only summarize previously published results, 
but we also present many new yet unpublished data.
Some experimental limitations are discussed in detail.
Those figures, which are here published for the first time, or where we
include new unpublished data are denoted by ``$*$'' in the figure caption.
All other figures (or data) are taken from the literature as stated in the
caption.
This review marks the end of a long term project 
that was supported over
more than 10 years by BMBF under project numbers 03{\sc si2mai}, 
03{\sc fu4dor4} and 03{\sc fu5dor1}.

\section{Orthoterphenyl} \label{Orthoterphenyl}
Every liquid may be solidified as a glass provided crystallization can
be bypassed.
Practically, atomic liquids cannot be supercooled into the region of interest
without immediate crystallization. 
The discovery of a concentration driven glass transition in colloidal
suspensions provided new model systems in which the dynamics on the length
scale of the nearest neighbours is probed by dynamic light scattering.
Indeed, very impressive confirmations of MCT predictions have been
provided by light scattering studies in such systems 
\cite{MeUn93a,MeUn93b,MeUn94}.

According to mode--coupling theory the dramatic increase of the relaxation
time with decreasing temperature or increasing pressure is due to 
relatively subtle but smooth changes in the static structure factor.
Since the theory was originally derived for ``simple liquids'',
it can be expected to be more reliable for simple systems without
internal degrees of freedom than for more complicated molecules.

Molecular liquids form stable supercooled liquids in favorable cases.
However, processes related to the molecular complexity 
which allows for supercooling should not screen phenomena that
are predicted by MCT, which, originally, has been developed for 
simple atomic  liquids with hard--core or Lennard--Jones interaction 
potentials without internal degrees of freedom.

The \vdW\ liquid Orthoterphenyl
(OTP: 1,2-diphenylbenzene, C$_{18}$H$_{14}$, boiling point 
$T_{\rm b}=605$\,K, melting point \Tm=329\,K)
is extensively studied for more than 40 years as
an archetypical glass former
by many methods and is representative for a whole class of glasses 
named fragile \cite{Ang95}, whose characteristic is an extremely 
non-Arrhenius behavior of the shear viscosity $\eta(T)$ 
\cite{McUb57,GrTu67a,LaUh72,CuLU73,SchKB98}.
OTP is a relatively simple compact, non--polar molecule 
with an overall shape that comes close to the spherical 
particles preferred by theory.
It consists of small, hopefully rigid molecular units which 
interact via weak non-directional van der Waals forces.
Therefore, we expect that the dynamics of supercooled
OTP represents structure independent, generic properties of
glass forming liquids and does not depend on microscopic structural 
peculiarities.

\begin{figure}[thb]
\begin{center}
\epsfxsize=120mm
\epsfysize=100mm
{\epsffile[200 296 413 495]{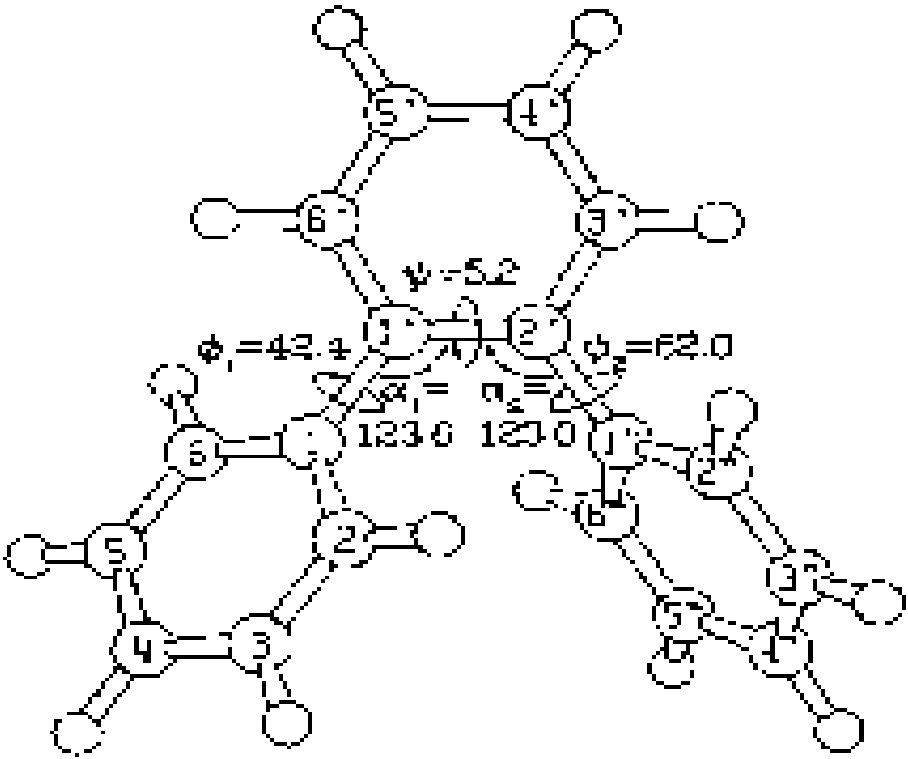}}
\end{center}
\caption{Molecular geometry of orthoterphenyl (C$_{18}$H$_{14}$)
as observed in the crystal structure. 
It is constituted by three phenyl rings, the two lateral rings being 
attached to the central one by covalent bonds.}
\label{otp-mol}
\end{figure}

The molecule consists of a central benzene ring and two lateral phenyl 
rings in ortho position.
For steric reasons, the lateral two phenyl rings are necessarily rotated
out of plane as shown in \fref{otp-mol}.
In addition, the overcrowding in the molecule leads to
significant bond-angle and out-of-plane distortions of the 
phenyl-phenyl bonds.
Such structural irregularities may explain why OTP can be supercooled
far easier than $m$- or $p$-terphenyl \cite{AnUb55}.

A number of structural features of the OTP molecule 
have their origin in the close proximity of the two bulky 
lateral phenyl rings. 
However, the details of the molecular structure may depend on 
whether it is in the crystalline, liquid, glassy or gas phase.

In the crystal, 
the angles for the out-of-plane rotation of the lateral phenyl rings 
are $\phi_1\simeq43^\circ$ and $\phi_2\simeq62^\circ$ \cite{BrLe79,AiMa78}.
For isolated molecules, an old electron diffraction study had suggested
$\phi_1=\phi_2=90^\circ$ \cite{KaBr44}, 
but newer experiments and calculations agree that 
in the gas or liquid phase
$40^\circ \lesssim \phi_1, \phi_2 \lesssim 65^\circ$ 
\cite{BaPo85,Gei94,MoLR00}.

The crystal structure belongs to the orthorhombic space group $P2_12_12_1$
with four molecules per unit cell and lattice parameters
$a=18.583$\,\AA, $b=6.024$\,\AA, $c=11.728$\,\AA\
at room temperature \cite{BrLe79,AiMa78}.
A sketch of the structure is given in figure~1 of Ref.~\cite{BrLe79}.
Experimental studies of the crystalline state 
have been performed mainly on polycrystalline powder samples or
on rather small single crystal needles.
However, single crystals of high quality and considerable size 
(several cm$^3$) can be grown out of a hot methanol solution 
either by very slow cooling or by evaporation over several months
\cite{ToZF00}.
The crystals grow preferentially along the shortest axis $b$.

Crystallization behavior of high purity OTP has been studied
extensively over a range of supercooling up to above the melting point
in connection with the glass forming ability 
\cite{MaHi73,ScUM74,OnUh80,YiUh81,YiUh82}.
For many glass formers the rate of cooling is taken as determining
whether or not a liquid will form a glass.
If unperturbed, OTP and some liquids of similarly shaped molecules
require years to nucleate \cite{MaHi73,ToSW97b}.
However, if abundant stable nuclei are artificially created in the 
sample, then the overall liquid $\to$ solid transformation occurs 
fairly rapidly on the time scale of seconds or minutes.
The crystal growth rate of OTP at atmospheric pressure passes through
a maximum value at 312\,K \cite{MaHi73}.
On the other hand, nucleation frequencies are difficult to measure 
and are usually calculated using classical nucleation theory.
For example, the steady state homogeneous nucleation frequency is highest at
251\,K \cite{YiUh82}.
The bulk heterogeneous nucleation frequency maximum lies between
260\,K and 275\,K for contact angles of 100$^{\circ}$ and 70$^{\circ}$,
respectively \cite{YiUh81}.
In most cases, nucleation takes place (heterogeneously) on surface
heterogeneities whereas homogeneous nucleation is rather exceptional
in glass forming liquids \cite{YiUh81}.
Heterogeneous nucleation at glass surfaces has not been studied
rigorously, even though it is the most common origin of devitrification.
Such nucleation may occur whether the surface of the glass is in contact 
with a container wall or exposed to atmosphere.
Heterogeneities characterized by small contact angles have a marked effect
on the cooling rates required for glass formation, with the required
cooling rate increasing with decreasing contact angle \cite{OnUh80,YiUh81}.
Nevertheless, with some care it is relatively easy to supercool OTP below
the melting point.
Depending on the cooling rate, the caloric glass transition
temperature \Tg\ lies between 242\,K \cite{ChBe72} and 245\,K 
\cite{GrTu67a}. 
Thus, OTP is rather easy to handle as the melting and the glass 
transition temperature are close to ambient temperature.

One further advantage particularly useful in neutron scattering studies
is that the molecule can be selelctively of even fully deuterated.
As a proton containing system the scattering is almost purely incoherent 
enabling us to study the single particle or tagged--particle dynamics.
Fully deuterated the scattering is mainly coherent giving rise to the 
observation of the pair correlation function.
Partial deuteration can be used to hide groups of atoms and to study
the intra-- and intermolecular motions separately.

Furthermore, we chose OTP for our investigations since it has
been extensively studied in the literature by many spectroscopic methods
as we will see in the next section (\ref{GT}).
Thus, as a function of temperature and pressure many data are 
available for comparison.
\section{Glass Transition in OTP} \label{GT}
\subsection{Macroscopic Quantities}
\begin{figure}[thb]
\begin{center}
\begin{minipage}[t]{75mm}
\epsfxsize=75mm
\epsfysize=55mm
{\epsffile[200 382 443 496]{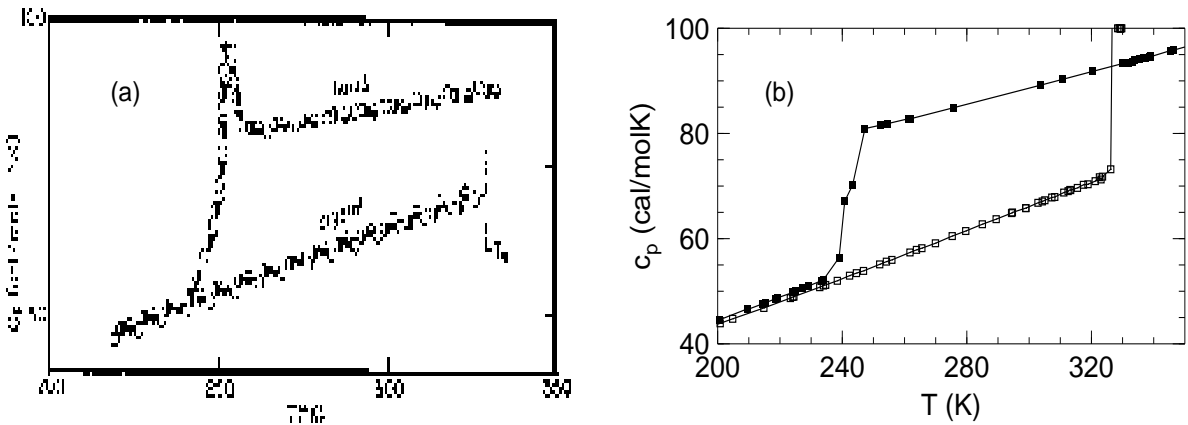}}
\end{minipage}
\hfill
\begin{minipage}[t]{75mm}
\epsfxsize=75mm
\epsfysize=55mm
{\epsffile[309 575 576 754]{Bilder/GT/cp-ChBe72.ps}}
\end{minipage}
\end{center}
\caption{Specific heat of liquid, supercooled liquid, glassy and crystalline 
OTP obtained by differential scanning calorimetry. 
The data in the left part (a) are obtained with a scanning speed of 0.3\,K/s 
(reproduced from \protect\cite{GrTu67b})
while for the data on the right (b) the scanning speed was 0.003\,K/s
(values taken from \protect\cite{ChBe72}).
}
\label{cp-GrTu67}
\end{figure}

The calorimetric techniques used for measuring the heat capacity $c_{\rm P}$
of glasses are the same as used for other solid materials.
In particular, differential scanning calorimetry (DSC) \cite{ChBe72,GrTu67b}
or differential thermal analysis (DTA) \cite{AtAn79} are used to determine
the glass transition temperature \Tg .
When the heat capacity $c_{\rm P}(T)$ of a supercooled liquid is measured as
a function of temperature (or pressure) one notices
a significant change occurring at a temperature \Tg\ 
(figure~\ref{cp-GrTu67}).

\begin{figure}[thb]
\begin{center}
\epsfxsize=120mm
{\epsffile[302 575 587 754]{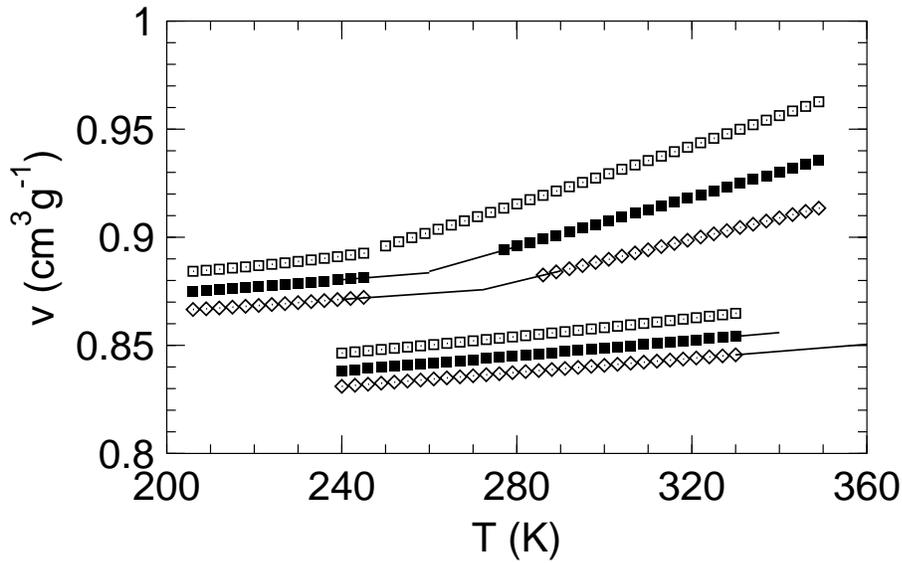}}
\end{center}
\caption{Varations of the isobaric specific volume of liquid, supercooled, 
glassy and
crystalline OTP as a function of temperature for various pressures 0.1\,MPa 
($\square$), 50\,MPa ($\blacksquare$) and 100\,MPa ($\lozenge$).
The solid lines are linear extrapolations from either side.
Reproduced using polynomial coefficients given in \protect\cite{NaKo89}.}
\label{v_T_P-NaKo89}
\end{figure}

A similar effect is observed in volume--temperature--pressure measurements
\cite{NaKo89}, reproduced in figure~\ref{v_T_P-NaKo89}.
A "discontinuity" in the thermal expansion coefficient
$\alpha=\partial \ln V/\partial T|_{\rm P}$,
the derivative of the volume $V$ at constant pressure $P$
can be deduced from figure~\ref{v_T_P-NaKo89}
for the glass while nothing particular happens in the crystal.
Although these features are characteristic of a second--order
equilibrium phase transition, \Tg\ cannot be related unambiguously
to an equilibrium function, such as the free energy.
In particular, the position and the sharpness of the transition near
\Tg\ {\it depends on the cooling rate} at which for example
the heat capacity is monitored.
In figure~\ref{cp-GrTu67} results with
two different cooling rates \cite{GrTu67b,ChBe72} are compared.
Conventionally, cooling rates of about 10\,K/min are used to determine \Tg .
Operationally, \Tg\ is defined as the center of the temperature range
of the large change in the heat capacity in a DSC run with cooling 
rate $\dot{T}$.
Note, that in this definition the glass transition is related with
a low--frequency excitation: e.\ g.\ for a typical rate of
$\dot{T}=10$\,K/min $\simeq 0.2$\,K/s and a width of the $c_{\rm P}$--jump
$\Delta T=10$\,K the excitation frequency is
$\omega=\dot{T}/\Delta T=0.02$\,s$^{-1}$, thus probing correlation
times of about 50\,s.
The glass transition temperature defined in this manner
can be understood as an isodynamic point 
which allows for a comparison of different glasses.

When pressure is applied the transition point shifts to higher temperatures
as can be seen in the volume--temperature--pressure relation in
figure~\ref{v_T_P-NaKo89}.
Note, that under pressure OTP shows a very strong density change. 
Calorimetrically, the glass transition temperature in OTP as a function
of pressure, $T_{\rm g}(P)$, has been measured by DTA by Atake and
Angell \cite{AtAn79}.
Up to at least about 100\,MPa the variation of $T_{\rm g}(P)$ is linear
with a slope ${\rm d}T_{\rm g}/{\rm d}P \simeq 0.26$\,K/MPa (260\,K/GPa).
These values for OTP and its isomeric mixtures are the highest yet observed
for molecular liquids (though some polymers exhibit higher values).
It was suspected that ${\rm d}T_{\rm g}/{\rm d}P$ is larger the more
closely the molecular substance behaves repulsively like a hard sphere.
These facts have been further motivations for us to use OTP as a model 
system also for pressure dependent neutron scattering experiments
(section~\ref{Experiments_under_Pressure}).

\subsection{Primary $\alpha$--Relaxation: Stretching and Scaling}
The fluctuation--dissipation theorem connects a linear response
with the time evolution or spectral distribution of spontaneous 
fluctuations.
Thus, relaxation in the equilibrium state can be studied
by a variety of spectroscopic or scattering methods.

\begin{figure}[thb]
\begin{center}
\epsfxsize=120mm
{\epsffile[200 -181 707 496]{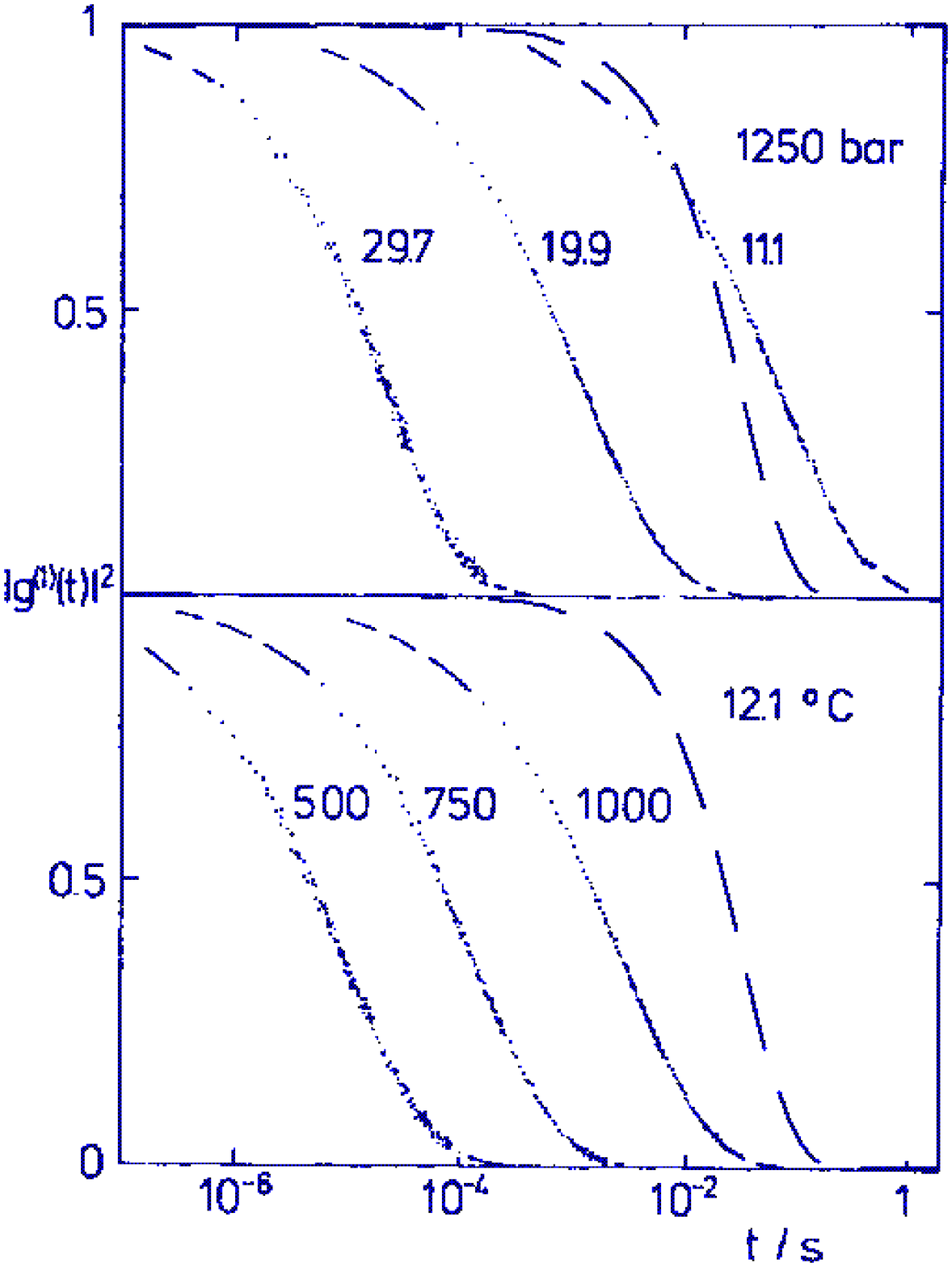}}
\end{center}
\caption{Normalized time--correlation function of liquid OTP,
obtained by photon correlation spectroscopy in depolarized (VH)
scattering geometry at constant pressure (top) and at constant 
temperature (bottom).
The Kohlrausch stretched exponential is fitted to the data (dots) 
with exponents $\beta$ between 0.53 and 0.62.
Reproduced from \protect\cite{FyDW83}.
For comparison, a simple exponential function was added (long dashed line).}
\label{pcs_t-FyWL81}
\end{figure}

Time--resolved measurements are performed  for instance by 
photon--correlation spectroscopy (PCS) which measures ultimately 
the evolution of density correlations $\Phi(t)$ \cite{FyWL81,FyDW83,HwSh99}.
The PCS data in figure~\ref{pcs_t-FyWL81}
illustrate nicely the most prominent features of structural relaxation:
stretching, scaling and a strong temperature and pressure
dependence of the mean relaxation time. 


The decay of the time correlations $\Phi(t)$ is called {\it stretched} 
since it extends over a much wider time range than a simple exponential 
process.
An exponential decay is expected when the dynamics consists of 
uncorrelated random steps.
This is the case for tagged--particle diffusion like motion in 
the long time and long distance limit \cite{WuCR96}.
On microscopic scales, however, the motion of molecules is a
highly collective process marked by memory effects.

In many cases the structural relaxation function $\Phi(t)$ is
well described by a Kohlrausch law \cite{Ko1854,Ko1863},
obtained by replacing $t/\tau$ in the Debye law by $(t/\tau)^{\beta}$:
\begin{equation}
 \label{KWW}
\Phi(t)=\exp(-(t/\tau)^{\beta}), \quad 0<\beta<1.
\end{equation}
For this function, the mean relaxation time $\langle \tau \rangle$
can be calculated as
\begin{equation}
\langle \tau \rangle = \int_{0}^{\infty} {\rm d}t \Phi(t)
                    =\tau\beta^{-1}\Gamma(\beta^{-1})
\end{equation}
where $\Gamma()$ is the gamma function.
The exponent $\beta$ is {\it not} (!) a material constant;
it differs for different correlation functions or observables which
are probed by different experimental techniques \cite{CuHL98} (see also 
figure~\ref{gt-beta}).
Such variations of $\beta$ demonstrate that an understanding of
structural relaxation will not be possible without accounting for the
different ways in which experimental probes couple to the microscopic
degrees of freedom.

\begin{figure}[thb]
\begin{center}
\epsfxsize=155mm
{\epsffile[24 627 573 803]{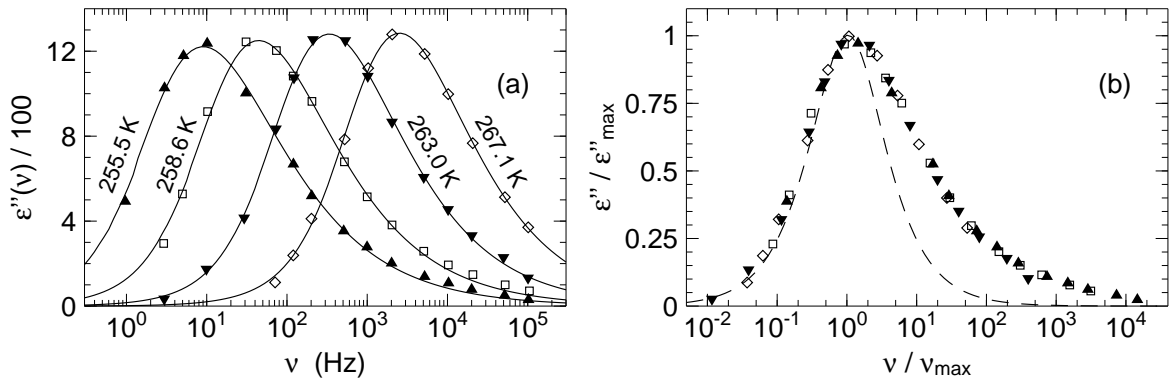}}
\end{center}
\caption{(a) Dielectric loss of 4.2 weight-\% anthrone
in OTP. Data are scanned from \protect\cite{WiHa71}
and re--fitted with a Fourier transformed Kohlrausch function.
The exponent $\beta$ seems to increase systematically from 0.52 at 255.5\,K
to 0.57 at 267.1\,K. ---
(b) Nevertheless, in a good approximation, all data can be condensed onto 
a master curve. Same symbols as in (a).
For comparison, the dashed line shows a Debye peak.}
\label{die_w-WiHa71}
\end{figure}

\begin{figure}[thb]
\begin{center}
\epsfxsize=120mm
{\epsffile[42 607 313 803]{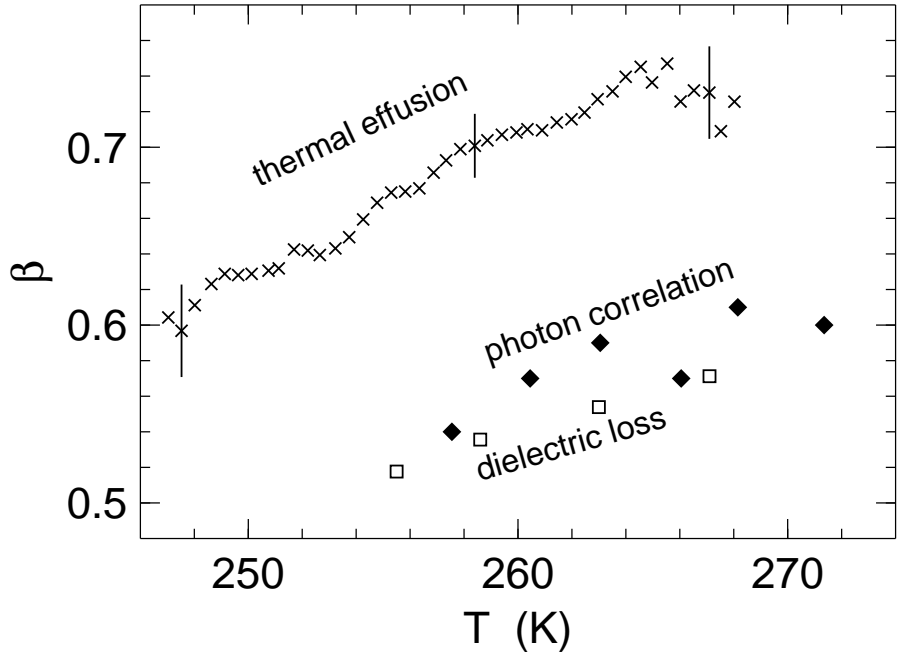}}
\end{center}
\caption{Stretching exponent $\beta$ from Kohlrausch fits to
thermal effusion ($\times$, OTP with 9\% o--phenyl--phenol 
\protect\cite{DiNa88}), photon correlation 
($\blacklozenge$ \protect\cite{FyWL81}),
and dielectric loss ($\square$, 4.2\% anthrone in OTP,
our fits to the data \protect\cite{WiHa71} of Fig.~\ref{die_w-WiHa71}).}
\label{gt-beta}
\end{figure}

On cooling or compressing a viscous liquid, $\tm$ increases
rapidly as can be seen in figures~\ref{pcs_t-FyWL81} and \ref{die_w-WiHa71}
while the line shape of $\Phi(t)$, expressed through $\beta$
varies only weakly.
This {\it scaling} property, known as the {\it time--temperature--pressure
superposition}, implies that in a first approximation the
measured individual correlations $\Phi(t;T,P)$ can be collapsed
onto a master curve 
\begin{equation}
  \label{mastercurve}
  \Phi(t;T,P)=\Phi(t/\tau(T,P))
\end{equation}
by rescaling the time axis to $\hat t = t/\tau(T,P)$.
This property turns out to be particularly useful in the analysis 
of neutron scattering experiments, to be 
discussed later in sections~\ref{t-arx} and \ref{p-arx}.

In many spectroscopic experiments, the response of a viscous liquid
is probed as a function of frequency.
A complex susceptibility $\chi(\omega)$ is related to the relaxation 
function $\Phi(t)$ through
\begin{equation}
 \chi(\omega)= \chi_{\infty} +
 (\chi_{\infty} - \chi_0) 
\int_{0}^{\infty} {\rm d}t e^{i\omega t} \dot{\Phi}(t).
\end{equation}
For high frequencies, the system reacts as a solid
$\chi(\omega \gg \tau^{-1})=\chi_{\infty}$;
for very low frequencies the fluid--like response is given by
$\chi(\omega \ll \tau^{-1})=\chi_0$.

For intermediate frequencies $\omega \simeq \tau^{-1}$ the Kramers--Kronig
relation relates the step in the real part $\chi^{\prime}(\omega)$ to
the maximum in the dissipative part $\chi^{\prime\prime}(\omega)$.
An exponential decay of $\Phi(t)$ leads to 
\begin{equation}
  \label{chi-Debye}
  \chi(\omega)= \chi_{\infty} +
  \frac{\chi_{\infty} - \chi_0}{1-i\omega\tau},
\end{equation}
which in the context of structural relaxation is often named 
after Debye, who derived it for Brownian motion of uncorrelated dipoles
in a viscous medium \cite{Deb29}.

For a stretched relaxation function, the step in $\chi^{\prime}(\omega)$
and the peak in $\chi^{\prime\prime}(\omega)$ are much broader than
a Debye peak.
To describe $\chi(\omega)$ one can either decorate (\ref{chi-Debye})
with one or two exponents \cite{CoCo41,DaCo51,HaNe67} or
use the Fourier transform of (\ref{KWW}) \cite{WiWa70}.

Frequency dependent dielectric susceptibilities for pure OTP have rarely 
been studied \cite{JoGo70,NaEM87,WuNa92,HaRi97,HaSB97}  
probably due to the extremely weak dipole moment.
Instead, in many studies the reorientation of polar host molecules is measured
\cite{WiHa71,WiHa72,ShWi73,DaHW73,NaTH74,NaKT74,NaMa83,DiWN90}.
Typical temperature dependent dielectric spectra and the corresponding 
master curve of 4.2 weight-\% anthrone in OTP are shown in 
figure~\ref{die_w-WiHa71}.
For comparison, a Debye peak is also shown.
Molecular reorientation of different probe molecules has also been studied by
optical \cite{HyEE90,CiBE95}, ESR \cite{AnSG97,EaMP97}
and NMR spectroscopy \cite{RoBT91}.


In figure~\ref{gt-beta} some published values of the Kohlrausch stretching
exponent $\beta$ are compared. 
For a given experimental probe, $\beta$ increases systematically with 
temperature in this temperature range.
Thus, the time--temperature superposition (\ref{mastercurve}) only
holds in a first approximation and in a limited temperature range.
In addition, for any given temperature, different experiments see different
exponents $\beta$ although approximate
values of $\beta$ seem to be correlated with other characteristics
of the glass transition dynamics \cite{BoNA93,Nga98}.

\subsection{Temperature and Pressure Dependence of the 
$\alpha$--Relaxation Time}
We have seen that the characteristic time (or frequency) of structural
relaxation strongly depends on temperature and pressure.
Cooling OTP by only 5\,K can slow down the decay of the correlation function
by a factor of ten or more, as shown in figures~\ref{pcs_t-FyWL81} and
\ref{die_w-WiHa71}.

\begin{figure}[thb]
\begin{center}
\epsfxsize=120mm
{\epsffile[308 575 576 757]{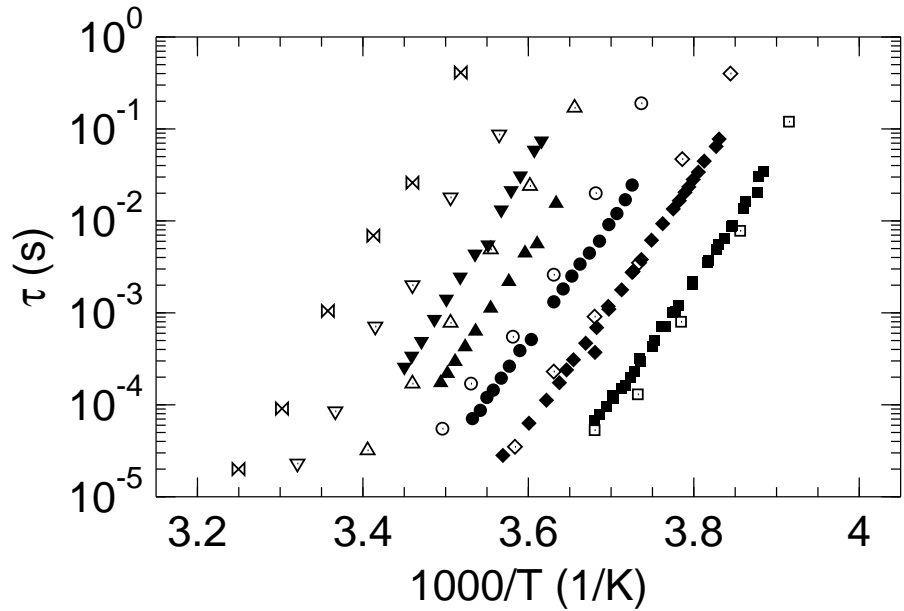}}
\end{center}
\caption{Temperature dependence of the isobaric relaxation time of 
viscous OTP for pressures between 0.1\,MPa and 125\,MPa from PCS 
(open symbols from right to left in steps of 25\,MPa) \protect\cite{FyDW83} 
and dielectric spectroscopy (closed symbols from right to left in steps of
19.6\,MPa) \protect\cite{NaEM87}.
}
\label{tau_T_P-FyDo83}
\end{figure}

Figure~\ref{tau_T_P-FyDo83} demonstrates that over many decades 
the characteristic time $\tau$ observed by different spectroscopic 
techniques follows the same temperature and pressure dependence, 
although at a given temperature or pressure the absolute values of 
$\tau$ may differ.
This finding is confirmed by measurements on many other glass formers.


The famous Maxwell relation $\eta = G_{\infty} \tm$
connects the shear viscosity $\eta$ with the relaxation time over 
the shear modulus at infinite frequency $G_{\infty}$.
Under the assumption that $G_{\infty}$ varies only weakly with external
control parameters like temperature and pressure,
it follows $\eta \propto \tm$.

Therefore, another operational definition of the glass transition 
temperature \Tg\ is given via the temperature, where the viscosity $\eta$ 
reaches $\eta(T_{\rm g}) = 10^{12}$\,Pa$\cdot$s.
With shear moduli of solids, typically of the order 
$G_{\infty} \simeq 10$\,GPa, this corresponds to a relaxation time of 
the order of 100\,s. 
This is the same time scale on that structural relaxation falls out of 
equilibrium in a caloric experiment (see above).

\begin{figure}[thb]
\begin{center}
\epsfxsize=120mm
{\epsffile[200 77 659 496]{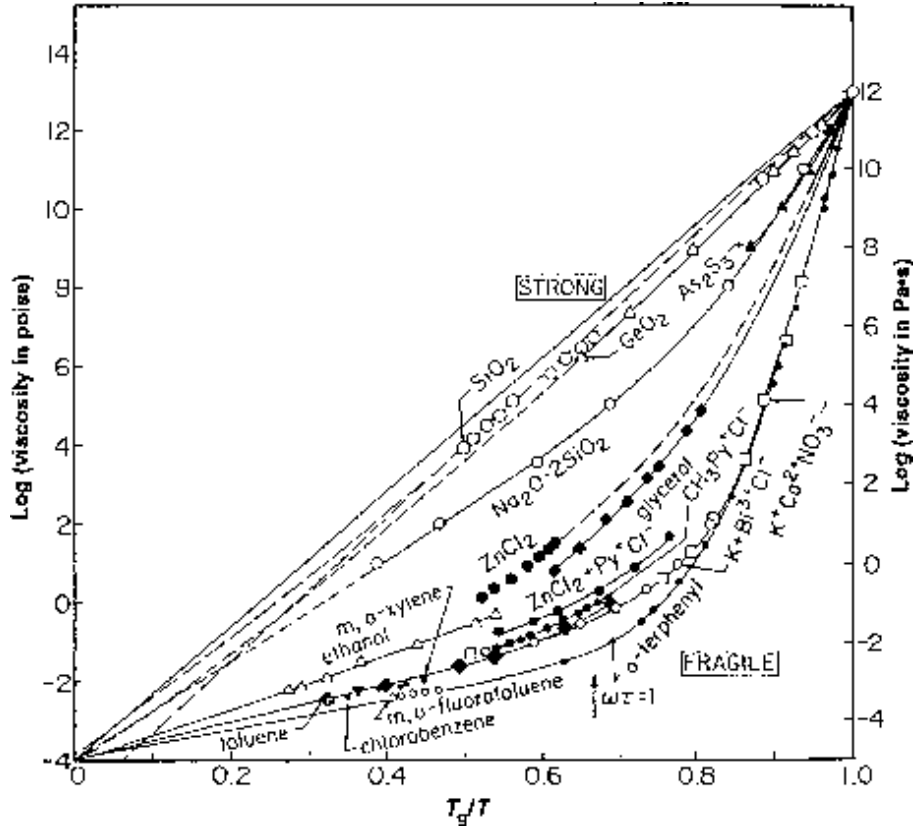}}
\end{center}
\caption{``Angell plot'': Viscosities of various glass forming liquids in 
an Arrhenius representation with reduced temperature scale $T_{\rm g}/T$.
This plot motivates Angells' "strong"---"fragile" classification in which
OTP is one of the most fragile liquids.
Reproduced from \protect\cite{Ang95}.
}
\label{Angell-plot}
\end{figure}

The Maxwell relation makes it possible to determine the temperature or 
pressure dependence of the structural relaxation time beyond the frequency
range of any spectroscopic technique.
In fact, the viscosity of OTP has been measured over more than 14 decades
as a function of temperature and pressure
\cite{McUb57,GrTu67a,LaUh72,CuLU73,SchKB98}.
The temperature dependent data at ambient pressure are shown in
figure~\ref{Angell-plot} together with viscosities of other glass formers
in a $\log\eta\ vs.\ T_{\rm g}/T$ representation (Angell plot) motivated by the
na$\ddot{\rm i}$ve expectation of thermally activated transport
\cite{Ang91}.
Although never observed in any real glass former the Arrhenius law
$\eta \propto \exp(-E_A/T)$ is taken as a reference to classify the
temperature dependence of $\eta(T,P)$.


This classification according to Angell is physical since the curvature in
the Arrhenius plot is correlated with the structure of a glass former:
{\it strong} liquids, which deviate only slightly from the Arrhenius law are
typically networks like SiO$_2$ or GeO$_2$ where the transport is
controlled by thermally activated breaking of covalent bonds.
Largest deviations are found for molecular liquids, ionic melts and some 
polymers. 
They are termed {\it fragile} liquids.
In this classification OTP is one of the most fragile liquids, which is
another motivation for its recurrent use as a model system.

From free-volume or entropy arguments \cite{CoTu59,CoGr79} 
it is possible to derive the popular Vogel--Fulcher--Tamman fit formula
\begin{equation}
\label{VFT}
\eta(T) \propto \exp \left( \frac{A}{T-T_0} \right)
\end{equation} 
with some arbitrary reference temperature $T_0 \ll T_{\rm g}$.
\begin{figure}[thb]
\begin{center}
\setlength{\unitlength}{1mm}
\begin{picture}(120,100)
\put(0,5){ \epsfxsize=120mm
{\epsffile[69 470 467 778]{Bilder/GT/vis-otp-T-MCT-fig.ps}}}
\put(75,38){ \epsfxsize=40mm
{\epsffile[200 240 395 495]{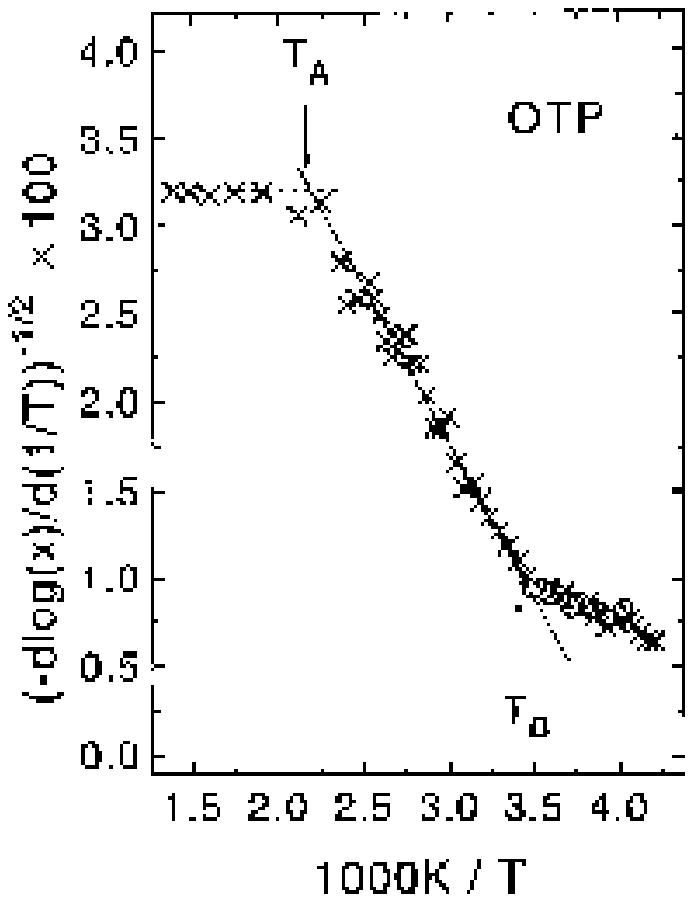}}}
\end{picture}
\end{center}
\caption{Temperature variation of $\eta(T)/T$.
The solid line is a free fit of a power law $(T-T_{\rm c})^\gamma$ 
expected from mode coupling theory yielding $T_{\rm c} \simeq 290$\,K
and $\gamma=2.59$.
The arrows indicate the caloric glass transition temperature 
$T_{\rm g}=243$\,K,
the critical temperature $T_{\rm c}=290$\,K and the melting temperature
$T_{\rm m}=329$\,K.
In the inset (taken from \protect\cite{HaSB97})
the temperature dependence of the maximum $f_{max}$
of the dielectric $\alpha$--process ($\circ$) and the viscosity $\eta(T)$ 
($\times$) is plotted as $[- {\rm d} \log (x)/{\rm d} (1/T)]$.
Such a representation linearizes a VFT--type temperature dependence 
(\protect\ref{VFT}) and an Arrhenius behavior appears as a horizontal
line.
$T_{\rm B}\simeq 290$\,K marks the temperature where the data begin
to deviate from a VFT--fit (line).
Above $T_{\rm A} \simeq 455$\,K $\eta(T)$ follows an Arrhenius law.}
\label{vis-otp-T-MCT}
\end{figure}

In any case, (\ref{VFT}) and similar functions are not able to fit $\eta(T)$
data over the entire temperature range, but they work much better when
applied to a restricted range.
A critical review of these and other approaches can be found in 
Ref.~\cite{CuLH97}.
The differential analysis of $\eta(T)$ in the inset of
figure~\ref{vis-otp-T-MCT} is one way to show that the temperature dependence
of $\eta(T)$ cannot be described by a single physical model
over the whole temperature range.
In particular, one recognizes a change of regime at about 290\,K.
Within mode--coupling theory this is interpreted as a cross--over from
liquid--like to hopping dynamics (section~\ref{Mode_Coupling_Theory}).
For OTP the value $T_{\rm c} \simeq 290$\,K was first determined by strict
elastic neutron scattering \cite{BaFK89} 
(section~\ref{t-squarerootsingularity}) 
and in the meantime it reappeared 
in many other measurements and simulations, 
as will be shown throughout this review.
In its simplest version, mode-coupling theory predicts for $T>T_{\rm c}$
a power law divergence
\begin{equation}
\eta \propto (T-T_{\rm c})^{\gamma}, \quad 1.76 < \gamma < 4.83.
\end{equation}
Figure~\ref{vis-otp-T-MCT} shows a three parameter fit of $\eta(T)/T$ 
yielding a critical temperature $T_{\rm c}=290$\,K and an exponent 
$\gamma=2.59$. 

\subsection{Diffusion and Decoupling of Time Scales}
It is interesting to compare the translational and rotational diffusion
coefficients, $D_t$ and $D_r$, which are integrals over single particle
correlation functions, with the shear viscosity $\eta$.
The shear viscosity depends upon intermolecular momentum transfer
through cross correlation functions of neighbouring molecules.
In simple \vdW\ liquids, the validity of the Stokes--Einstein and the Debye
relations,
\begin{equation}
 \label{Stokes-Einstein-Debye}
D_t=\frac{k_{\rm B}T}{6\pi\eta r_S}, \quad
D_r=\frac{k_{\rm B}T}{8\pi\eta r_D^3},
\end{equation}
originally derived for a Brownian particle in a viscous liquid, is well 
established in the fluid regime, although it is a non--trivial fact.
The lengths $r_S=0.21$\,nm \cite{McDF69} $\approx r_D=0.23$\,nm 
\cite{DrFK88,SchFS92} are temperature independent
and smaller than the geometrical \vdW\ radius $r_W=0.37$\,nm \cite{Bon64}.
It is most remarkable that $D_t$ and $D_r$ are proportional to $\eta^{-1}$
in the whole fluid regime.

\begin{figure}[thb]
\begin{center}
\epsfxsize=120mm
{\epsffile[200 256 542 496]{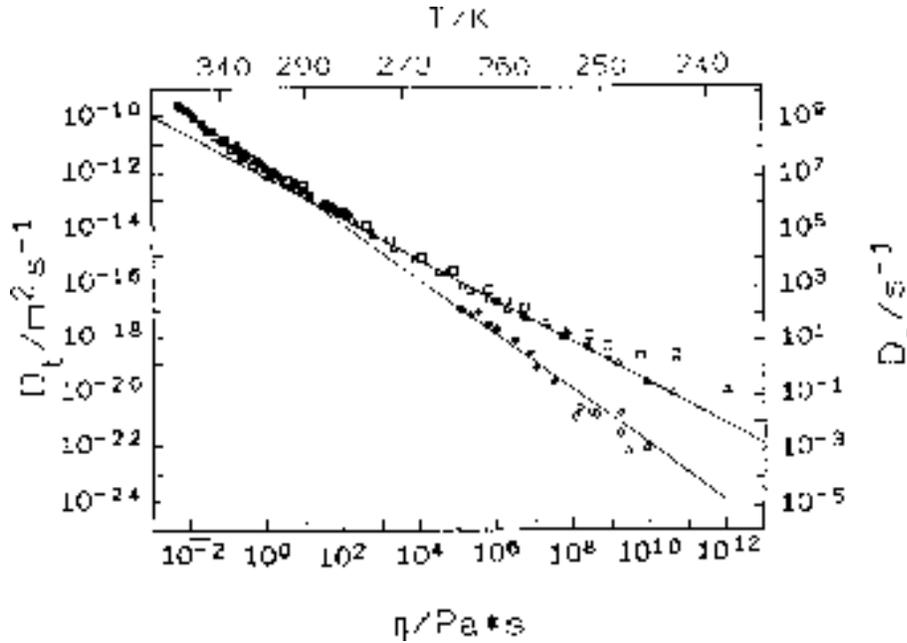}}
\end{center}
\caption{Log--log plot of the translational and rotational
diffusion coefficients $D_t$ and $D_r$ of OTP vs.\ viscosity $\eta$ and
temperature $T$: translational self diffusion ($\bullet$);
translational tracer diffusion (different photochromic molecules in OTP,
$\triangle, \square$);
rotational diffusion coefficient ($\blacklozenge, \lozenge$)
scaled along $D_t/D_r=4/3r^2$
(see (\ref{Stokes-Einstein-Debye}) using $r=0.22$\,nm).
The steep line represents the Stokes--Einstein relation
$D_t=k_{\rm B}T/(6\pi \eta r_S)$ using $r_S=0.22$\,nm and fits well the
$D_t(\eta)$ data above $T_{\rm c}\simeq 290$\,K and the $D_r(\eta)$ data.
The flat line that fits the $D_t(\eta)$ below $T_{\rm c}\simeq 290$\,K
has a slope $\eta^{-0.75}$ and crosses the $\eta^{-1}$ line at 290\,K.
Reproduced from \protect\cite{ChFG94}.}
\label{Decoupling}
\end{figure}

Rotational self diffusion in neat OTP has been measured by $^2$H--NMR 
stimulated echos \cite{Gei94,GeFS98}, translational self diffusion by 
$^1$H-NMR static gradient stimulated echos \cite{Gei94,FuGS92}.
Towards lower temperatures, tracer diffusion coefficients of two photochromic
tracer molecules having a size comparable to OTP and dissolved in OTP at 
low concentrations were measured by forced Rayleigh scattering \cite{LoSi91}.
Results are shown in figure~\ref{Decoupling}.

Rotational self diffusion could be measured almost down to $T_{\rm g}$ 
where Debyes' relation (cf.~\ref{Stokes-Einstein-Debye}) still holds.
Apparently, momentum transport is closely connected with molecular
reorientations.r
Translational diffusion, however, {\it decouples} from rotational
diffusion and viscosity and crosses over around
$T_{\rm c}=290\pm10$\,K to a weaker temperature dependence
approximately given by $D_t \propto \eta^{-0.75}$. 
Thus, $D_t$ and $\eta$ behave differently as the mechanism of molecular
motion changes from fluid to glass like behaviour.
On the other hand, $D_r$ remains proportional to $\eta^{-1}$ at temperatures
down to $T_{\rm g}$ indicating that rotational motion remains more
efficiently coupled to structural relaxation on approaching the glass
transition, at least, in fragile glass forming liquids like OTP,
toluene \cite{HiSi96}, salol, glycerin and others \cite{ChSi97}.

Indications for a cross--over between two distinct dynamic regimes
at a temperature $T_{\rm c}$ above the glass transition temperature
$T_{\rm g}$ have been found for many simple supercooled liquids 
\cite{Ros90}.
A discontinuity for the temperature coefficient of $\tau/\eta$ or
$(D\eta)^{-1}$ demonstrates the failure of the Stokes--Einstein relations
for $T < T_{\rm c}$.
The existence of the cross--over temperature $T_{\rm c}$ is further
substantiated by a plot of the correlation times on a reduced
temperature scale $T_{\rm g}/T$, which then shows a bifurcation of the
two relevant $\alpha$ and $\beta$ processes for $T < T_{\rm c}$.

\section{Experimental Method} \label{Experimental_Methods}
\subsection{Neutron Scattering}
In the previous section we have presented a variety of experimental
glass transition signatures observed with conventional 
spectroscopic methods.
What can we expect to learn about the glass transition from neutron 
scattering?

The energy of neutrons of the appropriate wavelength for structural and
dynamic studies corresponds to thermal energies for temperatures between
a few degrees K to well above room temperature.
In fact, neutrons produced from a reactor or spallation source have
their wavelength ``adjusted'' by the equilibration at a given 
temperature; thus ``cold'' neutrons have a wavelength of several \AA ,
``hot'' neutrons have a wavelength of a fraction of an \AA\ and ``thermal''
neutrons have a wavelength close to 1\,\AA .
Because of this wavelength--energy link, neutron inelastic scattering
allows the simultaneous study of spatial structure, on the one hand,
and dynamics, on the other, at the atomic and molecular level
\cite{Sch72,Squ78,Lov84,Bee88}.

Neutron scattering experiments on viscous liquids
are motivated by the following observations:
In amorphous and crystalline solids vibrational excitation frequencies
are generally of the order of 1\,THz;
long wavelength phonon frequencies of sound waves are found in the GHz to
THz region depending on the wave number $Q$ \cite{Squ78,Lov84}.
Furthermore, the mean relaxation time in normal fluids is typically of 
the order of 1\,ps \cite{Bee88}.
Thus, for the observation of the onset of the glass transition 
-- when relaxational motions decouple from vibrational ones -- we have to
look at the GHz to THz range which is traditionally the domain of scattering
experiments.
Such high frequencies are nowadays accessible with many experimental
techniques 
like e.\ g.\ dynamic light scattering, inelastic X--ray scattering,
quasielastic neutron scattering, dielectric spectroscopy or impulsive
stimulated light scattering (section~\ref{Other_Experimental_Approaches}).
However, not all of these techniques have been applied to OTP.

Neutron scattering remains one of the most versatile methods for the study
of the fast dynamics.
It can be used in almost every type of material since it does not depend on
optical transparency, dipole moments or polarizabilities \cite{Lov84}.
In neutron scattering one observes excitations at a well defined momentum
transfer $\bf Q$ of the order of several \Ar, {\it i.~e.} microscopic motion 
on interatomic and intermolecular distances can be resolved. 

The observed quantity is the so called dynamic structure factor
$S({\bf Q},\omega)$ and is a well defined correlation function
\cite{Sch72,Squ78,Lov84,Bee88}
\begin{equation}
 \label{Sqw-coh}
S({\bf Q},\omega)=
\frac{1}{N} \int \frac{{\rm d}\omega}{2\pi}e^{i \omega t}
\sum_{j,k=1}^{N} \langle e^{i{\bf QR}_j(0)}e^{-i{\bf QR}_k(t)}\rangle,
\end{equation}
which couples directly to the nuclear coordinates ${\bf R}_j(t)$.
It therfore provides an ideal link between experiment and theory as
$S({\bf Q},\omega)$ is the space--time Fourier transformation of the 
well know van Hove correlation function $G({\bf r},t)$ \cite{Hov54}.

Equation (\ref{Sqw-coh}) is based on the {\it coherent} superposition
of partial waves scattered by pairs of nuclei $j,k$.
The coherence can be diminished by spin flips except when both partial
waves are scattered by the same nucleus $j=k$.
This leads to an additional {\it incoherent} scattering
\begin{equation}
\label{Sqw_inc}
S_{\rm inc}({\bf Q},\omega)=
\frac{1}{N} \int \frac{{\rm d}\omega}{2\pi}e^{i \omega t}
\sum_{j=1}^{N} \langle e^{i{\bf QR}_j(0)}e^{-i{\bf QR}_j(t)}\rangle,
\end{equation}
which contains valuable information about tagged--particle motion.

In an actual experiment one measures always the sum of both contributions
\begin{equation}
  \label{DDCS}
S_{\rm exp}({\bf Q},\omega)=
 \sigma_{\rm coh} S_{\rm coh}({\bf Q},\omega)
+\sigma_{\rm inc} S_{\rm inc}({\bf Q},\omega).
\end{equation}
Incoherent and coherent scattering can be separated by a polarization 
analysis \cite{SchC93,Sch96} which is
based on the different scattering cross sections for
spin--flip $\sigma_{\uparrow \downarrow}=2/3 \sigma_{inc}$ 
and non--spin--flip $\sigma_{\uparrow \uparrow}=\sigma_{coh}+1/3 \sigma_{inc}$
\begin{equation} \label{sep}
\begin{array}{rcl}
S_{\uparrow \uparrow}({\bf Q},\omega) & = &
\sigma_{\rm coh} S_{\rm coh}({\bf Q},\omega)+
\frac{1}{3}\sigma_{\rm inc} S_{\rm inc}({\bf Q},\omega), \\
S_{\uparrow \downarrow}({\bf Q},\omega) & = &
\frac{2}{3}\sigma_{\rm inc} S_{\rm inc}({\bf Q},\omega). 
\end{array}
\end{equation}
Although this implies a considerable loss in neutron flux in 
high resolution experiments
such experiments are indeed feasible as shown below.

In practice, most often one concentrates on either incoherent or 
coherent scattering by choosing appropriate samples. 
In organic samples like OTP, the total cross section is largely dominated 
by incoherent scattering from hydrogen (table~\ref{cross-sections}).
Mainly coherent scattering can be measured on perdeuterated OTP-$d_{14}$.
Partial deuteration has been employed to hide the lateral phenyl rings
\cite{DeZB91} (section~\ref{t-squarerootsingularity}).

\begin{table} 
\caption{Neutron scattering cross sections 
for incoherent and coherent scattering and absorption cross section
in barn (=100\,fm$^2$) \protect\cite{Sea92}.}
\label{cross-sections} 
\begin{indented}
\item[]\begin{tabular}{lrrr}
\br
atom/molecule & $\sigma_{\rm inc}$ & $\sigma_{\rm coh}$ & $\sigma_{\rm abs}$   \\
\mr
H            & 80.27 & 1.76 & 0.33  \\
D            &  2.05 & 5.59 & 0.00  \\
C            &  0.00 & 5.55 & 0.00  \\
\mr
OTP-$d_0$    & 1123.8 & 124.5 & 4.62 \\
OTP-$d_{10}$ &  341.6 & 162.8 & 1.32 \\
OTP-$d_{14}$ &   28.7 & 178.2 & 0.00 \\
\br
\end{tabular}
\end{indented}
\end{table}


The spectrometers used for neutron scattering on OTP are listed in
table~\ref{instruments}.
They belong to basically five types \cite{Yel94} and are all
located at the Institute Laue--Langevin in Grenoble, France.
Three of them measure $S({\bf Q},\omega)$, one the so called intermediate
scattering function $S({\bf Q},t)$ which is the Fourier transform
of $S({\bf Q},\omega)$ and one spectrometer works without energy analysis:
\begin{itemize}
\item Triple--axis spectrometers (IN12) are instruments which can measure 
the four dimensional dynamic structure factor $S({\bf Q},\omega)$.
In single crystals the phonon dispersions relations $\omega({\bf Q})$ can 
explicitly be measured.
\item 
Time--of--flight (TOF) spectrometers with multichopper (IN5) or 
crystal (IN6, D7) monochromators where the energy of the scattered neutrons 
is measured by the time they need to travel a certain distance are used 
to measure vibrational dynamics as well as fast relaxation.
\item 
Back--scattering (BS) spectrometers achieve high energy resolution by 
(nearly) 180$^\circ$ Bragg scattering from both the monochromator and the 
analyzer crystals. 
They are used to measure the amplitude of elastic scattering and the onset
of the slow relaxation.
In order to measure full spectra the incident energy is changed either by
Doppler shifting the neutron wavelength (IN10, IN16) or by thermal expansion
of the monochromators (IN10, IN13). 
\item 
Neutron spin--echo (NSE) instruments (IN11) require by construction the use 
of  polarized neutrons.
They profit from the fact that the magnetic moment of the neutron undergoes 
Larmor precessions in a homogeneous magnetic field.
This is used to encode the neutrons' energy in the number of Larmor 
precessions.
The polarization analysis of the scattered neutrons amounts to an implicit
Fourier transform of the scattering function $\SQw$ and has a much larger
dynamic range as frequency domain spectrometers at the cost of a lower
contrast.
\item
Diffractometers (D7, D20) are used to measure the static structure factor 
of liquids and amorphous solids.
The integration over all energies is intrinsically performed.
In addition to a time--of--flight option, D7 can perform a polarization 
analysis and is thus able to measure
simultaneously incoherent and coherent scattering.
\end{itemize}
For isotropic samples like powders, liquids and glasses we can only access
$Q=|{\bf Q}|$.   

\begin{table}
\caption{Neutron spectometers used in our studies of the statics and dynamics
of OTP summarized in this review.
All instruments are located at the Institute Laue--Langevin in Grenoble and
use cold neutrons except IN13 which uses thermal neutrons.
Only IN12 is a single detector instument, on all other instruments
use multidetector banks which cover a wide range of scattering angles.
The energy resolution $\delta E$ varies with detector angle and chosen 
incoming wavelength $\lambda_i$.
Often quasielastic scattering data are analyzed by Fourier deconvolution
and the approximative time ranges are given.
The spin--echo spetrometer IN11 measures directly in the time regime and
can be operated with a single-- or a multidetector.}
\label{instruments}
\begin{indented}
\item[]\begin{tabular}{lllll}
\br
Instrument & type & $\lambda_i$ (\AA) & $\delta E$ ($\mu$eV) & Fourier time 
(psec)\\
\mr
IN5  & time--of--flight & 5.7--7.5 & 25--40 & 0.2 -- 35      \\
IN6  & time--of--flight & 5.1      & 80--120& 0.1 -- 10      \\ 
IN10 & back--scattering & 6.27     & 1-3    & 200 -- 2000 \\
IN11 & spin--echo       & 5.5--6.0 &  --    & 1 -- 2200   \\
IN12 & triple--axis     & 1.55 (k$_f$) & 300& --        \\
IN13 & back--scattering & 2.23     & 8--10  & 10 -- 200   \\
IN16 & back--scattering & 6.27     & 1      & 100 -- 2000 \\ 
D7   & time--of--flight & 4.8      & 150    & 0.2 -- 5       \\
\mr
D7   & diffractometer & 4.8    & -- & -- \\
D20  & diffractometer & 0.82   & -- & -- \\          
\br
\end{tabular}
\end{indented}
\end{table}


\subsection{Data Evaluation and Limitations}
To obtain the desired quantity $S(Q,\omega)$ from the measured 
scattering intensities several standard corrections have to be performed 
like normalization to the incoming flux and to the scattering of a
vanadium standard, subtraction of container scattering and background noise
and, if necessary, calculation of attenuation and absorption effects and 
conversion to energy.

\begin{figure}[thb]
\begin{center}
\epsfxsize=155mm
{\epsffile[45 568 572 746]{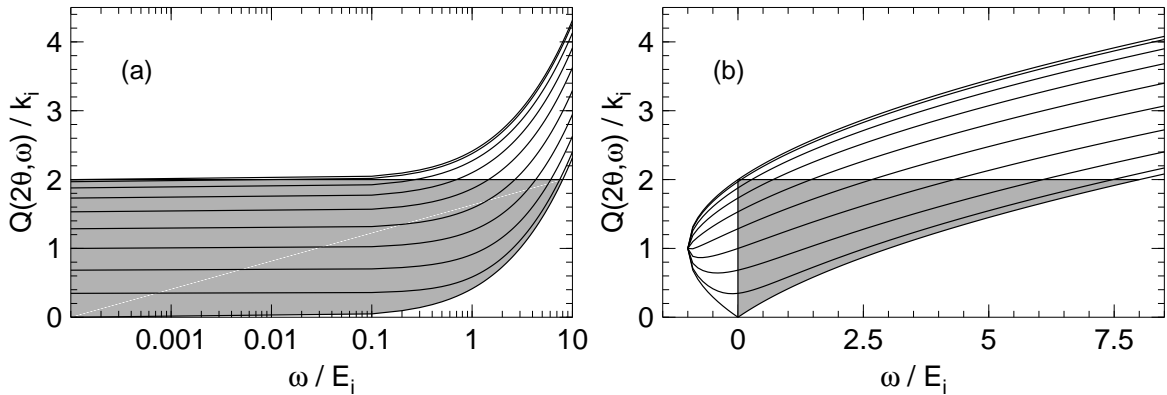}}
\end{center}
\caption{Maximum dynamic range of neutron scattering: (a) on a logarithmic
energy scale to emphasis the small energy transfer region typical for BS or
NSE instruments and (b) on a linear energy scale to emphasis the large
energy transfers typical for TOF instruments. 
The lines display the frequency dependence of the wave number 
$Q(2\theta,\omega)$ for several scattering angles ranging 
form $0^{\circ}$ to $180^{\circ}$ in steps of $20^{\circ}$
from bottom to top.
In any real experiment the subregion depends on the largest and 
lowest scattering angle accessible by each individual instrument.
Only the data along a $Q=const.$--line 
in the gray hatched subregion can be used 
when interpolated spectra $S(Q,\omega)$ are Fourier
transformed.
The vertical line $\omega=0$ is the lower energy bound as the
elastic channel is needed for the FT.
At a given wave number $Q$ the highest accessible energy is 
restricted and limited by the lowest scattering angle $\theta_{min}$.
}
\label{Q-of-w-Matte}
\end{figure}

Since all instruments measure at given scattering angles $\theta$ a 
conversion to the physically interpretable wave number $Q$ has to be done.
In figure~\ref{Q-of-w-Matte} we show the dependence of the wave number
$Q$ on the energy transfer $\omega$ for several scattering angles
$2\theta=0 \ldots 180^{\circ}$.

For backscattering and neutron spin--echo instruments, where the energy 
$E_i$ of the incident neutrons is much larger than the energy transfer 
$\omega$ the momentum transfer $Q$ at a given scattering angle is 
essentially constant (left panel).
However, when the energy transfer $\omega$ is of the order of $E_i$ 
-- for instance in time--of--flight spectroscopy -- then the dependence of 
$Q$ on $\omega$ cannot be neglected (right panel) although sometimes still 
done \cite{BuWR96}.

In any substance, one observes at intermediate temperatures significant
contributions to $S(2\theta,\omega)$ up to several 10\,meV. 
Before calculating a Fourier transform to get $S(Q,t)$, 
$S(2\theta,\omega)$ has to be interpolated to $S(Q,\omega)$.
As illustrated in figure~\ref{Q-of-w-Matte}, 
this imposes restrictions upon the 
accessible energy transfers.
The Fourier transform $S(Q,t)$ can only be 
evaluated if the frequency spectrum includes the elastic channel $\omega=0$.
The truncation problem at $\omega_{\rm max}$ is mathematically solved by a 
multiplication with the step function and leads to a lower bound in 
Fourier time $t$ below that 
the Fourier transform is not reliable \cite{WuPC95}.
The truncation effect may also appear in 
$S(Q,t=0)=2\int_0^{\infty}{\rm d}\omega S(Q,\omega)$ which is then less
than the true static structure factor $S(Q)$.
Calculating the density and tagged--particle correlation function 
$\Phi(Q,t)$ from the dynamic structure factors
\begin{equation}
  \label{corr-functions}
  \Phi(Q,t)=S_{\rm coh}(Q,t)/S_{\rm coh}(Q) \quad \mbox{and} \quad 
  \Phi^s(Q,t)=S_{\rm inc}(Q,t).
\end{equation}
might then lead to a wrong normalization.
For incoherent scattering this problem can be circumvented by measuring
$S(Q,\omega)$ at low temperatures where the scattering is totally elastic
and $S(Q)=1$.
For coherent scattering this is not possible as $S(Q)$ strongly depends on 
temperature and pressure. 
Note, that the truncation effect depends on $Q$ and on the dynamics of the
studied system itself and that two competing effects are operative:
With decreasing $Q$ the accessible energy range gets narrower (see
figure~\ref{Q-of-w-Matte}), but at the
same time the quasielastic broadening and the inelastic intensity
decrease as well.
Though, while it is  straightforward to measure a plateau or a decay
of the correlation function it is rather difficult to determine its 
absolute value within a few percent.
The truncation effect can be studied using several upper cut--off 
frequencies and is in general negligible at least for intermediate and
large wave numbers $Q$.

The determination of the dynamic structure factor in absolute units
is further complicated by the inevitable presence of multiple scattering.
Multiple scattering is basically a convolution of the dynamic
structure factor $S(Q,\omega)$ with itself.
The correction for it is far from being a standard data treatment.
In the inelastic spectrum, most multiple scattering intensity
comes from elastic--inelastic scattering histories \cite{Sea75}.
This contribution is nearly isotropic and therefore weakest at large
scattering angles. 
It tends to smear out the characteristic $Q$ dependence of the scattering,
and dominates at small angles where the single--scattering signal 
is expected to vanish as $Q^2$.

There are several different approaches to correct for this effect.
One can perform a Monte-Carlo simulation of the neutron histories
in an effective way \cite{WuKB93}.
In order to calculate the fraction of multiply scattered neutrons at one
point ($Q,\omega$) one has, in principle, to know the ideal scattering in 
the whole ($Q,\omega$)--space in absolute units.
As can be seen in figure~\ref{Q-of-w-Matte} only part of it is experimentally
covered.
Therefore, on needs a physical model for the ideal scattering including
$Q$-- and temperature dependences.
This makes the correction even more complicated for coherent scattering
where elastic, quasielastic and inelastic scattering are oscillating functions
of $Q$.

A new experimental, model independent, however very time consuming approach 
for multiple scattering correction in inelastic neutron scattering consists 
in the use of a set of different incident neutron wavelengths 
in connection with Monte Carlo simulations \cite{RuML00}.
Combining data at different wavelengths one can reliably establish an ideal, 
single scattering function in a wide $Q$ range which then serves as an input 
to the simulation followed by a self--consistent, iterative correction.

Another possibility at least for incoherent scattering
is to expand the scattering function 
$S(Q,\omega)=A(\omega) + Q^2 B(\omega) + \ldots$
which enables us to estimate the multiple scattering $A(\omega)$
by $Q^2 \to 0$ extrapolation
from the $S(2\theta(Q,\omega),\omega)$ spectra
however, loosing the elastic channels \cite{ToZF00}.
In practice, one often uses thin samples with high transmissions of the
order of 90\% and more.
Then multiple scattering is in general neglected at least at high scattering
angles \cite{WuKB93}.

Recently, an alternative approach has been worked out.
By extensive Monte Carlo simulations on very simple models
some generic trends of multiple scattering were identified \cite{Wut00a}:
it distorts the wave number dependence much more than the
frequency dependence of $S(Q,\omega)$.

Neutron scattering is always a compromise between 
the desired statistics, the wave number resolution and coverage,
the elastic energy resolution and the accessible energy window.
Apart from notoriously lacking statistics, one major problem in neutron
scattering is that practically all high resolution spectrometers lack dynamic
range.
It is known that in order to measure the dynamics of supercooled liquids and
glasses properly one has to cover several decades in frequency or
time.
Several approaches (partly) overcome this problem:
(i) the use of master functions obtained by scaling procedures like the
time--temperature superposition and / or 
(ii) the combination of data from different spectrometers with overlapping
time ranges.
Over the years we have been using both approaches.
Elastic resolution effects can be corrected in the $S(Q,t)$ representation
by division with the Fourier transformed resolution function (Fourier
deconvolution) that has been measured at very low temperatures either on
the sample itself or on a Vanadium standard.

Neutron spin--echo spectrometers which measure directly in time domain have
a rather large dynamic range of about 3 decades and more, depending on
the scattering angle and the incoming wavelength \cite{Yel94}.
However, the concomitant shortcuts are low countrates and low $Q$ resolution
compared to multidetector instruments.
Recently, a multidetector (42 detectors) has been installed on 
the NSE instrument IN11 covering about 40$^{\circ}$.
Now, by binning several detectors at least the statistical quality can be 
improved.
The $Q$ resolution is still implicitly limited by the transmission band 
of the velocity selector of
typically $\Delta \lambda/\lambda \simeq 10\,\%$ leading directly to
$\Delta Q/Q \gtrsim 10\,\%$.
One should be aware that in such a case one measures averages 
$\langle \Phi(Q,t) \rangle_Q$ over correlation functions. 
Increasing the $Q$ resolution by a narrower band width using for example
a graphite monochromator leads to an unacceptable reduction in intensity.

Improvements of neutron sources \cite{FRM2,ESS} to increase the incoming flux
and of instrumental performance \cite{Milleniumprogramm} to increase the 
efficiency are currently under way.
In order to test the new developments and extentions of the theory
(section~\ref{Extentions_of_MCT}) experimentally or to attack
even new questions the dynamic range
and concomitantly the $Q$--resolution and the $Q$--coverage 
have to be enhanced.

\begin{figure}[thb]
\begin{center}
\epsfxsize=155mm
{\epsffile[93 442 579 746]{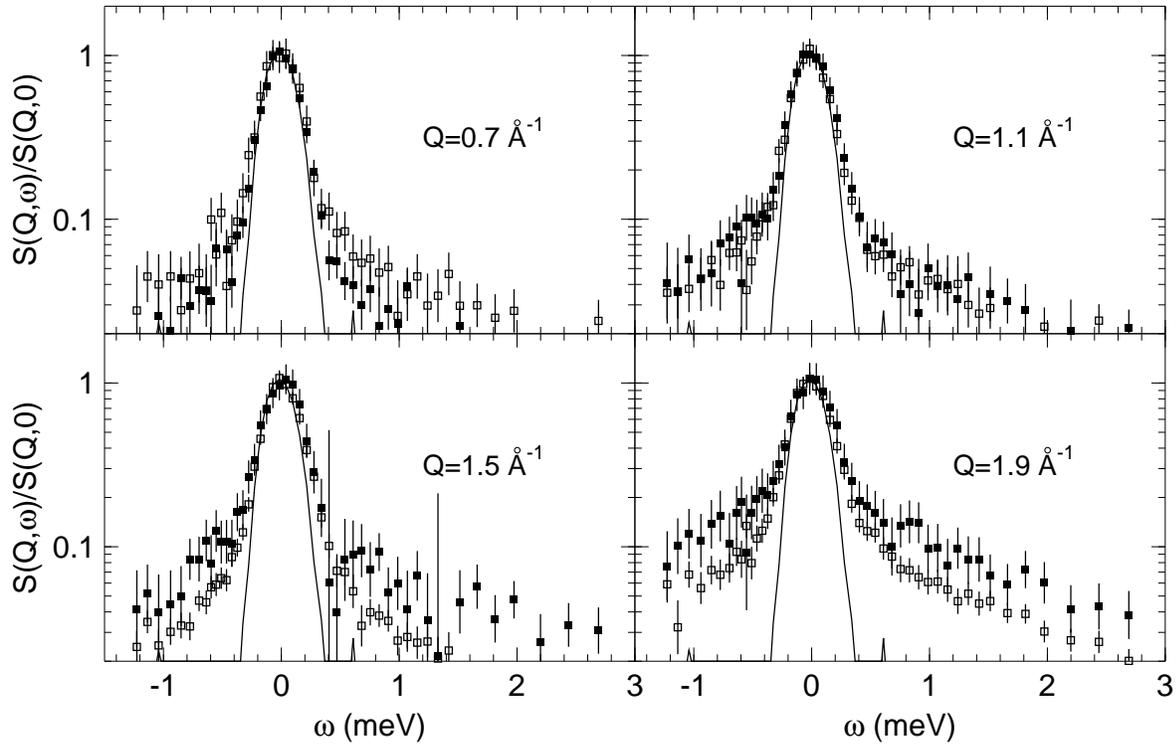}}
\end{center}
\caption{(*) Coherent (open symbols) and incoherent (full symbols) 
dynamical structure factor $S(Q,\omega)$ of OTP for several values of $Q$
at 340\,K as measured on D7 with polarization analysis.
Note the different quasielastic scattering for different $Q$.}
\label{Coh-inc}
\end{figure}

Let us demonstrate to which extent it is desirable to push 
improvements for a real progress. 
As an example we show in figure~\ref{Coh-inc} the tagged--particle and the 
density dynamic structure factor of OTP measured in the only proper way
by polarization analysis on the instrument D7 at the ILL.
Currently, D7 is the only multidetector instrument in the world which is able
to separate the energy resolved incoherent and coherent scattering
from one sample.
Some triple--axis spectrometers can do polarization analysis
as well, however, are equipped with a single detector only.
On each other instrument one always measures both contributions which are
differently weighted depending on the spectrometer type. 

This possibility is particularly interesting for some important questions
not only in glass science which have not been studied up to now.
For example, it would be possible to study exclusively the center--of--mass 
(COM) motion of OTP.
Measuring the incoherent scattering of OTP-$d_{10}$ where only the lateral 
phenyl rings are deuterated gives access to the tagged--particle correlation
function of the COM motion.
Conversely, measuring the coherent scattering of central ring deuterated
OTP-$d_{4}$ would lead to the density correlation function of
the COM motion.
At the same time, important information can be gained from the 
coherent scattering and the incoherent
scattering of the lateral phenyl rings in OTP-$d_{10}$ and OTP-$d_{4}$, 
respectively, about e.~g.\ relaxational modes and their time scales
or vibrational excitations of intramolecular motions.

Furthermore, it would also be possible to compare both structural relaxation
times and their $Q$ dependence in the context of reccurent used models
like ``de Gennes'' narrowing \cite{Gen59}. 
The advantage is evident: such studies could be performed on one and the 
same sample under exact identical experimental conditions.

The spectra shown in figure~\ref{Coh-inc} were obtained under fairly
optimized experimental conditions.
Nevertheless, although the transmission of 67\% was extremely low and the 
energy resolution rather moderate data accumulation took about 42\,h!
Such test experiments make it clear that, for example, increasing the 
incoming flux by a factor of two while leaving the instruments
as they are is far too less.
The total efficiency which is a product of the source strength, the 
transport efficiency and the instrument efficiency has rather to be 
increased by orders of magnitudes \cite{Milleniumprogramm}.

\subsection{Other Experimental Approaches} 
 \label{Other_Experimental_Approaches}
In section~\ref{GT} we have already presented many experimental techniques
probing the glass transition phenomena and the slow structural relaxation
of OTP.
Therefore, we limit ourselfs to a short enummeration of those other
experimental approaches that probe the fast dynamics in the GHz -- THz 
band and have been applied to OTP.

Light scattering spectroscopy has been often employed in OTP 
and many other supercooled liquids and glasses.
Raman scattering has primarily been performed in connection 
with the Boson peak and the study of the vibrational dynamics 
of the glassy and polycrystalline state \cite{StZP94,CrBA94,KiVP99}.
Brillouin scattering explores the longitudinal (LA) and transverse (TA)
acoustic modes in the GHz range and their interaction with the 
structural relaxation \cite{HiWa81,WaZS87,MoCF98b}.
Inelastic X-ray scattering extends the $Q$--$\omega$ range of the 
conventional Brillouin technique and makes it possible to study
sound waves on intermolecular distances and in the THz range 
\cite{MaMR98,MoMR98,MoFM99}.
Depolarised light scattering using geometries in which the usual
LA and TA Brillouin components are forbidden is the most often used
method to study the sub GHz to THz dynamics 
\cite{CuHL98,StPM92,StMP93,StPG94,PaSM94,CuLD97}.
Although the scattering mechanism and the connection to density fluctuations
are still debated \cite{PaSN97} the experimental results have often
been compared to mode--coupling theory.
A promising development covering about 5 orders of magnitude in time
is actually taking place using the optical Kerr effect in terms of a 
pump--and--probe experiment \cite{HiBG00a,HiBG00b}.


Computer simulation yield $\Phi(t)$ as primary output.
In recent Monte--Carlo \cite{Bas94} and molecular dynamics 
\cite{KoAn94,GaST96} studies the time range
was extended to such long times, that they now deal with the same
windows as achieved by the above mentioned modern experiments.
In molecular dynamics simulations many different observables can be 
studied which are not accessible by any other experimental methods.
The data quality is so good that the results can be compared 
with more sophisticated mode--coupling models and with numerical 
solutions of the mode--coupling equations, 
where the static structure factor is the only input.
However, simplified models have to be used approximating the real 
molecule \cite{MoLR00,LeWa93,LeWa94,KuWi95,WaLe97,ScTa99}.
The need for extending the simulations to times long enough to follow the 
relaxation processes in the supercooled and glassy state still necessitates
a compromise with respect to the complexity of the model and
the number of interaction sites.
Lewis and Wahnstr\"om used a three--site model mimicking OTP, 
where each phenyl ring was treated as a single interaction site 
\cite{LeWa93,LeWa94,WaLe97}.
Although simple, this model essentially reproduces the dynamics of OTP, 
in particular the anomalous behaviour of the Debye--Waller (\Mob--Lamb)
factor and the existence of a fast motional process.  
Later, Scortino et al.\ extended the calculations of Lewis and Wahnstr\"om
by more than one decade towards longer times \cite{ScTa99}.
In addition to the translational correlation functions the rotational ones
were calculated and compared with theory.
However, the planar nature of model makes it easier for the molecules
to slide past one another.
The 18--site, three--ring model of Kudchadkar and Wiest \cite{KuWi95}
and later of Mossa \etal \cite{MoLR00} takes into account the 
three--dimensional nature of the OTP molecule and 
may provide a better representation of the reorientational dynamics.
In addition it retains some internal torsional motions of the lateral 
rings with respect to the central one.
However, with the extended complexity only a few number of molecules
could be simulated and the time range is rather limited.
 

\section{Mode--Coupling Theory} \label{Mode_Coupling_Theory}
\subsection{Aim}
The dynamics of normal simple liquids is governed mainly by
excluded volume and binary collisions between molecules which, at
high temperatures or low density become more and more statistically 
independent from each other.
It can be described by generalizing the kinetic equation of
Boltzmann.
When the liquid is densified it is clear that this description breaks
down at some point, as the binary collision events become highly
correlated.
This gives rise to two well known phenomena:
the back--flow and the cage effect.
Suppose a tagged--particle is trapped in a cage formed
by its nearest neighbours which are caged themselves.
On a short time scale the only possible motion is a rattling in this cage.
The particle cannot escape and move over distances larger than
the next neighbour distances. 
These correlated collisions in the cage lead to a strong damping of
the motion and correspond to an increasing friction.
On longer time scales and for transport over longer distances the cage
must open which requires a cooperative rearrangement of many particles.
For large densities the cage forming particles are subject to the same 
dynamics as the particle in the cage, {\it i.~e.} all fluctuations have to
be treated on the same level.

It is the essence of the mode--coupling theory of the glass transition 
(MCT) that it indeed deals with the two phenomena cited above.
Its advent has highly stimulated many discussions and experiments.
MCT describes the evolution of structural relaxation in simple liquids.
The cage effect is responsible for the slowing down of the
dynamics of supercooled liquids and drives the system towards
an ergodic--nonergodic transition at some critical temperature
$T_{\rm c}>T_{\rm g}$ or some critical density $n_{\rm c}<n_{\rm g}$.
A smooth cross--over from vibrations to structural relaxation
is predicted as a new dynamic feature -- the $\beta$--relaxation.
At some point the back--flow effect 
counterbalances the cage effect and allows the system to stay
ergodic until it drops finally out of equilibrium at the caloric 
glass transition.

In the following we briefly recall the analytical solutions of
the mode--coupling theory as derived for the idealized case
as far as necessary for the interpretation of the experimental
data in sections~\ref{Temperature_Dependent_Experiments} and
~\ref{Experiments_under_Pressure}.
It is expected that the essential features of the idealized theory 
above $T_{\rm c}$ (or below $n_{\rm c}$) prevail in real systems 
\cite{Schi94}.
While the analytical solutions for the idealized theory provide
handy fit formulars, numerical solutions of the mode--coupling theory
can only hardly be used to analyze the experimental data.
The most adequate for our purpose is the formulation of the results 
in the time.
\subsection{Equations of Motion}
Starting point of the original (referred to as simplified or 
idealized) version of MCT is a generalized Langevin equation of 
motion for the normalized density fluctuation correlator
$\Phi_Q(t)=\langle \rho_Q(0) \rho_{-Q}(t) \rangle /
           \langle \rho_Q(0) \rho_{-Q}(0) \rangle$,
where $\rho_Q(t)$ is the Fourier component of the density fluctuation:
\begin{equation}
 \label{eq-of-motion}
\partial_t^2\Phi_Q(t) + \nu_Q\partial_t\Phi_Q(t) + \Omega_Q^2\Phi_Q(t)
+ \Omega^2_Q\int_0^t dt' m_Q(t-t')\partial_{t'}\Phi_Q(t')=0.
\end{equation}
The set of integro--differential equations can be closed 
using Kawasakis factorization approximation.
The kernel $m_Q(t)$ is expressed in terms of products of correlation 
functions (called the mode--coupling approximation):
\begin{equation}
 \label{m2}
m_Q(t)=\sum_{k_1+k_2=Q}V(Q;k_1,k_2)
                          \Phi_{k_1}(t)\Phi_{k_2}(t)
      ={\cal F}_Q(V,\Phi_k(t)).
\end{equation}
The internal control parameters $V$ representing the strength of
the coupling of different density modes are controlled by the external
control parameters like temperature $T$ or density $n$ via the static
structure factor $S(Q;n,T)$.
They can be calculated from the interaction potentials and vary smoothly 
with wave number $Q$ and all external control parameters.
Changing the temperature or the pressure, a dynamic phase 
transition from a liquid to a glass occurs at a critical value 
$V_c$ called the glass transition singularity. 
In the vicinity of a particular value $V_c$ a characteristic slowing down 
of $\Phi_Q(t)$ is found \cite{Leu84,BeGS84}.
As it is not accompanied by marked structural changes we deal with a
purely dynamic transition.

In most real materials the knowledge of the interaction potentials
is too inaccurate to compute the MCT solutions explicitly.
However, in the vicinity of the glass transition singularity $V_c$ 
a number of generic features of the solutions exist and are independent 
of the detailed form of the coupling constants.
Most of the predictions of MCT are obtained from asymptotic
expansions in the vicinity of the singularity.

\subsection{Solutions of Mode--Coupling Equations}
(i)~{\it The Square Root Singularity} \\
Denoting the distance from the singularity $V_c$ by the 
{\it separation parameter} $\sigma$, it follows for the 
Debye--Waller factor or the nonergodicity parameter
$f_Q=\lim_{t \to \infty} \Phi_Q(t)$ in leading order \cite{Got85,Got91}
\begin{equation}
\label{fq-eff}
  f_Q(\sigma)= f_Q^c + \left\{
    \begin{array}{l @{\quad : \quad} l}
      {\cal{O}}(\sigma) & (\sigma < 0, ergodic)  \\
      h_Q \sqrt{\sigma} + {\cal{O}}(\sigma)
               & (\sigma > 0, nonergodic)
    \end{array} \right. 
\end{equation}
$h_Q$ is called the critical amplitude.
Let $x$ be one external control parameter like $T$ or $n$, then for the
glass transition we have $\sigma \propto (x-x_c)/x_c$.
This square root cusp in $f_Q(x)$ is an important signature
of $x_c$ in the MCT.

The Debye Waller factor can easily be measured by elastic neutron scattering.
In practice, due to the finite instrumental resolution $\delta E$ 
it is impossible to measure strictly elastic scattering $\omega=0$.
All scattering events within $\delta E$ appear elastic.
Therefore, an effective nonergodicity parameter $f_Q$ is used
which corresponds to the replacement of $f_Q \delta(\omega)$ by
the area under the $\alpha$ peak or equivalently
by the plateau height of the correlation function
$\Phi_Q(t\simeq\hbar/\delta E)$.

\noindent
(ii)~{\it The $\beta$--Relaxation} \\
Let us now turn to the dynamic behaviour of the density correlation function
$\Phi_Q(t)$ in the vicinity of the dynamic phase transition.
The existence of a \brx\ with a critical spectrum is probably the most 
important result of MCT.
It concerns the mesoscopic time range
between the microscopic short time dynamics determined by $\Omega_Q$
and the \arx\ dynamics at long times.
This time range corresponds to the frequency range around the
minimum in the susceptibility spectrum $\chi^{\prime\prime}(\omega)$
separating the microscopic peak from the low frequency \arx\ peak.
In the ideal version of MCT the shape of $\Phi_Q(t)$ in this region is
completely determined by a separation parameter $\sigma$
and a material dependent constant $\lambda$.

Strong predictions are made in the $\beta$--relaxation regime,
the external control parameters being sufficiently close to the 
critical point.
For $t_0=\Omega_Q^{-1} \ll t \ll \tau$ and for $|x-x_c|$ sufficiently
small the correlators $\Phi_Q(t)$ stay close to the plateau value
$f_Q^c$.
The equation of motion (\ref{eq-of-motion}) can then be asymptotically
developed in terms of $|\Phi_Q(t)-f_Q^c|$.
Note, thereby one implicitly defines a time interval, where
$\Phi_Q(t)$ is close to $f_Q^c$ called the $\beta$--regime.
One obtains in leading order \cite{Got91,Got93}
\begin{equation}
  \label{beta-scaling-law}
  \Phi_Q(t)=f_Q^c + H_Q
                    g_{\pm}(t/t_{\sigma}),\quad 
(\sigma \gtrless 0).
\end{equation}
where the amplitude and the cross--over time
\begin{equation}
\label{hq}
H_Q=h_Q c_0 |\sigma|^{1/2} \quad \mbox{and} \quad
t_{\sigma}=t_0 |\sigma|^{1/2a}
\end{equation}
depend critically on the distance from the transition point.
$c_0$ is a material dependent dimensionless constant describing
the proportionality between the theoretical and the experimental
control parameters.

It is a remarkable property that $\Phi_Q(t)$ factorizes into
a purely $Q$--dependent amplitude $H_Q$ and a purely time--dependent
function $g_{\pm}(t/t_{\sigma})$.
Applied to the particle density $n({\bf r},t)$ 
this means that the variations of the density in space are uncorrelated
with those in time. 

The scaling function $g_{\pm}(\tilde t)$ depends only on the 
line shape parameter $\lambda$ and can be calculated from the 
expansion \cite{Got90} 
\begin{equation}
  \label{expansion}
  g_{\pm}(\tilde{t})=\left\{ \begin{array}{l @{\quad : \quad} l}
\tilde t^{-a} \pm A_1\tilde t^{a} + A_2\tilde t^{3a} \pm \ldots 
         & \quad (t_0 \ll t \ll t_{\sigma}, 
\sigma \gtrless 0 ) \\
-B\tilde t^{b} + B_1/(B\tilde t^{b}) + \ldots 
         & \quad (t_{\sigma} \ll t \ll \tau, \sigma <0). \end{array} \right.
\end{equation}
The exponents $a$ and $b$ as well as the expansion coefficients $A_i$
and $B_i$ and the cross--over time are tabulated functions of $\lambda$.
Starting from short times $t \to t_{\sigma}$, $g_{\pm}$ slowly
decays to zero.
For times $t$ longer than $t_{\sigma}$ it changes its sign for
$\sigma < 0$ and initiates the further decay of the correlations.

The two non--universal critical exponents $a$ and $b$
are related to the line shape parameter $\lambda$
via the transcendental equation
\begin{equation}
\frac{\Gamma^2(1-a)}{\Gamma(1-2a)}=\lambda=
\frac{\Gamma^2(1+b)}{\Gamma(1+2b)}, \quad 0<a<1/2, \quad 0<a<b<1
\end{equation}
from which follows that $1/2<\lambda<1$.

On leaving the plateau region, we obtain from (\ref{expansion})
the asymptotic power law
\begin{equation}\label{vSchw}
\Phi_Q(t)=f_Q - B h_Q |\sigma|^{1/2}(t/t_{\sigma})^{b}
         =f_Q - B h_Q (t/\tau)^{b}
\end{equation}
with a new universal time scale
\begin{equation}
  \label{alpha-scale}
  \tau=t_0|\sigma|^{-\gamma}, \quad \gamma=\frac{1}{2a}+\frac{1}{2b}
\end{equation}
which leads up to the final decay of correlations in the $\alpha$ process.
Its fractal time dependence is the key to stretching \cite{Got85}.
Eq.~(\ref{vSchw}) is called the von Schweidler law and
its region of validity is often rather limited.
However, it gives a theoretical explanation
of the often observed time--temperature--pressure principle
\begin{equation} 
  \label{alpha-scaling-law}
  \Phi_Q(t;X)=\Phi_Q(t/\tau(X)), \quad X=T, P, \ldots.
\end{equation}
This means in general, that the line shape may still depend on $Q$ but 
not on $X$.
A change of the external control parameter only causes a rescaling
in time.

MCT predicts that the relaxation times $\tau$ of all
correlators that couple to density fluctuations should show a divergent
behaviour near the critical point in the form of a power law with the same
universal exponent $\gamma$.
This property is called the $\alpha$ scale universality.
The exponent $\gamma$ can be calculated once the line shape parameter
$\lambda$ is know and vice versa.

The final decay of the correlations is often parameterized by the Kohlrausch 
stretched exponential (\ref{KWW}).
Note, that the Kohlrausch law is not a general solution of the MCT equations.
However, it is an asymptotic result in the limit of large wave numbers
\cite{Fuc94}.

\noindent
(iii) {\it Q--dependence} \\
In the framework of MCT the $Q$ dependence of $f_Q, h_Q, A_Q, \tau_Q$
and $\beta_Q$ is determined solely by the static properties of the system
and have so far been determined from numerical
solutions of mode--coupling equations for systems like hard spheres
\cite{BeGS84,FuHL92}, Lennard--Jones \cite{Ben86a}
spheres, and binary mixtures of hard spheres \cite{BoTh87} and
Lennard--Jones spheres \cite{NaKo97}.

In the incoherent case, {\it i.~e.} the tagged--particle correlations, $f_Q^s$
reflects the single particle oscillations and it closely
corresponds to a Gaussian
distribution in space $f_Q^s=\exp(-Q^2r_s^2)$ \cite{BeGS84,FuGM98}.
The half width of $f_Q^s$ is inversely proportional to the root--mean--square
displacement of a particle.
The amplitude $h_Q^s$ starts to increase with a $Q^2$ behaviour that is
also expected from harmonic theory of vibrational excitations
in the high frequency low temperature limit.

For the coherent case, {\it i.~e.} the density correlations, the amplitudes 
$f_Q$ and $h_Q$ reflect the oscillatory character of $S(Q)$ and are found to
oscillate in phase and out of phase with the static structure factor 
$S(Q)$, respectively.

Theory does not claim that the short time expansion of (\ref{KWW})
matches exactly (\ref{vSchw}).
However, it makes it plausible that the $\alpha$-relaxation time~$\tau_Q$ 
follows the $Q$~dependence of $f_Q/h_Q$.
The \arx\ time of the density correlations $\tau_Q$ is found to vary
strongly around $Q_0$, the position of the first structure factor
maximum \cite{FuHL92}.
Such oscillations of~$\tau_Q$ are in a more general context
known as de Gennes narrowing \cite{Gen59}.
Since in general $\beta_Q \neq b$ oscillations in $f_Q$ and 
$h_Q^{-1}$ should also lead to oscillations in~$\beta_Q$.

The \arx\ time of the tagged--particle correlations $\tau^s_Q$
decreases monotonously close to a $Q^{-2}$ law while $\beta^s_Q$
varies monotonically from 1 in the $Q\to 0$ limit to a constant value in the
large $Q$ limit \cite{FuHL92}.

For both, tagged--particle and density correlations the lineshape parameter
$\lambda$ and the crossover time $t_{\sigma}$ are identical and
$Q$--independent.
\subsection{Extentions of Mode--Coupling Theory} 
\label{Extentions_of_MCT}

In the idealized theory below $T_{\rm c}$ the \arx\ time $\tau$
and the cross--over time $t_{\sigma}$
are predicted to diverge at $T_{\rm c}$ with the viscosity $\eta$.
However, in reality they stay finite at this temperature.
The divergence is due to the overestimation of the cage effect in the 
approximation used for the memory kernel $m_Q(t)$ (\ref{m2}).
In the extended theory \arx\ and cross--over evolve smoothly
with decreasing temperature below $T_{\rm c}$.
Including the coupling to density currents, {\it i.\ e.} products of type
$\Phi_Q(t){\dot \Phi}_Q(t)$ or ${\dot \Phi}_Q(t){\dot \Phi}_Q(t)$
in (\ref{m2}), lifts the divergence at $T_{\rm c}$ and the system
remains ergodic even below the critical temperature.
This so called "phonon assisted hopping" processes counterbalance
the cage effect and restore ergodicity at $T_{\rm c}$.
Thus, the ideal glass transition is replaced by a change of mechanism of 
motion from liquid--like diffusional motion
above $T_{\rm c}$ to solid--like jump processes below $T_{\rm c}$.

The equations and solutions of the extended version are rather involved
and we expect that the essential features of the idealized version
concerning the \brx\ prevail.
For several glassy systems, in particular OTP, we will see that nature comes
very close to the ideal theory.

Equations (\ref{expansion})
are the predictions of the theory about the leading asymptotic behaviour
for the time and the temperature (or density) dependence of a generic 
correlator.
Recently the next leading order corrections to the asymptotic solutions 
have been calculated \cite{FuGM98,FrFG97a}.
These corrections lead to some modifications in the {\it early}
$\beta$--regime for $t_0 \ll t \ll t_{\sigma}$, namely 
\begin{equation}
\label{early-correct}
\Phi_Q(t)=f_Q^c + h_Q (t_0/t)^a \{1+[K_Q + \Delta] (t_0/t)^a \}
\end{equation} 
with a $Q$--independent constant $\Delta$ and a $Q$--dependent constant 
$K_Q$; both are temperature independent.
In the {\it late} $\beta$--regime the correlation function is now predicted to
behave like
\begin{equation}
\label{late-correct}
\Phi_Q(t)= f_Q^c - h_Q (t/\tau)^b \{1-[K_Q + \Delta^{\prime}] (t/\tau)^b \}. 
\end{equation}
The corrections are the second terms in the curly brackets in 
(\ref{early-correct}) and (\ref{late-correct}).
The important result is that the $Q$ dependent correction constant $K_Q$ is
the same in both equations.
In addition, the constanst $\Delta$ and $\Delta^{\prime}$ are
related with each other.
Indications of the existance and the correctness of these corrections
have been found in molecular dynamics simulations \cite{GlKo99}.

The success of mode--coupling theory for simple liquids has stimulated 
a considerable amount of theoretical work to extend it to molecular
liquids.
Recent contributions include the extention to the rotational dynamics 
of a linear molecule immersed into a system of spherical 
particles \cite{FrFG97b} and
the generalization to molecular liquids of linear and rigid 
molecules \cite{ScSc97}.
Using the projection operator formalism an equation of motion is derived
for the correlators $S_{lm,l^{\prime}m^{\prime}}(Q,t)$ of the tensorial
one particle density $\rho_{lm}(Q,t)$, which contains also the 
orientational degrees of freedom for $l>0$.
Now, the translational and the orientational degrees of freedom 
freeze in in a different manner and at different critical values and
demonstrate a hierarchy for the freezing: the orientational degrees of
freedom can never freeze in before the translational degrees of freedom
are frozen.

MCT has also been extended to treat the dynamics of full molecular
systems using a site--site representation \cite{CoHi98}.
A further step has been undertaken to describe the orientational and
translational dynamics of liquids composed by rigid molecules
of arbitrary shape \cite{FaLS99}. 

As in neutron scattering we are unable to unravel the reorientational
correlation functions we do not go into detail here.
The interested reader is referred to the cited references.

Generally, the mode--coupling approximations (\ref{m2}) are difficult to 
control as there is no obvious smallness parameter.
Thus, MCT predictions have to be checked by comparison with experimental
and numerical results.
This will be done in the next sections.

\section{Temperature Dependent Experiments} 
          \label{Temperature_Dependent_Experiments}
\subsection{Experimental Considerations}
As already mentioned in section~\ref{Orthoterphenyl} 
the crystallization tendency of OTP
requires some care.
Our first experiments have been performed with aluminum
containers where the temperature range between 245\,K and 290\,K was
inaccessible due to crystallization.
Later, the crystallization tendency could be significantly reduced
using high purity samples in clean glass capillaries with 
very smooth container surfaces. 
For incoherent scattering experiments with OTP-$d_0$ several hundred
soda lime glass capillaries of 0.32\,mm inner diameter and 0.04\,mm wall
thickness were used.
For the coherent scattering experiments with fully deuterated
OTP-$D_{14}$ below $T_{\rm m}$ soda lime glass capillaries with an inner
diameter of 1.2\,mm and a wall thickness of only 10\,$\mu$m were used.
The sealed capillaries were arranged on a circle of an appropriate sample
holder whose diameter was adapted to the neutron beam size mimicking
a hollow cylinder geometry.
This geometry keeps multiple scattering and 
self--shielding effects relatively isotropic.
Furthermore, the use of capillaries allows to check visually for any 
traces of crystallization.
Another positive side effect not considered before:
The glass tubes can be used for scattering experiments beyond
$Q\ge2.7$\,\AA$^{-1}$ as no Bragg reflection occurs
contrary to Al sample holders.

High purity samples were obtained by repeated recrystallization out of a 
hot methanol solution and subsequent vacuum distillation.
Before filling the capillaries were flushed with distilled water
to remove dust and baked out at about 850\,K for several hours.
After filling, the samples were tempered at about 350\,K for several days.
When left undisturbed at room temperature no crystallization occured for 
more than 2 years.

The measured transmission of a collimated beam of the samples was always 
higher than 90\,\% which is generally regarded as a good compromise
between the conflicting requirements of high single scattering
and low multiple scattering. 
Therefore, no attempt has been made to correct for multiple scattering
leading to some reservations in particular at low wave numbers, say
$Q \lesssim 0.5$\,\AA$^{-1}$.

\subsection{Static Structure Factor} \label{Static_Structure_Factor}
\begin{figure}[thb]
\begin{center}
\begin{minipage}[t]{75mm}
\epsfxsize=75mm
{\epsffile[200 334 409 496]{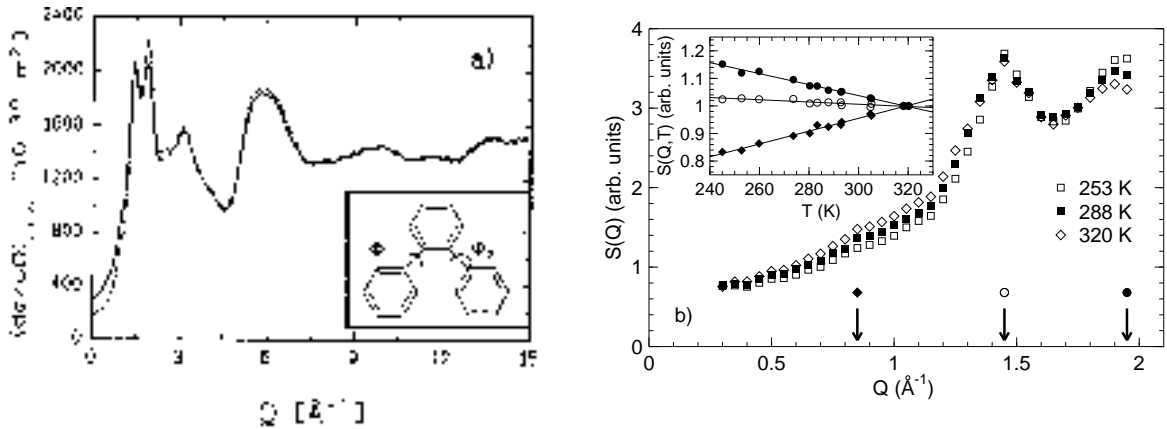}}
\end{minipage}
\hfill
\begin{minipage}[t]{75mm}
\epsfxsize=75mm
{\epsffile[68 297 534 651]{Bilder/T-Exp-structure/Sq-IN5.ps}}
\end{minipage}
\end{center}
\caption{(a) Static structure factor of fully deuterated OTP at 314\,K
(full line) and 255\,K (dashed dotted line) as measured on the neutron 
diffractometer D20 of the ILL, reproduced from {\protect\cite{BaBC93}}.
(b) Static structure factor in the low $Q$ region of deuterated OTP from IN5.
Note the appearance of a clear shoulder around $Q=0.85$\,\AA$^{-1}$
The inset shows the temperature dependence of $S(Q)$ at selected $Q$
values which are indicated by the arrows in the figure.
$S(Q)$ shows strong temperature variations at $Q=0.85$\,\AA$^{-1}$
and $Q=1.95$\,\AA$^{-1}$. Taken from \protect\cite{ToSW97b}.}
\label{Sq-Bartsch}
\end{figure}

Although OTP is a relatively simple molecule 
the static structure factor $S(Q)$ is rather involved \cite{BaBC93}.
A straightforward connection of the structure factor peaks to single 
molecular units is no trivial task since the static structure factor 
$S(Q)$ is a weighted sum of atomic correlations.
$S(Q)$ of supercooled deuterated OTP 
as measured on the diffractometer D20 of the Institute Laue
Langevin (ILL), Grenoble, is shown in figure~\ref{Sq-Bartsch}.

In contrast to atomic systems the main peak of the static structure factor
is splitted into two maxima at about 1.4\,\AA$^{-1}$ and 1.9\,\AA$^{-1}$.
With decreasing temperature a slight shift of the peak positions to 
higher $Q$ values due to the increase in density $n$ 
and in addition an increase of the peak height at 1.9\,\AA$^{-1}$ 
is observed.
The peak height at 1.4\,\AA$^{-1}$ is slightly reduced
which is probably connected to the decrease of the isothermal
compressibility $n\chi_T k_{\rm B} T = S(Q=0)$.

Due to experimental limitations the $Q$ region below
about 1\,\AA$^{-1}$ was not accessible in the D20 experiment.
Therefore, we show in the right part of figure~\ref{Sq-Bartsch}
the static structure factor $S(Q)$ in the low $Q$ region
as measured on the time--of--flight spectrometer IN5 of the ILL
\cite{ToSW97b}.
The double peak structure is resolved and the temperature dependence
of $S(Q)$ agrees well with the one measured on D20.
One recognizes a clear shoulder around 0.85\,\AA$^{-1}$ with an
increasing peak height with increasing temperature.
In inelastic X-ray scattering experiments a real prepeak
at $Q\simeq 0.85$\,\AA$^{-1}$ was observed \cite{Mas98}
together with a main peak at $Q\simeq 1.45$\,\AA$^{-1}$ while
nearly nothing is found around $Q\simeq 1.9$\,\AA$^{-1}$.
Note that in neutron scattering on fully deuterated OTP both Deuteron
and Carbon atoms are seen with about the same scattering cross sections 
(cf.\ table~\ref{cross-sections}), while in X--ray scattering only the 
distribution of Carbon atoms has to be taken into account.

\begin{figure}[thb]
\begin{center}
\epsfxsize=120mm
{\epsffile[200 335 392 496]{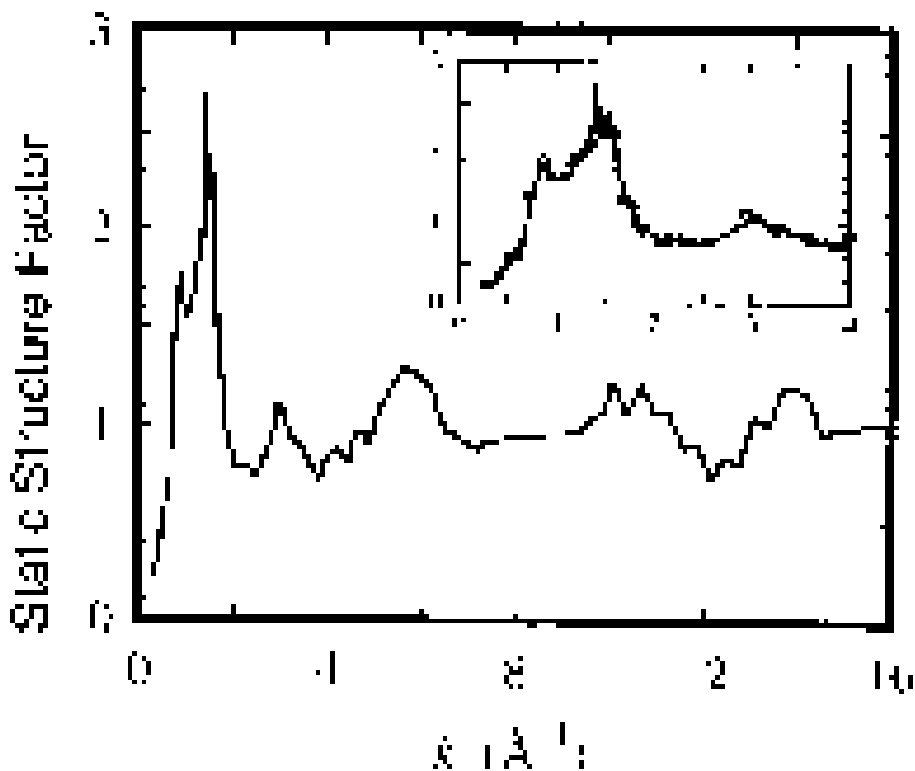}}
\end{center}
\caption{Calculated static structure factor for the 18-site model of OTP.
The inset displays the small $Q$ region in more detail.
Note the shoulder around $Q=0.8$\,\AA$^{-1}$.
Reproduced from \protect\cite{KuWi95}.}  
\label{Sq-Kudchadkar}
\end{figure}

In simulations simplified models are used for the complex molecule 
to keep the computing time reasonable.
Lewis and Wahnstr\"om employed a three site complex, each site playing the
role of a whole benzene ring \cite{LeWa94}.
The static structure factor of the center of mass shows a maximum 
around $0.8 \ldots 1$\,\Ar\ while the one of the sites is solely peaked at
$1.5$\,\Ar.
A refinement of this model with 18 sites on 3 rings was later studied by
Kudchadkar \cite{KuWi95}.
The static structure factor shown in 
figure~\ref{Sq-Kudchadkar} now well reproduces the experimental one, 
in particular at higher $Q$ including the hump at $3$\,\AA$^{-1}$
and the prominent peak at  $6$\,\AA$^{-1}$.  
For smaller $Q$ it yields, besides a double peak structure 
at $1.4$\,\AA$^{-1}$ and $1.5$\,\AA$^{-1}$, 
also a clear shallow peak around $0.8$\,\AA$^{-1}$ as illustrated in the
inset of figure~\ref{Sq-Kudchadkar}.
The fact that the second peak of the double peak is at lower $Q$
than observed experimentally is probably due to the fact of
``fusing'' the carbon and the attached hydrogen atoms into one
``united atom''.
In a recent work on a flexible molecule model \cite{MoLR00}
a good agreement among the molecular dynamics structure factor and that 
measured by neutron scattering was obtained in the high--$Q$ region.
A clear shoulder was observed at 0.9\,\AA$^{-1}$, a diffraction peak at
1.5\,\AA$^{-1}$ but no double peak structure, only a small bump 
around 1.8\,\AA$^{-1}$.
Their analysis in terms of a calculated structure factor on the molecular
and ring center of mass clearly support our simplified picture that follows
below.

In this context we point out that
the structure factors of fully deuterated benzene
and hexafluorobenzene, that can be considered as unconnected subunits 
of OTP  are nearly indistinguishable 
from the one of OTP both showing the double peak 
at the right positions as well as the hump at $3$\,\AA$^{-1}$
and $6$\,\AA$^{-1}$ \cite{BaBC85}. 
Similar double-peak structures are found even in similarly simple 
phenyl systems like m--fluoroanilin \cite{MoAB97}.

In this context we point out that
the structure factors of fully deuterated benzene
and hexafluorobenzene, that can be considered as unconnected subunits 
of OTP  are nearly indistinguishable 
from the one of OTP both showing the double peak 
at the right positions as well as the hump at $3$\,\AA$^{-1}$
and $6$\,\AA$^{-1}$ \cite{BaBC85}. 
Similar double-peak structures are found even in similarly simple 
phenyl systems like m--fluoroanilin \cite{MoAB97}.

The following simplified picture may therefore be developed:
The peak at $Q_2 \simeq 1.9\,$\AA$^{-1}$ is probably built up mainly by 
correlations within the phenyl rings, while the maximum at 
$Q_1 \simeq 1.45\,$\AA$^{-1}$ might be associated with correlations
between phenyl rings. 
The shoulder around $Q_0 \simeq 0.85\,$\AA$^{-1}$ is presumably the most 
direct manifestation of intermolecular correlations 
\cite{MoLR00,LeWa94,KuWi95}.
Its position is indeed compatible with the inverse van der Waals
radius of $r_W=3.7$\,\AA \cite{Bon64}.
We don't believe that the shoulder in OTP is due to pure orientational
correlations that can lead to a prepeak at low $Q$ \cite{ThSch99}.

\subsection{Square Root Singularity} \label{t-squarerootsingularity}
\begin{figure}[thb]
\begin{center}
\epsfxsize=120mm
{\epsffile[324 575 576 754]{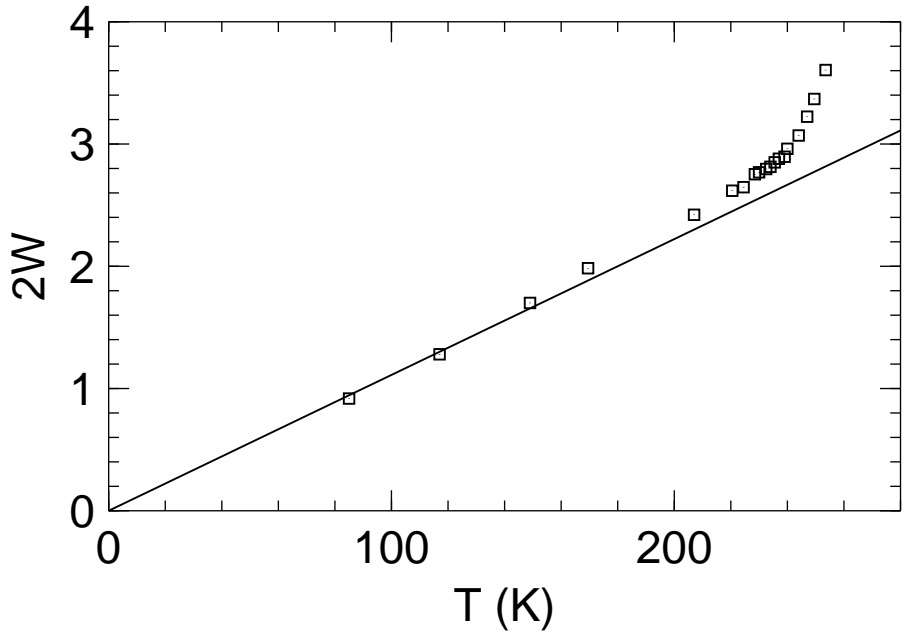}}
\end{center}
\caption{Exponent of the Debye--Waller factor of $^{57}$Fe in ferrocene 
dissolved in OTP as a function of temperature.
Based on values given in \protect\cite{VaFl79}.
The straight line is a linear fit to the lowest two values and zero
to give a consistent picture with the results discussed in 
section~\protect\ref{Vibrational_dynamics}.
Note a nearly linear temperature dependence at low temperatures  
followed by a strong anomalous decrease above T$_{\rm g}$.}
\label{DWF-VaFl79}
\end{figure}

The first experimental indications of an anomalous behaviour of the 
Debye--Waller factor $f_Q=\exp(-2W)$ have been observed long before MCT 
of the glass
transition was developed in \Mob--spectra of $^{57}$Fe in ferrocene dissolved
in butylphthalate \cite{RuZF76} and OTP \cite{VaFl79}.
The results for OTP at $Q=7.3$\,\AA$^{-1}$ are shown in 
figure~\ref{DWF-VaFl79}.
One notices a linear temperature dependence at low temperatures.
Above $T_{\rm g}$ a rather strong increase of $2W$ is observed.

\begin{figure}[thb]
\begin{center}
\begin{minipage}[t]{75mm}
\epsfxsize=75mm
{\epsffile[200 299 462 496]{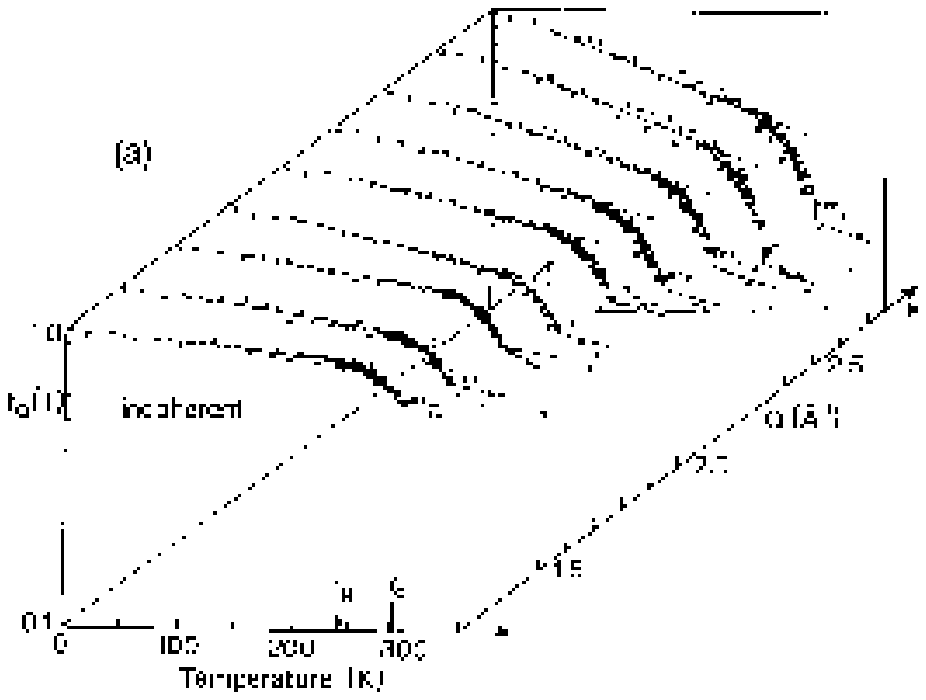}}
\end{minipage}
\hfill
\begin{minipage}[t]{75mm}
\epsfxsize=75mm
{\epsffile[65 458 467 776]{Bilder/T-Exp-squareroot/DWF-otp-d14-T.ps}}
\end{minipage}
\end{center}
\caption{(a (*)) Incoherent \Mob--Lamb factor of protonated OTP
for several $Q$--values \protect\cite{PeBF91}. 
A harmonic behaviour at low temperature is followed by
the onset of an anomalous decrease around \Tg.
Note the cusps in the temperature dependence at $T_{\rm c}$.
A square--root law fit gives a common $T_{\rm c} \simeq 290$\,K for all $Q$.
(b (*)) Temperature dependence of the Debye--Waller factor 
of the density correlations for various values of $Q$.
Lines are fits with a square root law leading to a critical 
temperature of $T_{\rm c} \simeq 290$\,K.
Towards higher temperatures a weak linear temperature dependence is
assumed.
The inset demonstrates the square root law in linearised form.}
\label{DWF-T}
\end{figure}

The predicted square root singularity (\ref{fq-eff}) was extensively tested
by neutron scattering experiments 
\cite{ToSW97b,BaFK89,DeZB91,PeBF91,BaFL95,BaFL96,Fuj93}.
At low temperatures $f_Q$ can be easily measured by the elastic intensity 
in a BS experiment or by determining the plateau of the $\Phi(Q,t)$ vs.\
$\log t$ curve.
In figure~\ref{DWF-T} we give results for temperature dependent incoherent
and coherent scattering.
For low temperatures we simply follow the harmonic evolution of the mean
square displacement ($\ln f_Q(T) \propto T$).
The anomalous decrease connected with the glass transition
starts somewhere around $T_{\rm g}$ in the
glassy phase due to anharmonic motion on a
microscopic length scale.
It is clear that the structural relaxation which at $T_{\rm g}$ is of the
order of seconds cannot be responsible for this effect and therefore
$T_{\rm g}$ as conventionally defined has nothing to do with it.

On approaching $T_{\rm c}$ the sharp transition predicted by the idealised
version of the theory is smeared out through thermally activated hopping
processes, which are present in any molecular liquid.
Clearly, between the caloric glass transition temperature \Tg\ and 
$T_{\rm c}$ structural relaxation is present.
When its time scale falls into the experimental window quasielastic
broadening appears and its total intensity can no longer be determined from
strict elastic scattering alone.
Instead, the integral over the whole $\alpha$ peak has to be taken.
The results are presented in figure~\ref{DWF-T} showing all features
predicted by (\ref{fq-eff}).

\begin{figure}[thb]
\begin{center}
\epsfxsize=120mm
{\epsffile[200 353 408 496]{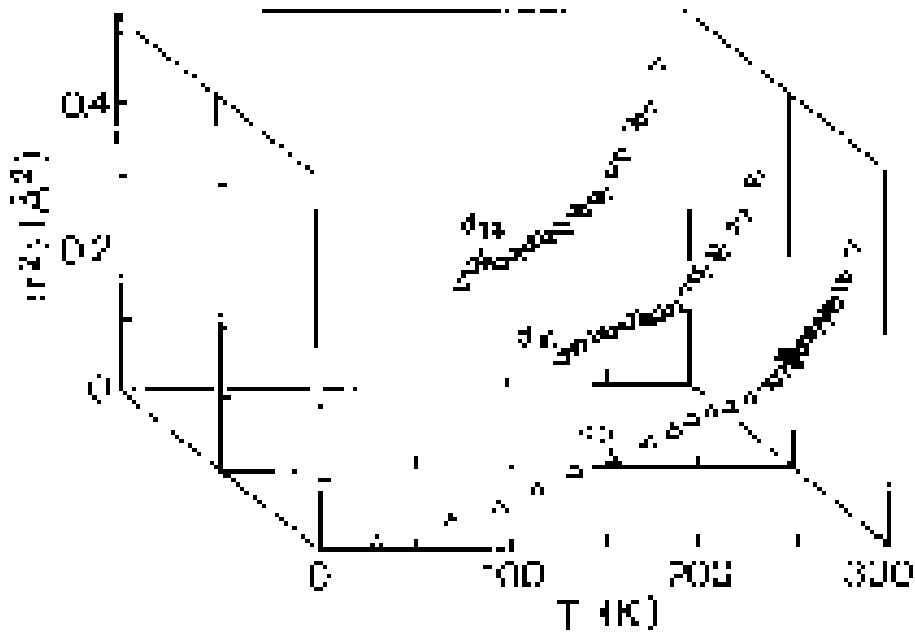}}
\end{center}
\caption{Mean square displacement of three isotopically substituted OTPs
obtained from elastic scans on IN13.
Within experimental error the three data sets are identical.
Taken from \protect\cite{DeZB91}.}
\label{MSD-isotops}
\end{figure}

An important question to answer is whether the anomaly of the Debye--Waller
factor may be caused by any intramolecular motion such as the librational
dynamics of the lateral phenyl rings relative to the central one.
This can be checked by measurements of the Debye--Waller factor in
isotopically substituted OTP systems, namely fully protonated OTP-$d_0$,
fully deuterated OTP-$d_{14}$ and selectively deuterated at the lateral
phenyl rings OTP-$d_{10}$.
Assume two types of motion, a center--of--mass (COM) motion and an
intramolecular (IM) motion of the lateral phenyl rings.
Then the central ring only contributes to the COM motion, while the lateral
rings contribute to both the COM and the IM motion.
From the total scattering cross sections (cf.\ table~\ref{cross-sections}) for
the nuclei participating in each motion it follows, that in particular in
OTP-$d_{10}$ neutron scattering is nearly insensitive to the lateral rings
(they are "hidden") as the dominat scattering steems from the central ring
protons.
Consequently, if the $\beta$--process, which is responsible for the anomalous 
decrease of $f_Q(T)$ is due to intramolecular motion, it should be much less 
pronounced in OTP-$d_{10}$.
As the $Q$ dependence of the DWF is essentially the same for the 
three isotopically substituted OTPs \cite{DeZB91} and can be approximated 
by a Gaussian $f_Q=\exp[-\frac{1}{3}Q^2 \langle r^2 \rangle]$ we can deduce 
the mean square displacement shown in figure~\ref{MSD-isotops}.

From the comparison of the observed anomaly in OTP-$d_0$ and OTP-$d_{10}$, 
we can exclude any intramolecular phenyl ring dynamics as the dominant 
mechanism for the $\beta$ process.
The data thus support MCT proposing a center--of--mass motion as precursor
of the glass transition.
Note, that although the values for $\langle r^2 \rangle$ of OTP-$d_{14}$ 
conicide with those of OTP-$d_0$ and OTP-$d_{10}$, the comparison is 
biased as in OTP-$d_{14}$, which is a dominant coherent scatterer, we do not 
observe the single particle mean square displacement.
In addition, with a higher $Q$--resolution one can clearly observe
oscillations in $f_Q$ (see figure~\ref{fq-hq}) leading to deviations from the 
simple Gaussian behaviour.

The anomalous behaviour and cusp of the nonergodicity parameter $f_Q$ has 
also been observed in computer simulations \cite{LeWa93,LeWa94}.
The cited critical temperature $T_{\rm c}=280$\,K is in reasonable 
agreement with the one determined from neutron scattering.
X-ray results allowed for the identification of a  
temperature $T_x \simeq 290$\,K and a cusp from the
temperature dependence of a high energy propagating 
soundlike mode \cite{MaMR98}.

\subsection{Critical Correlations}
On BS spectrometers only indirect information can be gained about the
existance of a fast process.
In order to study the $\beta$ process directly shorter times have to be
accessed which is possible on TOF spectrometers.
Whether data are analyzed in frequency or time is a question of 
convenience.

\begin{figure}[thb]
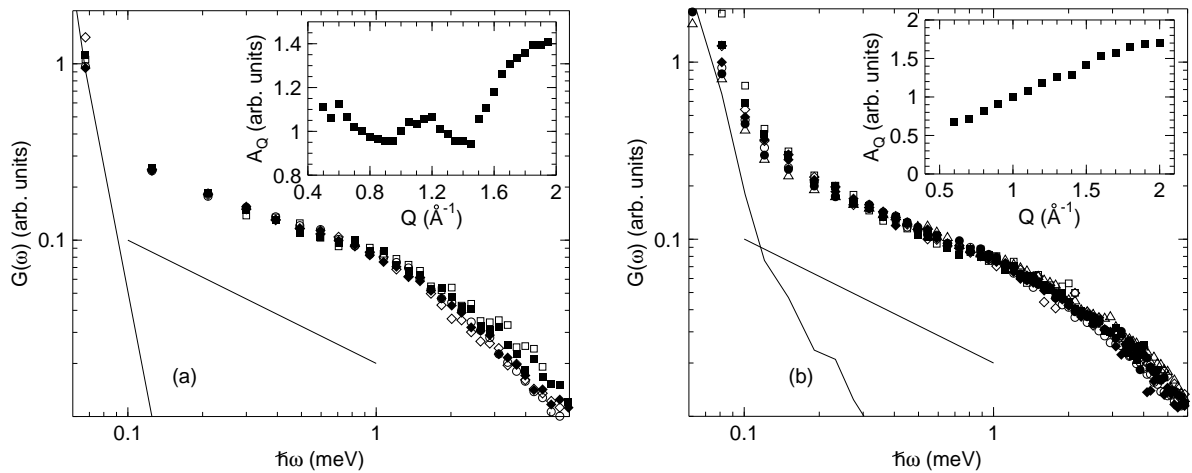

\begin{center}
\begin{minipage}[t]{75mm}
\epsfxsize=75mm
{\epsffile[54 356 537 746]{Bilder/T-Exp-beta/Factorisation-coh.ps}}
\end{minipage}
\hfill
\begin{minipage}[t]{75mm}
\epsfxsize=75mm
{\epsffile[54 356 535 746]{Bilder/T-Exp-beta/Factorisation-inc.ps}}
\end{minipage}
\end{center}
\caption{Test of the factorization property for coherent
at 293\,K (a) and incoherent scattering data at 324\,K and 60\,MPa 
(b (*)).
The rescaled intensities $S(Q,\omega)/S(Q)A_Q=G(\omega)$ 
for different wave numbers ranging from 0.4\,\Ar\ to 2\,\Ar\
are shown to fall onto a common curve in the energy range
between 0.1\,meV and 1\,meV.
For comparison a power law $\omega^{1-a}, a=0.3$
is shown which is the proposed line shape from theory.
The steeper lines are the measured resolution functions and set the 
lower bounds.
The insets show the scaling factors $A_Q$ which are proportional to the
\brx\ amplitude $h_Q$.
Note for coherent scattering the two minima corresponding to the prepeak and 
the first maximum of the static structure factor.}
\label{Factorisation}
\end{figure}

One remarkable prediction of MCT is the factorization property of the
\brx.
In a certain frequency and temperature or pressure range, all observables
are expected to have the same spectral distribution.
This implies that the dynamic structure factor \Sqw\ factorises
into a $Q$-- and a $\omega$ dependent part as a direct consequence of the
cage effect.
Figure~\ref{Factorisation} demonstrates the validity of the factorization
property for incoherent and coherent scattering: the rescaled intensities
$S(Q,\omega)/S(Q)A_Q=G(\omega)$ fall onto a common master curve for
different wave numbers $Q$ \cite{ToSW97b,WuKB93,KiBD92}.

\begin{figure}[thb]
\begin{center}
\begin{minipage}[t]{75mm}
\epsfxsize=75mm
{\epsffile[208 366 576 811]{Bilder/T-Exp-beta/Sqt-coh.ps}}
\end{minipage}
\hfill
\begin{minipage}[t]{75mm}
\epsfxsize=75mm
{\epsffile[209 366 576 811]{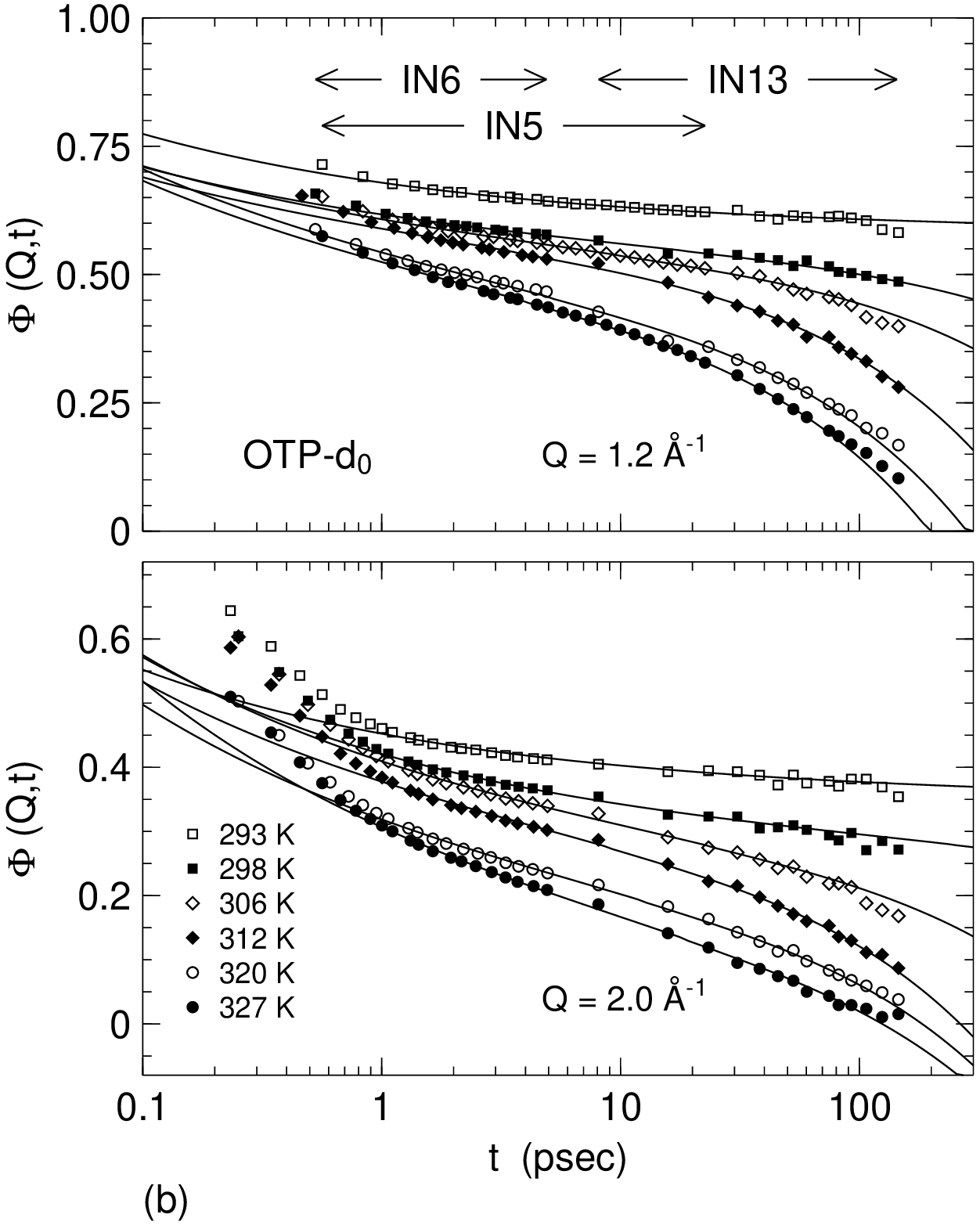}}
\end{minipage}
\end{center}
\caption{Normalized intermediate scattering functions $\Phi(Q,t)$ 
for density (a (*)) and tagged particle fluctuations 
(b) (taken from \protect\cite{PeWu95}) for two different 
wave vectors at various temperatures.
The ranges of the instruments used are indicated.
Solid lines are fits with the MCT scaling law (\protect\ref{beta-scaling-law}).
The line shape parameter $\lambda=0.77$ was kept 
fixed for the incoherent data while for the coherent data is was fitted freely.
The decay of the correlations is described over several 
decades.
For the coherent data we included KWW fits (\protect\ref{KWW}) to the long 
time tail of the correlation functions (dotted lines).}
\label{Sqt-T}
\end{figure}

For a detailed study of the \brx\ we Fourier deconvolute the dynamic
structure factors as the quasielastic broadening is of the order of the 
instrument's resolution. 
For an analysis in time domain we combine data from different 
spectrometers with overlapping time ranges to enlarge the dynamic range.
Some results for the tagged--particle \cite{WuKB93} and the 
density correlations \cite{ToSW97b,BaFL95} at
different temperatures are shown in figure~\ref{Sqt-T}.
In Ref.~\cite{KiBD92} a second Fourier transform was hereafter performed to
analyze the scaling behaviour of the dynamic suceptibility 
$\chi^{\prime\prime}(\omega)$ around its minimum.

\begin{figure}[thb]
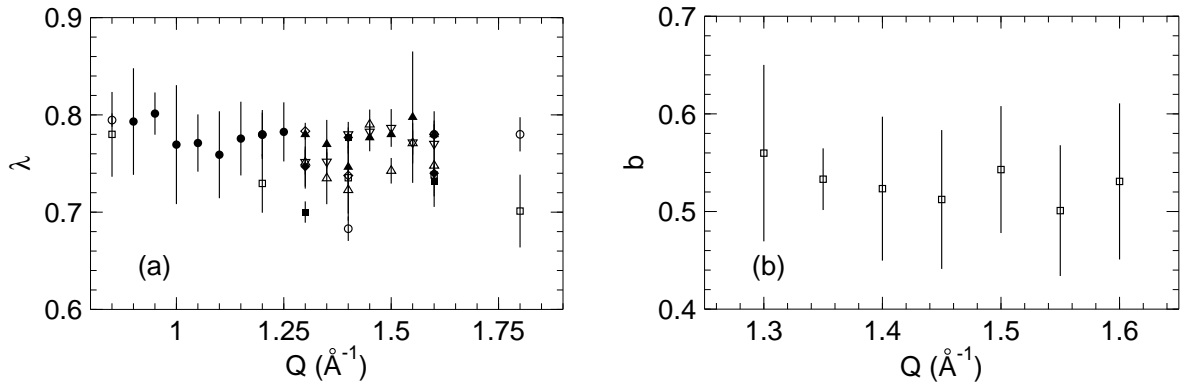

\begin{center}
\begin{minipage}[t]{75mm}
\epsfxsize=75mm
{\epsffile[312 575 576 754]{Bilder/T-Exp-beta/lambda-fig.ps}}
\end{minipage}
\hfill
\begin{minipage}[t]{75mm}
\epsfxsize=75mm
{\epsffile[313 575 576 753]{Bilder/T-Exp-beta/b-VS.ps}}
\end{minipage}
\end{center}
\caption{(a (*)) Line shape parameter $\lambda$ from fits to the density 
correlation functions (figure~\ref{Sqt-T}) which have been obtained by 
combining up to four different instruments (IN5,IN11,IN13,IN16).
The values lie between 0.7 and 0.8 and scatter statistically around 
a mean value of 0.77 which is consistent with the one determined by
other means.
Note, no significant temperature (different symbols) or $Q$--dependence
is observed in accordance with the prediction of MCT.
(b (*)) Von Schweidler exponent $b$ for density correlations from fits to
the master curves. 
The data scatter statisctically around a mean of 0.53 which is consistent
with $\lambda=0.77$.}
\label{lambda}
\end{figure}

For fits with (\ref{beta-scaling-law}) we use the tabulated expansion
coefficients \cite{Got90} of the scaling function $g_{\lambda}(t/t_{\sigma})$
which is completely determined by the exponent parameter $\lambda$.
Meaningful four-parameter fits are often not warranted by the dynamic range
covered, here in particular for the incoherent data shown in 
figure~\ref{Sqt-T} on the right.
However, as the line shape is not very sensitive to small variation of
$\lambda$ it was kept fixed at the value $0.77$ previously derived from a
scaling analysis of the long time asymptote of the correlation function
$\Phi(Q,t)$ \cite{PeBF91,BaDF91}. 
This value is compatible with the one from the power-law fit to the
viscosity $\eta(T)/T$ in figure~\ref{vis-otp-T-MCT} \cite{BaDF91}. 
For the coherent data by the combination of time--of--flight (IN5),
spin--echo (IN11) and back--scattering (IN13, IN16) data a time range 
of now more than four decades is covered as can be seen in the left part 
of figure~\ref{Sqt-T}.
IN5, IN11 and IN13 have a broad overlap and their $\Phi(Q,t)$ 
show very good agreement after the absolute scales were adjusted by
typically 5\,\%.
IN16 data couldn't be obtained in absolute units and were matched
to IN11 or IN13 data.
Note, a combination of spin--echo data with TOF and BS data has rarely been 
attempted \cite{KnMF88} possibly due to the different weighting of 
coherent and incoherent scattering in (\ref{sep}).
However, when the coherent scattering largly dominates the signal,
for example around the static structure factor maximum
such a combination is reasonably possible.
Now a four-parameter fit seems more promising and the results for
the line shape parameter $\lambda$ is shown in figure~\ref{lambda}.
The values for different temperatures $T$ and wave numbers $Q$ indeed scatter
statistically around the mean value $\bar{\lambda}=0.77$
validating our previous analysis \cite{ToSW97b,WuKB93}. 


The \brx\ scaling is expected to hold only in a wave number and temperature
dependent time window $[t_{\rm min}(Q,T),t_{\rm max}(Q,T)]$ 
around the cross--over time $t_{\sigma}$.
As the scaling law is a long--time expansion where details of the vibrational
motion are neglected it should fail to hold for times shorter as or of the 
order of a typical vibrational period, say a psec.
Towards longer times the \arx\ which itself strongly depends on $Q$ and $T$ 
starts to merge with the late $\beta$ regime and renders the
identification of a fit range rather ambiguous. 
As the predictions of MCT are based on expansions around $T_c$ we expect the
scaling law to hold best close to the critical temperature. 
There, on the other hand, additional transport processes not included in
MCT's elementary version may obscure it.

From figure~\ref{Sqt-T} we can see that the fits with (\ref{beta-scaling-law})
are able to decribe the decay of tagged--particle as well as density
correlations over several decades in time.
It might come as a suprise that the correlators at the highest temperatures
are described down to about $f_Q/2$ as the leading order solutions
$\Phi(Q,t)$ (\ref{beta-scaling-law}) have been obtained by an expansion in
the distance from the plateau $|\Phi(Q,t)-f_Q|$.

\begin{figure}[thb]
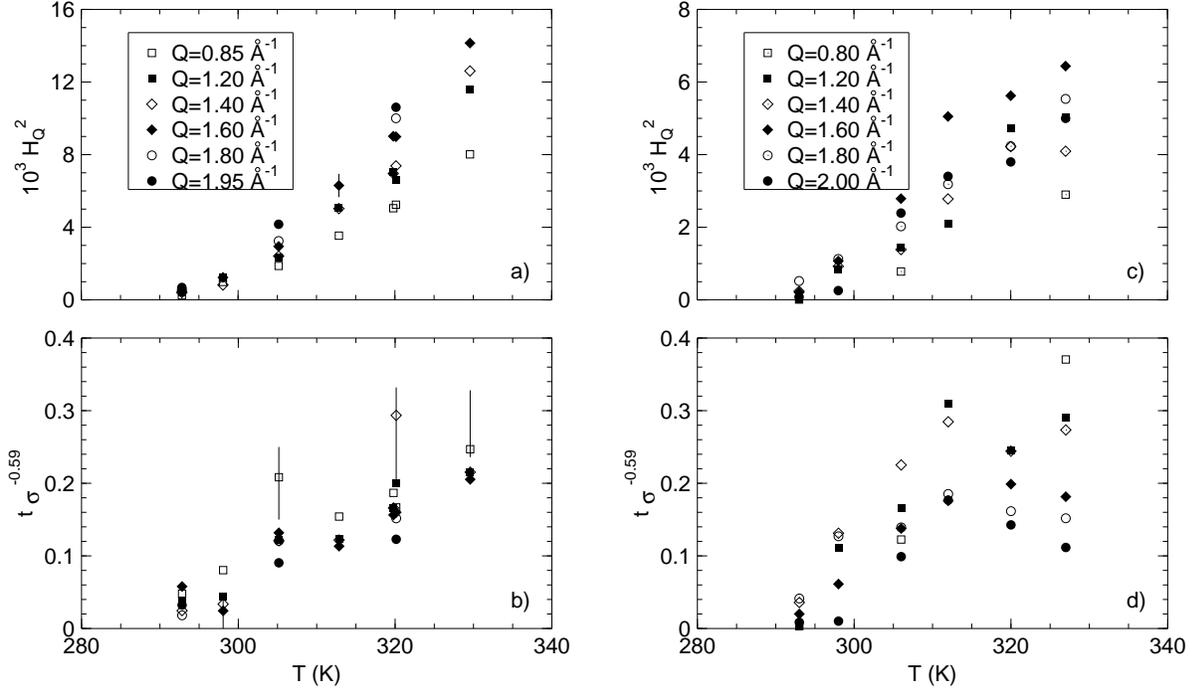

\begin{center}
\begin{minipage}[t]{75mm}
\epsfxsize=75mm
{\epsffile[89 300 463 755]{Bilder/T-Exp-beta/hq_ts-T-coh.ps}}
\end{minipage}
\hfill
\begin{minipage}[t]{75mm}
\epsfxsize=75mm
{\epsffile[89 300 463 755]{Bilder/T-Exp-beta/hq_ts-T-inc.ps}}
\end{minipage} 
\end{center}
\caption{Temperature dependence of the amplitude $H_Q$ and the cross--over
time \t_s\ as obtained from fits to the correlation functions $\Phi(Q,t)$.
The parameters are plotted such that a linearized $T$ dependence is expected 
according to (\ref{hq}).
Coherent (a,b) as well as incoherent (c,d) data extrapolate consistently to 
$T_c \simeq 290$\,K.
While the absolute values of both, the incoherent and the coherent cross--over
times \t_s\ are quite the same the amplitudes $H_Q$ differ.
Taken from \protect\cite{ToSW97b,WuKB93}.}
\label{hq-ts-T}
\end{figure}

The physical meaning of such fits has to be judged from the comparison of
the obtained parameters with the theoretical predictions (\ref{hq}).
Such comparisons should comprise not only temperature $T$ and / or 
pressure $P$ but also the wave number $Q$ dependence.
The temperature dependence of $h_Q$ and \t_s for tagged--particle and density
correlations is show in figure~\ref{hq-ts-T} confirming the asymptotic
prediction (\ref{hq}).
Extrapolation of the $T$--dependence gives consistently $T_c \simeq 290$\,K
for {\it all} data sets coinciding with the critical temperature 
determined from the square root anomaly of the Debye--Waller factors and 
the divergence of the viscosity.

\begin{figure}[thb]
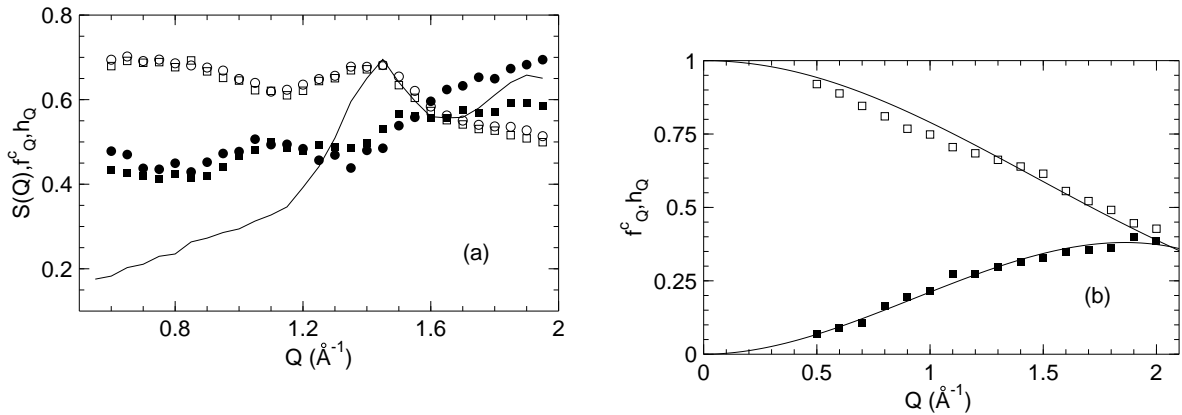

\begin{center}
\begin{minipage}[t]{75mm}
\epsfxsize=75mm
{\epsffile[88 514 455 754]{Bilder/T-Exp-beta/fq_hq_Q-coh.ps}}
\end{minipage}
\hfill
\begin{minipage}[t]{75mm}
\epsfxsize=75mm
{\epsffile[82 542 452 783]{Bilder/T-Exp-beta/fq_hq_Q-inc.ps}}
\end{minipage} 
\end{center}
\caption{(a (*)) Wave number dependence of the non--ergodicity 
parameter $f_Q^c$
(open symbols) and the \brx\ amplitude $h_Q$ (closed symbols) for the 
density fluctuations compared to the static structure factor (line).
The results stem from the fits to the correlation function (squares)
and to the DWF (circles).
$f_Q^c$ varies in phase, $h_Q$ out of phase with the static structure
factor.
(b (*)) Same as in (a) for the tagged particle fluctuations.
The lines are predictions of theory redrawn from \protect\cite{BeGS84}
with the abscissa scaled to a hard sphere diameter 
$\sigma \simeq 7.5$\,\AA\ which corresponds to the 
prepeak of the static structure factor $S(Q)$.}
\label{fq-hq}
\end{figure}

Also the wave number dependence shown in figure~\ref{fq-hq} at least
qualitatively agrees with results from numerical solutions of mode--coupling
equations \cite{BeGS84,FuHL92}.
For incoherent scattering $f_Q^{s}$ reflects the single particle oscillations
and it corresponds closely to a Gaussian distribution \cite{PeBF91,BaDF91}.
For coherent scattering the plateau height $f_Q^c$ and the amplitude
$h_Q$ are obtained from a fit by (\ref{fq-eff}) to the DWF shown in
figure~\ref{DWF-T} and from a fit of (\ref{beta-scaling-law}) to $\Phi(Q,t)$
shown in figure~\ref{Sqt-T}.
Both results are nearly identical and oscillate in and out of phase
with the static structure factor $S(Q)$, respectively.
Remember, the scaling factors $A_Q$ in the insets of 
figure~\ref{Factorisation} are proportional to $h_Q$, which independently
gives strong support for our findings.


When the abscissa of the theoretical curves \cite{BeGS84,SjGo89} is scaled
to a hard sphere diameter of $\sigma \simeq 7.5$\,\AA\ even a more 
quantitative agreement is found for both, the coherent and incoherent results.

The factorization property has also been verified for the
coherent intermediate scattering function in an appropriate
time regime \cite{ToSW97b}.
According to (\ref{beta-scaling-law}) the quantity $(\Phi(Q,t)-f_Q^c)/h_Q$
is independent of $Q$.
This has been tested using the fit parameters $f_Q^c$ and $h_Q$
for wave numbers between 0.8\,\AA$^{-1}$ and 2\,\AA$^{-1}$.
The cross--over time $t_{\sigma}$ and the time range
over which the factorization is fulfilled decrease with increasing
temperature.
This factorization is only possible if the line--shape parameter
$\lambda$ and the cross--over time $t_{\sigma}$ of the $\beta$--correlator
are both independent of $Q$.
A fit of $g_{\lambda}(t/t_{\sigma})$ to the combined data
gives $\lambda=0.78$ in excellent accord with the values determined 
by other means \cite{ToSW97b}.

\begin{figure}[thb]
\begin{center}
\epsfxsize=120mm
{\epsffile[240 475 576 755]{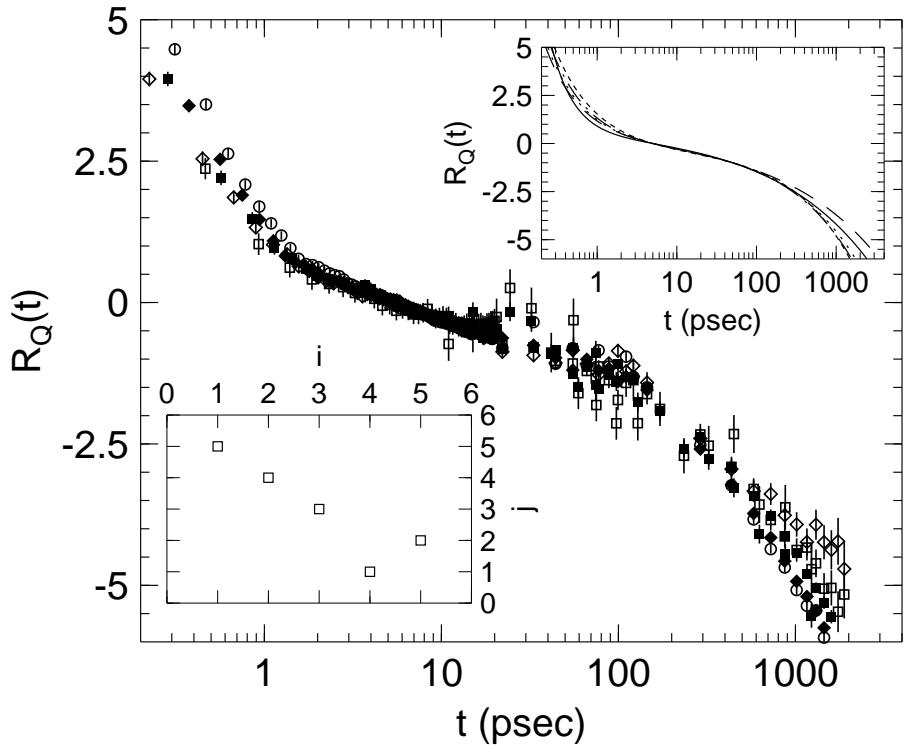}}
\end{center}
\caption{(*) Time dependence of the ratio $R_Q(t)$ 
given in (\protect\ref{R-t}) demonstrating the factorisation 
property for T=293\,K.
In the upper right inset $R_Q(t)$ for fitted curves is shown.
For the lower left inset see text.}
\label{Rexp-t}
\end{figure}

Here we want to follow another approach.
As already mentioned in section~\ref{Mode_Coupling_Theory} in the context
of equation (\ref{beta-scaling-law}), the whole time dependence of any 
arbitrary time--correlation function $\Phi_Q(t)$
which couples to density fluctuations is given by the $Q$--independent 
function $g_{-}(t/t_{\sigma}) (\sigma < 0)$.
Introducing the following function \cite{GlKo99}
\begin{equation} 
\label{R-t}
R_Q(t)=
\frac{\Phi_Q(t)-\Phi_Q(t^{\prime})}
     {\Phi_Q(t^{\prime\prime})-\Phi_Q(t^{\prime})}
\end{equation}
one can test this prediction even without fitting with 
(\ref{beta-scaling-law}).
Here $t^{\prime} \neq t^{\prime\prime}$ are abitrary times in the
\brx\ regime.
From (\ref{beta-scaling-law}) we immediately recongnize that the function 
$R_Q(t)$ is independent of the correlator, {\it i.\ e.\ } of $Q$.
To test this property we take the density correlation function at 293\,K of 
figure~\ref{Sqt-T}.
For the five different $Q$ 
(0.85\,\AA$^{-1} \le$ 1.8\,\AA$^{-1}$) we determined $R_Q(t)$ choosing
arbitrarily $t^{\prime}=5$\,ps and $t^{\prime\prime}=50$\,ps.
Figure~\ref{Rexp-t} shows that in the \brx\ regime the different 
$R_Q(t)$ indeed collapse onto a master curve demonstrating the 
validity of the factorization property.
This is not a trivial result as outside the 
\brx\ regime the curves fan out.

One can even try to go further.
Regarding the next order corrections (\ref{early-correct}) 
and (\ref{late-correct}) it can easily be shown that for the 
function $R_Q(t)$ the order of the corrections in the early \brx\ regime and 
in the late \brx\ regime are related in the following way:
Counting from top to bottom the first (second, $\ldots$) curve
at short times (label $i$) is also the first (second, $\ldots$) 
at large times (label $j$) which means that $i=j$.
Both times have to be taken slightly outside the range where 
(\ref{beta-scaling-law}) holds.
Counting at $t\simeq1$\,ps and at $t\simeq1000$\,ps gives us the 
function $j(i)$ displayed in the inset of figure~\ref{R-t}.
Although only five values are considered here there seems to be no clear 
correlation $i=j$ but rather an anticorrelation.
This is in contrast to results from molecular dynamic simulations 
\cite{GlKo99}.
As the experimental data scatter quite large we fitted the 
correlation functions with a power law similar to the (\ref{beta-scaling-law}) 
however without any constraints to have a good smooth representation 
of the data in the time range from about 0.2\,ps to about 2000\,ps.
We then calculate $R_Q(t)$ for the fitted curves which are displayed in the 
upper right inset of figure~\ref{Rexp-t}.
Identical results are obtained concerning the factorization 
and the function $j(i)$ as from the experimental data. 
Here again the limitations of neutrons scattering show up:
To tackle this problem properly, high energy transfers (very short times) 
in combination with high energy resolution (long times) 
(preferably on one spectrometer) would be needed in addition 
to a good $Q$ resolution and a large wave number range 
to have a considerable amout of different correlation functions.

The fast dynamics has also been experimentally observed by depolarized
light scattering \cite{CuHL98,StPG94,CuLD97}.
Comparisons with idealized MCT yields a critical temperature
$T_{\rm c} \simeq 288 - 292$\,K with line shape parameters of $a=0.33,
b=0.65, \lambda=0.7$ \cite{StMP93,StPG94} and equally
$T_{\rm c} \simeq 288 - 292$\,K with $a=0.32, b=0.61, \lambda=0.72$
\cite{CuLD97}.
The numbers for $T_{\rm c}$ and $\lambda$ are consistent within the
experimental uncertainies with the values used to describe the neutron
scattering results.
Depolarized light scattering has a rather large dynamic range
covering about 4 decades and high statistics such that a comparison with the
extended MCT became possible \cite{CuLD97}.
The analysis yields $T_{\rm c}\simeq 276$\,K with $a=0.3, b=0.54,
\lambda=0.76$.

The dynamics below the critical temperature $T_{\rm c}$ was investigated
less extensively.
Below $T_{\rm c}$ the idealized MCT predicts a nontrivial behaviour of the
generalized susceptibility $\chi^{\prime\prime}(\omega)$:
a white--noise spectrum at very low frequencies
$\chi^{\prime\prime}(\omega) \propto \omega$ should cross--over to
a power law $\chi^{\prime\prime}(\omega) \propto \omega^a, a<0.4$
resulting in a "knee" in $\chi^{\prime\prime}(\omega)$ at intermediate
frequencies.
Rather counterintiutive is the nontrivial prediction for the temperature
dependence of the position of the "knee"--frequency:
$\omega_{\beta}^{-1}$ should {\it increase} with increasing temperature and
diverge at $T_{\rm c}$.
In our neutron scattering experiments in the temperature range
$200 < T < T_{\rm c}$ no indication for such a cross--over is
observed in the frequency range $\omega \gtrsim 0.1$\,meV (25\,GHz).
The same observation was made in careful light scattering experiments
\cite{GaSP99}.
However, this does not necessarily mean that the "knee" does 
not exist as the influence
of the hopping processes {\it i.\ e.\ } its strength is unkown at present.
If the hopping is sufficiently small the ideal spectrum is rather undisturbed
and the "knee" should appear, however, if the hopping term is large
the inital $\alpha$ decay can make it dissapear.

So far, the only support of the existence of such a diverging time
scale below $T_{\rm c}$ comes from PCS experiments on a hard
sphere colloidal liquid \cite{MeUn93a,MeUn93b,MeUn94} and on
a polymer micronetwork colloidal liquid \cite{BaFB97}
where no microscopic dynamics and no hopping processes disturb the
$\beta$--process.

\subsection{$\alpha$--Relaxation}\label{t-arx}
\begin{figure}[thb]
\begin{center}
\epsfxsize=155mm
\epsfysize=60mm
{\epsffile[72 589 567 753]{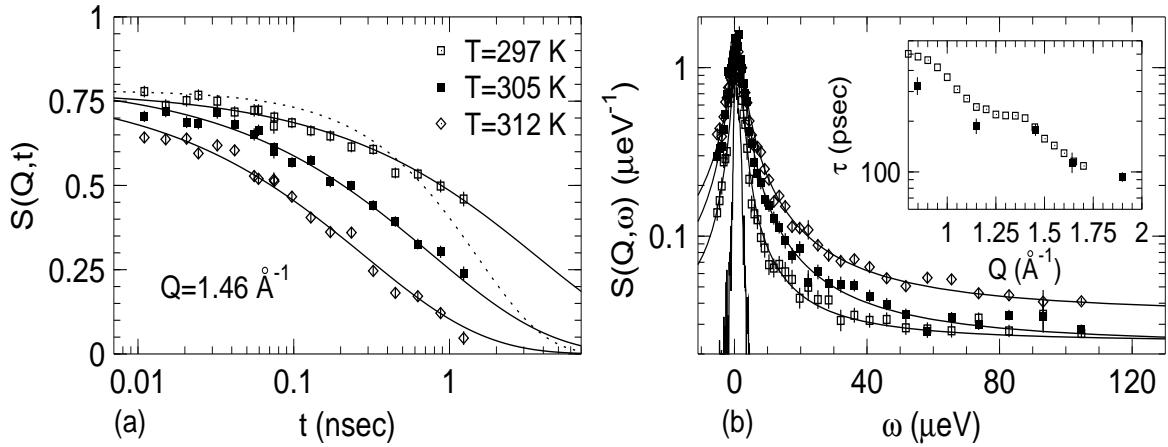}}
\end{center}
\caption{(a (*)) High resolution spectra of OTP-$d_{14}$ at 
$Q=1.46$\,\AA$^{-1}$
for several temperatures as measured on the SE instrument IN11C
using the multidetector.
The fit curves demonstrate that the line shape of the structural relaxation
is well described by a KWW function according to (\ref{KWW}) with 
$\beta \simeq 0.57$ and not by a simple Debye process drawn for 
comparison (dashed line).
(b (*)) Dynamic structure factor $S(Q,\omega)$ at 320\,K for
$Q=0.85$\,\AA$^{-1}$ ($\square$),
$Q=1.45$\,\AA$^{-1}$ ($\blacksquare$) and
$Q=1.90$\,\AA$^{-1}$ ($\lozenge$)
as measured on the BS instrument IN10B with a heated monochromator.
Lines are fits with a Fourier transformed Kohlrausch function with
$\beta_Q=0.57$ fix for all $Q$.}
\label{Sqt-IN11C}
\end{figure}

The \arx\ can be directly studied e.\ g.\ either by energy resolved BS or NSE
experiments.
For instance in figure~\ref{Sqt-IN11C} we show the coherent scattering 
function $\Phi(Q,t)$ of OTP-$d_{14}$ as measured on the SE instrument
IN11C using the multidetector set--up.
For comparison, $S(Q,\omega)$ as measured on IN10B with a heated
monochromator is shown on the right hand side of figure~\ref{Sqt-IN11C}
for several wave numbers $Q$.
In both cases the line--shape cannot be described by a single Debye 
function but can be well fitted by a Kohlrausch function.


The validity of viscosity scaling and time--temperature superposition is
still an ongoing debate.
We have seen that it is a very good first approximation at least in
a narrow temperature range.
Statistics and the limited dynamic range of one instrument hardly allows for
a serious test.
However, consistent results with acceptable error bars can be obtained by
combination of measurements from different temperatures (or pressures).
Assuming that viscosity scaling holds over the covered $T$--range 
we employ the following interative procedure
to construct master curves $\hat\Phi(Q,\hat t)$:
We use published viscosities \cite{McUb57,GrTu67a,LaUh72,CuLU73,SchKB98}
to rescale experimental times to
\begin{equation}\label{t-T-resc}
\hat t = t[\eta(T_0)/T_0]/[\eta(T)/T] 
\end{equation}
with an arbitrary normalization $T_0=290$\,K.
From Kohlrausch fits to $\hat\Phi(Q,\hat t)$ we obtain first approximations
for the line shape parameter $\beta_Q$ and the relaxation time $\tau_Q(T_0)$.
By means of this $\tau_Q(T_0)$ the unscaled $\tau_Q(T)$ of the individual
temperatures are calculated and are then kept fixed together with
the line shape parameter $\beta_Q$ in fits to the 
individual $\Phi(Q,t)$.
The resulting set of amplitudes $A_Q$ is used to obtain reduced correlators
$\hat\Phi(Q,\hat t)/A_Q$ and the procedure is repeated.
It is numerically stable and converges after a few 
iterations \cite{PeBF91}.
To get the temperature dependence of each parameter the individual 
$\Phi(Q,t)$ at each temperature can then be fitted
with one free parameter while the other two are kept fixed at the
values determined from the iterative procedure.

\begin{figure}[thb]
\begin{center}
\begin{minipage}[t]{75mm}
\epsfxsize=75mm
{\epsffile[46 471 467 789]{Bilder/T-Exp-alpha/Mqt-alpha-coh-T.ps}}
\end{minipage}
\hfill
\begin{minipage}[t]{75mm}
\epsfxsize=75mm
{\epsffile[200 299 458 496]{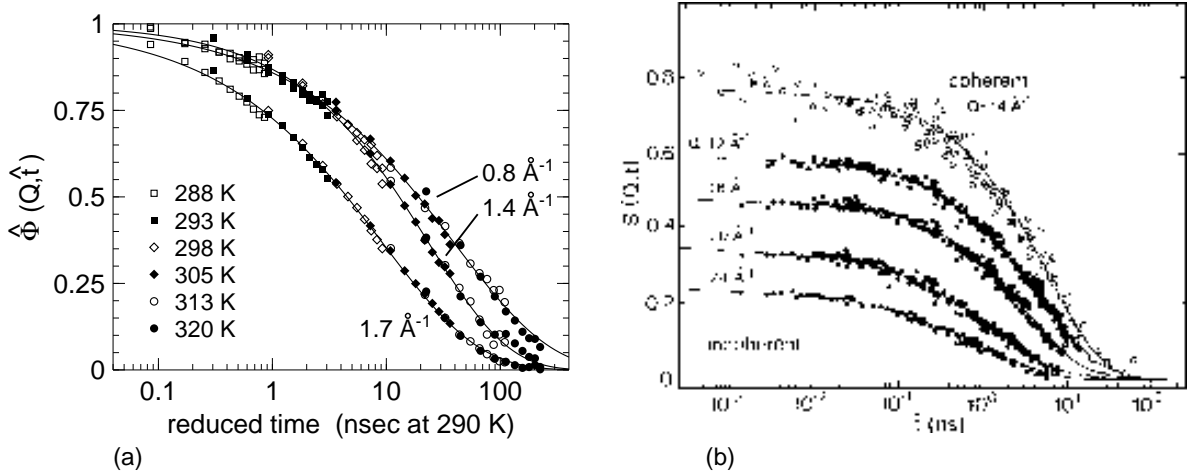}}
\end{minipage}
\end{center}
\caption{(a) Master curves $\hat \Phi(Q,\hat t)/A_Q(T)$ for the 
$\alpha$--relaxation of
the density fluctuations at different temperatures,
obtained by rescaling times with viscosity $\eta(T)/T$.
From \protect\cite{ToWS98}.
(b) Typical master curves obtained using the scaling assumption
for the tagged particle fluctuations. From \protect\cite{PeBF91}.}
\label{Mqt-T}
\end{figure}

The results in figure~\ref{Mqt-T} confirm the validity of viscosity scaling 
for temperature dependent tagged--particle and density correlations.
Note, that the scaling is a necessary but not sufficient condition for the
applicability of MCT.


The temperature dependence of the relaxation time $\tau_Q$ of the density
correlations at the structure factor maximum is shown in figure~\ref{tau-otp}.
Combination of data from several spectrometers using the iterative procedure 
described above makes it possible to cover nearly five decades in relaxation 
time.
The structural relaxation time follows nicely the viscosity confirming the
validity of viscosity scaling.
\begin{figure}[thb]
\begin{center}
\epsfxsize=120mm
{\epsffile[44 471 467 789]{Bilder/T-Exp-alpha/tau-otp-d14-Tq14.ps}}
\end{center}
\caption{(*) Temperature dependence at ambient pressure of the \arx\ 
time $\tau_Q$ at the structure factor maximum $Q=1.4$\,\Ar\ 
obtained from the scaling analysis of BS ($\square$), NSE 
($\lozenge$: single detector, $\blacklozenge$: multidetector)
and TOF ($\blacksquare$) data.
The solid line is the scaled viscosity interpolated
from literature data \protect\cite{McUb57,GrTu67a,LaUh72,CuLU73}.
The inset shows $\tau_Q^{-1/\gamma}$ vs.\ $T$ from BS and TOF data
at three different $Q$--values using $\gamma=2.59$.
A linear behavior extrapolating to zero at $T_c \simeq 292$\,K
is observed for all $Q$ in accordance with the prediction of MCT.}
\label{tau-otp}
\end{figure}


In the inset of figure~\ref{tau-otp} we test the divergence of the $\alpha$
scale predicted by MCT (\ref{alpha-scale}).
Assuming $|\sigma| \propto (T-T_c)$ a plot of $\tau_Q^{-1/\gamma}$ vs.\ $T$
should produce a straight line extrapolating to zero at $T_c$.
Indeed, a linear behaviour is observed for all wave numbers studied
extrapolating to zero at $T_c \simeq 292$\,K.
This also supports the $\alpha$--scale universality.
The deviations from this law in a narrow temperature range close to
$T_{\rm c}$ may be due to thermally activated hopping processes not
contained in MCT's ideal formulation and smearing out the sharp transition.
As a consequence, testing the power law only in a narrow range around the
critical point could give a wrong idea of, e.\ g., $\gamma$.

\begin{figure}[thb]
\begin{center}
\begin{minipage}[t]{75mm}
\epsfxsize=75mm
{\epsffile[313 324 576 746]{Bilder/T-Exp-beta/VS-coh.ps}}
\end{minipage}
\hfill
\begin{minipage}[t]{75mm}
\epsfxsize=75mm
\epsfysize=120mm
{\epsffile[200 245 363 496]{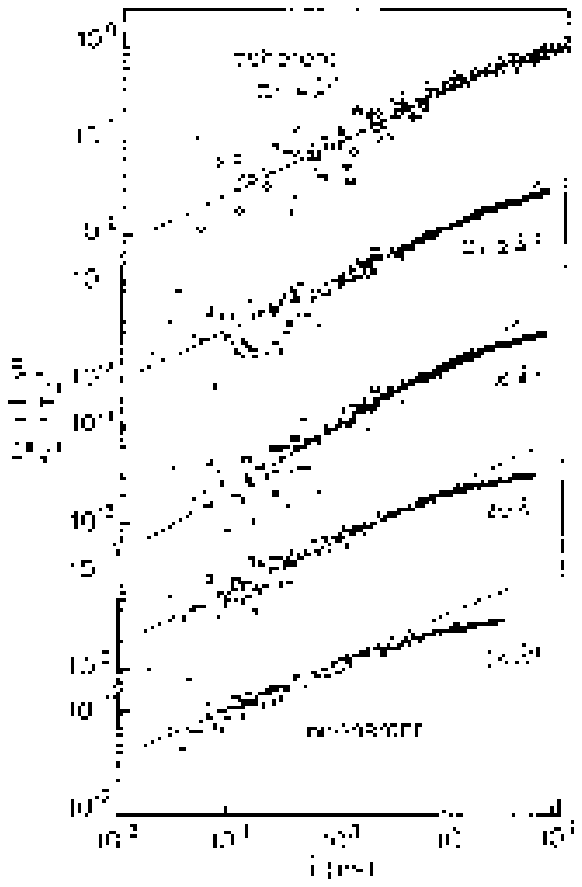}}
\end{minipage} 
\end{center}
\caption{Fits of the von Schweidler law (\protect\ref{vSchw})
to some linearized master curves for coherent scattering (a (*)) and
incoherent scattering (b) (from \protect\cite{PeBF91}) in order to show the 
fractal time dependence $t^b$.
The slopes $b$ scatter around a mean value of 0.53 for both coherent and
incoherent data.
This value is in excellent accord with the value imposed by the line shape 
parameter $\lambda$=0.77.}
\label{VS}
\end{figure}

The cross--over from the $\beta$-- to the $\alpha$--process is described
by the von Schweidler law (\ref{vSchw}).
Thus, it should be detectable in the short time behaviour of the 
$\alpha$--process. 
By plotting the master curves in the form 
$\log [1-\Phi(Q,\hat t)/f_Q]\,vs.\, \log \hat t$ 
this should yield straight lines with slope $b$ 
for short rescaled times.
The result for coherent and incoherent data is shown in figure~\ref{VS}.

Both data sets show a von Schweidler behaviour
at short times while at long times deviations occur.
In accordance with the theory the values of the von Schweidler exponent $b$ 
do not show any systematic $Q$ dependence (see figure~\ref{lambda}).
They scatter around a mean value of ${\bar b}=0.53 \pm 0.03$.

In order to investigate the $Q$ dependence of the parameters describing
the density correlations of the \arx\ a high $Q$ resolution is necessary.
For the question studied here spin--echo seems less appropriate than high
resolution multidetector instruments as the $Q$ resolution is implicitly
limited by the transmision band of typically 
$\Delta \lambda/\lambda \simeq 10$\,\%.
Note, increasing the $Q$ resolution by use of a narrower bandwidth
or a graphite monochromator leads to an unacceptable reduction in
intensity in particular if the experimental beam time comprises several
$Q$ and $T$ scans.

The big advantage of spin--echo, the large dynamic range of about
three decades, can be compensated either by combining data from 
time--of--flight and backscattering spectrometers or by the use of master 
curves.
On the BS instruments IN16 (figure~\ref{Mqt-T}) 
and IN10B (figure~\ref{Sqt-IN11C})
the $Q$ resolution was improved by covering
part of the analyzer crystals by large cadmium shields \cite{ToWS98}.
This reduces considerably the number of detectors which
can be used in the analysis.
Multidetector TOF instruments intrinsically have a good 
$Q$ resolution.
However, now a compromise between wave number coverage and energy 
resolution has to be chosen.
On IN5 the elastic energy resolution (fwhm) was set to 
25\,$\mu$eV restricting the
accessible wave number range to $Q\le 1.7$\,\AA$^{-1}$.
As 25\,$\mu$eV is a rather broad resolution for a study of the \arx\ 
higher temperatures are required to make the dynamics appear in the 
experimental energy window.

\begin{figure}[thb]
\begin{center}
\epsfxsize=155mm
{\epsffile[19 419 566 635]{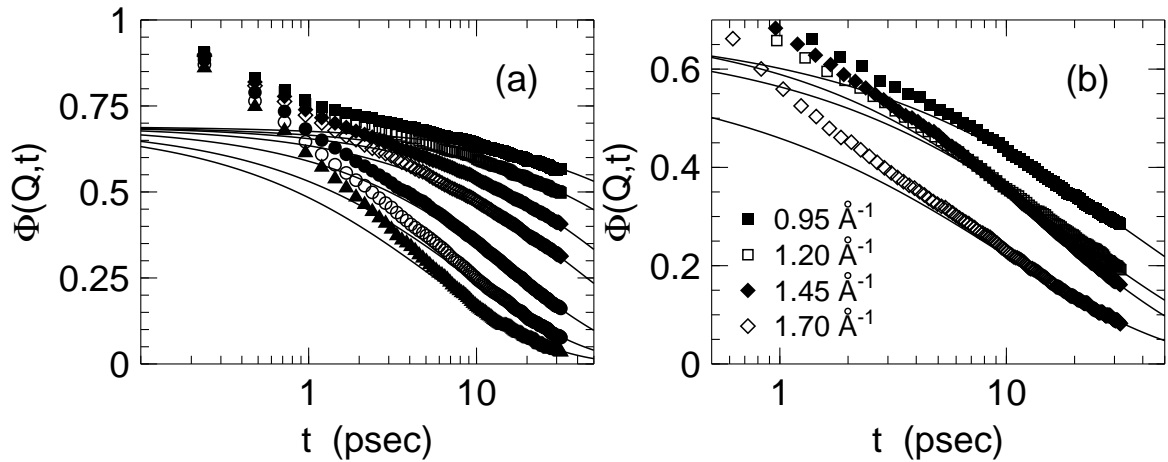}}
\end{center}
\caption{Density correlation function,
obtained by Fourier deconvolution of coherent neutron scattering spectra
measured on the time-of-flight spectrometer IN5.
(a) for different temperatures at the structure factor maximum 
$Q_1=1.45$\,\AA$^{-1}$. Temperatures are from top to bottom: 313\,K, 
320\,K, 330\,K, 340\,K, 360\,K, 380\,K, 400\,K.
(b) on an enlarged scale for 360\,K at different wave numbers $Q$. 
Solid lines are Kohlrausch stretched exponentials
as determined from a master curve analysis.
Without any fitting one recognizes that correlations decay steeper at 
$Q_1$ than at other wave numbers.
Taken from \protect\cite{ToWS98}.}
\label{Sqt-coh-IN5}
\end{figure}


Some of the data are shown in figure~\ref{Sqt-coh-IN5}a.
Up to high temperatures, far above the {\it melting} point, correlations
decay in two steps.
The second step is strongly stretched and temperature dependent as
expected for the \arx.
In figure~\ref{Sqt-coh-IN5}b the long time tail belonging to the \arx\ 
is plotted on an enlarged scale for different wave numbers $Q$.
Not only the relaxation time, but also the line shape exhibits a pronounced
$Q$ dependence.
At $Q_1=1.45$\,\AA$^{-1}$, the position of the first maximum of the static
structure factor, the correlations decay steeper, {\it i.\ e.\ } with
{\it less} stretching, than at any other wave number.
Such a behaviour can also be seen in the master curves displayed in
figure~\ref{Mqt-T}.
As one can easily imagine from the data in figure~\ref{Sqt-coh-IN5}a the 
individual correlation functions $\Phi(Q,t;T)$ do not allow for a three 
parameter fit of (\ref{KWW}): Either, for low temperatures the \arx\ is 
outside the accessible time
window of the spectrometer, or at high temperatures, where the decay is fully
in the window it is heavily disturbed by the \brx.
Then a separation of both contributions is not obvious and the fit range 
is somewhat arbitrary.
Note, that even at the highest temperature the decay of the correlation
function cannot be fully described by a single Kohlrausch law.
Moreover, a free fit with (\ref{KWW}) yields unphysical results like a 
strongly decreasing stretching to values as low as 0.3 and an increase of 
the amplitude with increasing temperature.

\begin{figure}[thb]
\begin{center}
\epsfxsize=155mm
{\epsffile[49 471 560 811]{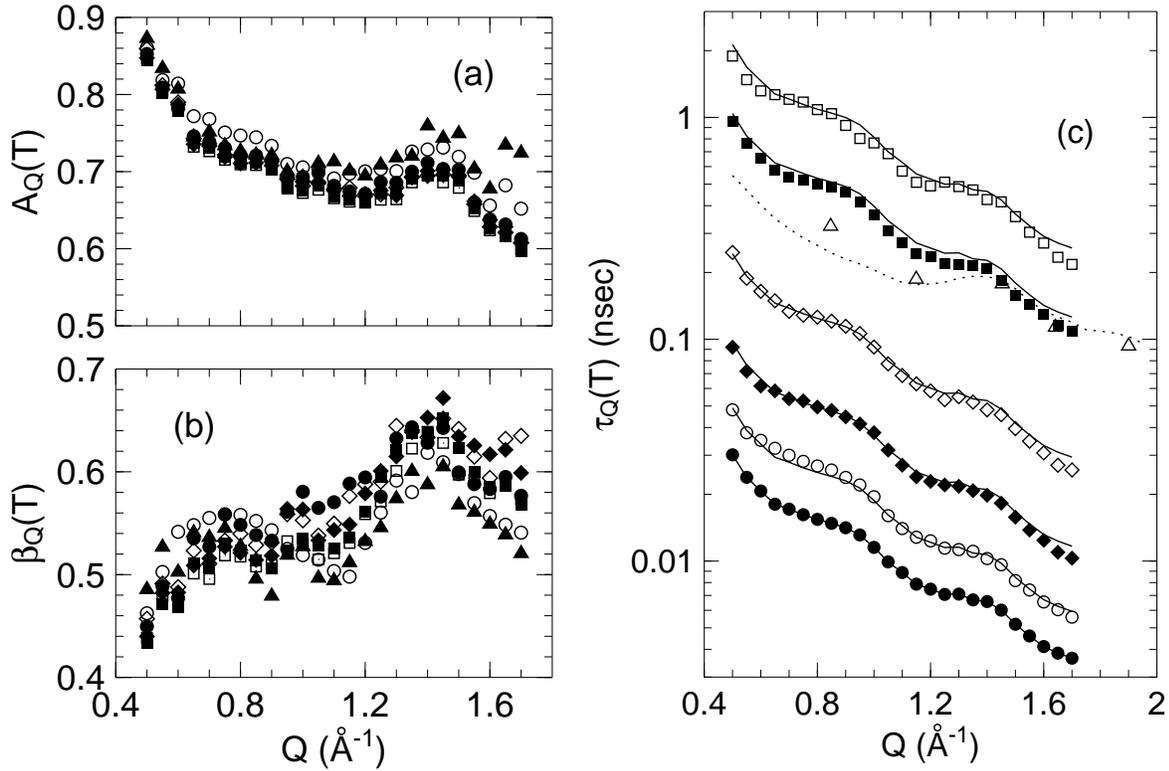}}
\end{center}
\caption{Kohlrausch parameter at seven different temperatures ranging from
313\,K to 400\,K.
In each of these fits, one parameter was free while the two others were
kept fixed at the values obtained from the master curve analysis.
(a) The amplitude $A_Q$ and (b) the stretching exponent $\beta_Q$ are 
temperature independent confirming our scaling approach.
(c) The \arx\ time $\tau_Q$ follows for all $Q$ the viscosity and exhibits a 
generic $Q$ dependence.
The data ar 330\,K have been omitted for clarity as at 320\,K the results 
from the BS instrument IN10B are included ($\opentriangle$).
The dotted line represents the quantity $S(Q,320\,{\rm K})/Q^2$ know as de 
Gennes narrowing.
Based on \protect\cite{ToWS98}.}
\label{Atb-Q-T}
\end{figure}

For a more quantitative analysis we employ the time--temperature superposition
to obtain master curves.
Indeed, without further adjustment, the data converge towards a temperature
independent long--time asymptote \cite{ToWS98}.
The results from this scaling analysis are shown in figure~\ref{Atb-Q-T}.
The amplitude factor $A_Q$ and the line shape parameter
$\beta_Q$ are quite independent of temperature in the range from 313\,K
to 400\,K in accord with results from light scattering \cite{CuLD97}.
In contrast to an often held belief, the stretching exponent does not increase
with temperature towards a diffusion limit $\beta=1$ in the normal liquid 
phase.
Even 70\,K above the melting point the correlation functions $\Phi(Q,t)$ 
remain stretched.
The structural relaxation time follows the viscosity and has a 
generic $Q$--dependence which confirms the $\alpha$ scale universality.

All parameters describing the density correlations exhibit pronounced
characteristic oscillations in phase with the 
static structure factor $S(Q)$ (figure~\ref{Atb-Q-T}).
The amplitude $A_Q$ is in almost quantitative agreement with the
Debye--Waller factor $f_Q$ previously obtained from the \brx\ plateau
(figure~\ref{fq-hq}).
Note, $f_Q$ and $A_Q$ not necessarily have to be identical.
The relaxation time $\tau_Q$ shows two plateaus around 0.8\,\Ar\ and
1.4\,\Ar\ on top of an overall decrease.
More important, the stretching exponent $\beta_Q$ shows a systematic 
variation in phase with $S(Q)$ being largest at the structur factor 
maximum.
Even around 0.8\,\Ar\ a slight variation is suggested from our data.
At 1.4\,\Ar\ the results are obvious from the raw data by plotting the 
$Q$ dependent decay curves in one diagram (see figures.~\ref{Mqt-T} and 
\ref{Sqt-coh-IN5}).
They are fully confirmed by fits of (\ref{KWW})
to numerical solutions of MC equations for the hard spheres system
\cite{FuHL92} and a soft spheres mixture \cite{FuLa93}.

Some reservation is necessary for the smallest wave numbers 
$Q \lesssim 0.5$\,\Ar\ where $A_Q$ starts to tend to 1.
Even for 99\% deuteration the incoherent scattering amounts to 
about 15\,\% to the total scattering cross section.
In addition, in the small $Q$ region the coherent scattering itself
should be weak as it is roughly proportional to $S(Q)$ which tends towards a 
constant level.
Therefore, the relative incoherent contribution might be even much stronger
than 15\,\%.
This effect can be studied in more detail by a separation of coherent and 
incoherent scattering by polarization analysis, cf.\ figure~\ref{Coh-inc}.
From this experiment we learn that the coherent $f_Q$ or the \arx\ 
amplitude $A_Q$ must be smaller than the incoherent one at 
$Q \lesssim 0.7$\,\AA$^{-1}$ and remains larger for higher $Q$ values.
This behaviour is indeed observed in figure~\ref{fq-hq} and we can conclude
that at the lowest $Q$ values contributions from multiple 
scattering cannot be excluded.
They are now even more complicated to treat as coherent and incoherent 
scattering are a weighted mix.
A separation of coherent and incoherent scattering from very thin samples 
{\it i.\ e.\ } with high transmision seems the most straightforward
access to solve this problems.
However, as outlined above quantitative results are still out of reach 
within an acceptable measuring time with the presently available neutron flux
and spectrometers.

In a recent work on Toluene, the structural relaxation time $\tau_Q$ 
followed nearly perfectly $S_{\rm COM}(Q)/Q^2$, where $S_{\rm COM}(Q)$
is the center--of--mass static structure factor calculated by Monte--Carlo
simulation \cite{AlTM00}.
Such a dependence, in a more general context known as de Gennes narrowing
\cite{Gen59},
indicates the dominance of the center--of--mass motion, which is of
purely translational nature.
We therefore included in figure~\ref{Atb-Q-T} $S(Q)/Q^2$.
Around $Q=1.4$\,\AA$^{-1}$ the variation of $\tau_Q$ with $Q$ is indeed
correclty described by $S(Q)/Q^2$.
The simulated center--of--mass static structure factors for OTP is peaked
around 0.8\,\AA$^{-1}$ \cite{KuWi95}.
Therefore, the plateau in $\tau_Q$ around 0.8\,\AA$^{-1}$ strongly supports 
our simplified picture that was drawn from $S(Q)$ in 
section~\ref{Static_Structure_Factor}.   

The wave number dependence of the tagged--particle correlations 
measured by incoherent scattering has already been discussed extensively
\cite{PeBF91,BaDF91}:
all parameters depend monotonously on $Q$:
$\beta_{inc}(Q) \approx {\rm const.}$, $A_Q \propto f_Q$ decreases like 
$\exp{(-cQ^2)}$ and $\tau_{inc}(Q)$ decreases close to a $Q^{-2}$ law
which is prefectly confirmed by a recent molecular dynamic simulation
\cite{MoLR00}.

\section{Experiments under Pressure}
       \label{Experiments_under_Pressure}
\subsection{Experimental Considerations}
For practical reasons most investigations on the glass transition
concentrate on temperature effects at ambient pressure $P_0=0.1$\,MPa
although there is a clear interest in studying supercooled liquid
dynamics in the full $T,P$ parameter space.
We emphasize the difference between the effect of pressure and temperature.
The primary effect of pressure is to change interatomic distances;
changes, for example in atomic vibrations are a secondary effect
via anharmonicity of the potential.
In contrast, the primary effect of temperature is to alter
the atomic vibrations; changes in interatomic distances are now a 
secondary effect.
Pressure and temperature are thermodynamically independent variables
even though their effects on one set of observables might be the same.
Pressure experiments have demonstrated that the density is not the only
driving force for the glass transition \cite{MaLi65,CoKH94,LeSD95,LiKO95}.
In the microscopic approach of MCT, both pressure and
temperature control the dynamics via variations of the static structure
factor $S(Q)$ which has been measured on fully deuterated OTP-$d_{14}$.
All other experiments have been performed with fully protonated
OTP, thereby measuring the tagged--particle dynamics.

While low temperature is regularly used in neutron scattering,
the use of high pressure is less standard as experiments under pressure 
are much more difficult to perform.
There are special requirements for high--pressure cells used in neutron
scattering.
For example, the flux rates of current neutron sources require large
sample volumes in order to get acceptable signal--to--noise ratios 
in reasonable time.
On the other hand the cell body in the neutron beam gives rise to 
unwanted scattering and absorption which demands a minimum of 
material in the beam.

High pressure experiments are possible only in a limited 
temperature--pressure range characteristic for a certain 
material \cite{SpPa77}.
The high pressure limit of a cylindrical cell is set
by the material strength and the ratio of outer--to--inner diameter 
$r_a/r_i$.
In order to use the full size of a neutron beam 
the inner diameter has to be of the order of some centimetres.
This easily results in large wall thicknesses leading not only to a strong
background but also to an attenuation of the incoming beam and to rather
difficult corrections.
The background signal, however, not only depends on on the wall thickness, 
{\it i.\ e.} from the number of scatterers in the beam but also from the
 scattering cross--sections of the material.
Corrections are difficult to perform if the attenuation of the beam is
caused by the energy dependent absorption.
Consequently, small ratios are used to keep these effects as 
small as possible.

High pressure cells based on high strength Aluminum alloys have many of 
the required/desired properties.
Vessels with small wall thicknesses withstanding high pressures produce
a rather low background and nearly no absorption due to the low
scattering and absorption cross--section.
Bragg scattering of Al occurs only for wave numbers $Q\ge 2.7$\,\AA$^{-1}$.

The experiments reported here were performed in a standard high pressure 
cell made of the Aluminium based alloy Al7049.T6 of the ILL.
The ratio of outer--to--inner diameter of 2 allows to attain pressures
up to 250\,MPa at temperatures up to about 340\,K.
The composition and scattering cross--sections of the high pressure
material are listed in table~\ref{Al7049.T6}.

\begin{table}[!ht]
\caption{Composition, scattering and absorption cross--sections
 of the high pressure material Al7049.T6 of the vessel.}
\label{Al7049.T6}
\lineup
\begin{indented}
\item[]\begin{tabular}{lllllll}
\br
Element    & wt\% & $\sigma_c$/[barn] & $\sigma_i$/[barn] & $\sigma_t$/[barn] & 
$\sigma_a$/[barn] \\
\mr
$_{26}$Fe & \00.3  & 11.22     &  0.40      & 11.62      &  2.56       \\
$_{14}$Si & \00.4  & \02.16    &  0.004     & \03.49     &  0.171      \\
$_{24}$Cr & \00.15 & \01.66    &  1.83      & \04.35     &  3.05       \\
$_{30}$Zn & \07.6  & \04.05    &  0.077     & \04.131    &  1.11       \\
$_{12}$Mg & \02.5  & \03.63    &  0.08      & \03.71     &  0.063      \\
$_{29}$Cu & \01.5  & \07.49    &  0.55      & \08.03     &  3.78       \\
$_{13}$Al & 87.55  & \01.49    &  0.008     & \01.503    &  0.231      \\
\mr
Al7049.T6 & 100  &  \01.865    &  0.026     & \01.892    &  0.358      \\
\br
\end{tabular}
\end{indented}
\end{table}

\begin{figure}[thb]
\begin{center}
\epsfxsize=120mm
{\epsffile[178 514 545 755]{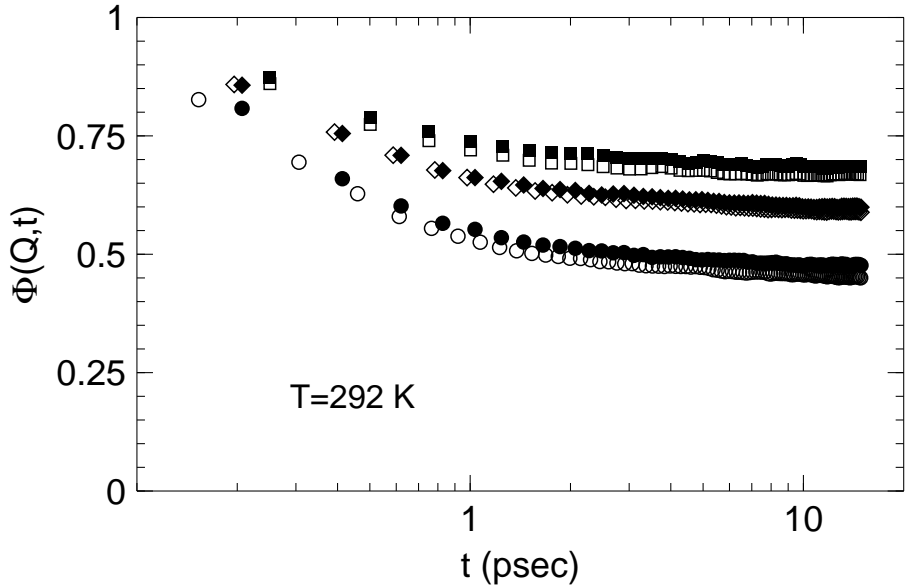}}
\end{center}
\caption{(*) Comparison of the correlation functions obtained without
(open symbols) and in the high pressure cell (closed symbols)
at $Q=1.2$ ($\square$,$\blacksquare$), $Q=1.5$ ($\lozenge$,$\blacklozenge$) 
and $Q=1.8$ \AA$^{-1}$  ($\circ$,$\bullet$).
The curves are nearly identical which validates the simplified approach 
of treating the high pressure cell like an ordinary container.}
\label{Signal-comparison}
\end{figure}

The transmission of the pressure cell was about 85\,\% and it was
handled like a usual container.
Restricting ourself to large scattering angles implying
large wave numbers $Q$ multiple scattering effects were neglected.
The approximate correctness of this approach is demonstrated in 
figure~\ref{Signal-comparison} where the correlation functions measured 
with and without pressure cell are compared.
Apart from a tiny difference in the plateau height the curves 
are identical.

Due to crystallization tendencies of OTP we had to use $^4$He gas as
pressure transmitting medium.
Any other attempt to use a liquid or OTP itself failed.
With a scattering cross--section $\sigma_{\rm coh}$=1.34\,barn
$^4$He gas is nearly transparent to neutrons.
In addition the analysis is not complicated by the fact that a liquid 
pressure transmitting medium will show the same effects of dense liquid
dynamics one is looking for in the sample.
Clearly, $^4$He diffuses into OTP with a time constant of several hours.
However, although the solubility increases with temperature and increasing
pressure according to Henrys law, with a typical solubility of about
10$^{-5}$\,mol/(gMPa) in aromates there is less than one He atom in the first
coordination shell.


\subsection{Static Structure Factor}
\begin{figure}[thb]
\begin{center}
\epsfxsize=155mm
{\epsffile[42 495 571 803]{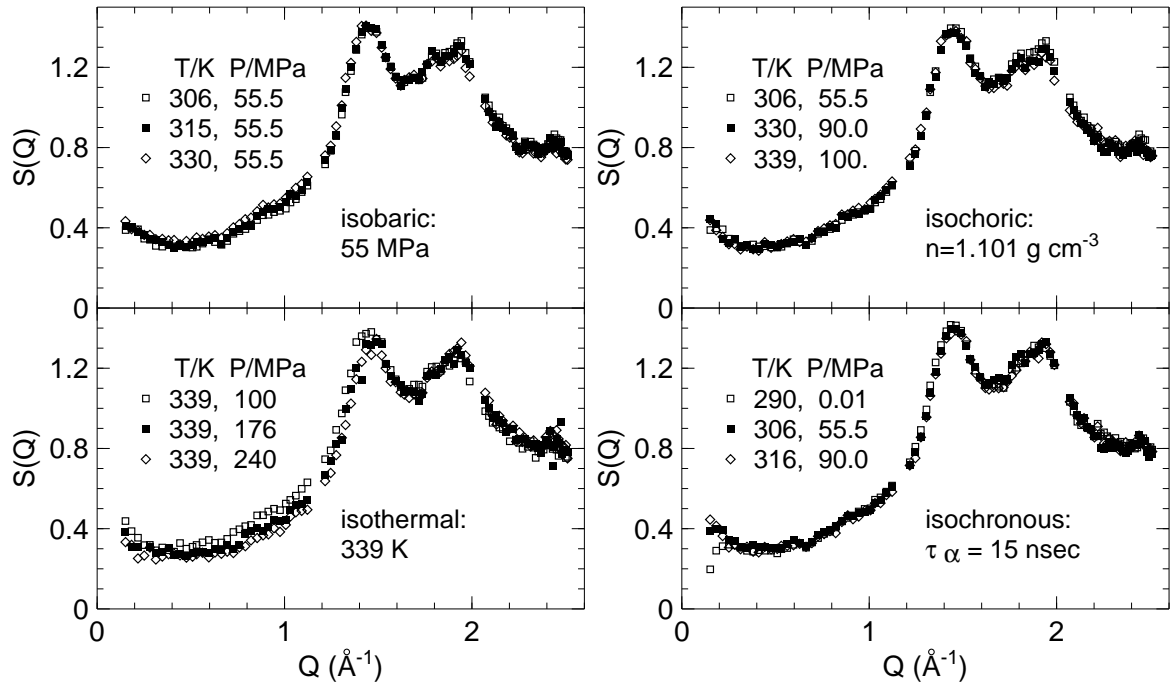}}
\end{center}
\caption{(*) Static structure factor $S(Q)$ 
under isobaric ($P=55$\,MPa),
isothermal ($T=339$\,K), isochoric ($n=1.101$\,gcm$^{-3}$) and
isochronic ($\tau_{\alpha}\simeq 15$\,nsec) conditions.}
\label{Sofq}
\end{figure}

The static structure factor of OTP-$d_{14}$ was measured on the
instrument D7, now used as a diffractometer without energy analysis.
The pressure and temperature dependence of 
$S(Q)$ is shown in figure~\ref{Sofq}.
$S(Q)$ was obtained in absolute units by comparison with a vanadium standard.


Under temperature variation at constant pressure (isobaric)
the peaks behave as at ambient pressure.
With decreasing temperature the scattering intensity in the $Q \to 0$ region 
reduces due to the decreasing isothermal compressibility $\chi_T$.
$S(Q= 0)=n \chi_T k_{\rm B} T$ compares well with the values calculated from
temperature and pressure dependent specific volumes \cite{NaKo89}. 
The peak positions shift slightly to higher $Q$
values due to an increase in density $n$.
Concerning the peak heights the maximum at 1.4\,\AA$^{-1}$ is not much 
affected, the maximum at 1.9\,\AA$^{-1}$ is enhanced and the
prepeak is lowered when cooling.

The effect of increasing pressure at constant temperature (isothermal)
on $S(Q)$ is similar to the one of decreasing temperature:  
$S(Q\to 0)$ decreases and the peak positions move to higher $Q$.
The intensity of the prepeak and the first maximum decreases while that of
the second maximum increases.
This behaviour is in agreement with results on other materials
\cite{MoDP97,AlMF98}.

Two other types of measurements are interesting in the light of MCT
since the static structure factor is considered to be the trigger of the 
slowing down of the \arx:
isochoric measurements of $S(Q)$, {\it i.\ e.\ } at $(T,P)$ combinations 
leading to the same bulk density and isochronous measurements, 
{\it i.\ e.\ } at $(T,P)$ combinations where the \arx\ time is identical.
Such measurements are included in figure~\ref{Sofq}.
While along isobars and isotherms the static structure factors $S(Q)$
evolve continuously, it is nearly identical along an isochor and an isochron.

\subsection{Square Root Singularity}
\begin{figure}[thb]
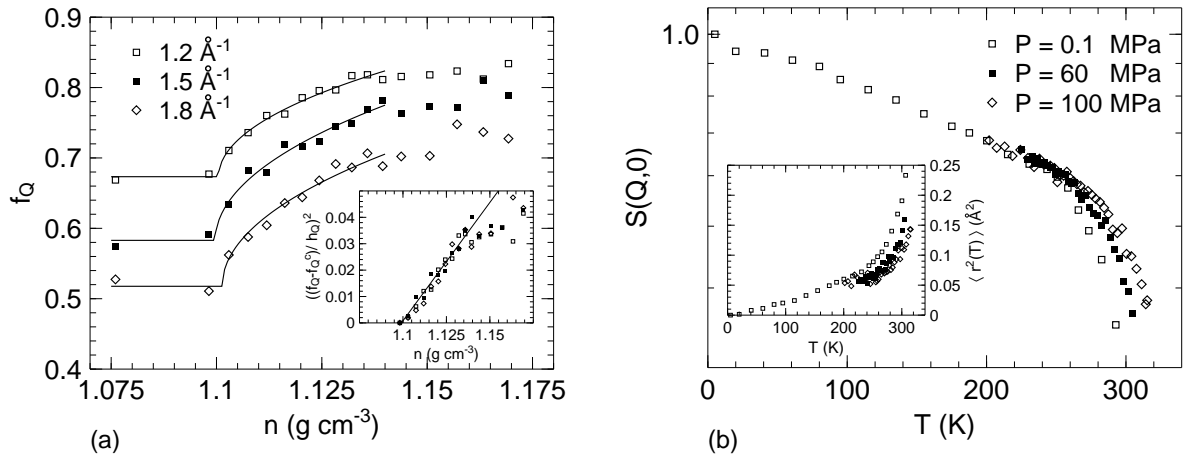

\begin{center}
\begin{minipage}[t]{75mm}
\epsfxsize=75mm
{\epsffile[62 452 476 776]{Bilder/P-Exp-squareroot/DWF-otp-d0-T301-rho.ps}}
\end{minipage}
\hfill
\begin{minipage}[t]{75mm}
\epsfxsize=75mm
{\epsffile[62 465 467 777]{Bilder/P-Exp-squareroot/DWF-otp-d0-TP-q18.ps}}
\end{minipage}
\end{center}
\caption{(a (*)) Plateau height $f_Q$ of isothermal tagged particle 
correlations as a function of density at constant temperature
$T=301$\,K for several $Q$--values.
Solid lines are fits with a square root law leading to a critical density
of $n_c \simeq 1.100$\,gcm$^{-3}$ which corresponds to a critical pressure
of $P_c \simeq 42$\,MPa.
Towards smaller densites the integrated $\alpha$ intensities
are independent of pressure.
The inset demonstrates the square root law in linearized form,
based on \protect\cite{ToSW98}.
(b (*)) Incoherent elastic scattering intensity (\Mob\ factor) as a
function of temperature for different pressures at $Q=1.8$\,\Ar.
The linear behaviour at low temperatures is due to classical
phonons $\exp(-Q^2 \langle r^2(T) \rangle)$ with 
$\langle r^2(T) \rangle \propto T$.
With increasing pressure the anomalous decrease shifts to higher 
temperatures.
Above a temperature $T_c(P)$ quasielastic broadening from
structural relaxations comes into play and the integrated
$\alpha$ intensity has to be taken into account.
The inset shows the temperature dependence of the
mean sqaure displacement which contains the
complete $Q$ information.}
\label{DWF-P}
\end{figure}

When the pressure dependence of the elastic scattering at constant 
temperature is monitored a behaviour slightly different as compared to the
temperature dependent experiments is observed.
In figure~\ref{DWF-P} left the isothermal DWF for incoherent scattering 
(\Mob--Lamb factor) as a
function of density is shown for several wave numbers $Q$.
At high pressures or densities, in the glassy phase, the elastic intensity
remains rather constant.
Thus, no harmonic evolution had to be taken into account 
as in the case of the temperature dependent measurements.
Releasing the pressure, an anomalous decrease of the elastic intensity
sets in and indicates the presence of a faster relaxation.


When constant pressure is applied the anomalous decrease shifts to
higher temperatures with increasing pressure as apparent in
figure~\ref{DWF-P} right which shows the isobaric DWF as a function of
temperature for three different pressures at $Q=1.8$\,\AA$^{-1}$.

Inspecting the isobaric decrease of $f_Q$ in figure~\ref{DWF-P} 
at high temperatures more closely it seems that the slope of the 
decrease diminishes with increasing pressure and the transition range 
becomes broader.
This slope is related to the strength of the fast $\beta$ relaxation.
One might speculate that the effect vanishes under extreme conditions.
However, when the $Q$ dependence of the elastic scattering is 
analyzed in terms of the mean square displacement 
$\langle r^2(T) \rangle$ extracted from the Debye Waller
factors in a Gaussian approximation (inset of figure~\ref{DWF-P})
this effect disappears.
While around 200\,K the $\langle r^2(T) \rangle$ are almost identical
the anomalous increase shifts to higher temperatures.
The pressure dependence of the mean square displacement can even be scaled
to a master curve taking the thermodynamic shift d$T_{\rm g}/$d$P$
as the scaling parameter.
Similar observations were made in the polymer polybutadiene 
\cite{FrAH97,FrAl99} while in polyisoprene the application of pressure 
separates an additional dynamic process \cite{FrAl99}.

\subsection{Critical Correlations}
We have already seen in the previous section that a kink in the Debye--Waller
factor reveals the onset of the \brx.
The anomalous increase of the DWF above a certain density indicates that the
$\beta$ process is also strongly influenced by the pressure or density.
This fast process is responsible for the decrease of elastic intensity 
on costs of inelastic intensity.
Again the \brx\ can be accessed directly on a TOF spectrometer like IN6.
The spectral density below about 2\,meV varies strongly with pressure.
In this region the dynamic structure factor satisfies the factorization rule
for all pressures as already demonstrated in figure~\ref{Factorisation}.

The \brx\ has been studied at variable pressure up 240\,MPa
\cite{ToSW98,ToSW97a}.
The tagged--particle correlations in the compressed liquid also decay in two
steps and can be described by the MCT scaling laws.
Unfortunately, the used instruments IN16 and IN6 do not overlap 
(see table~\ref{instruments}) and
the dynamic range of IN6 alone is clearly not sufficient to warrant
a four parameter fit.
To get the pressure dependence of the \brx\ amplitude $h_Q$ and the
cross--over times \t_s\ we use the results from the \arx\ analysis
to be presented in the following section in more detail.
Besides the DWF $f_Q$ we fix the line shape parameter to the known 
value $\lambda=0.77$, which is compatible
with the frequency dependence $\omega^{1-a} (a=0.3)$ shown in
figure~\ref{Factorisation}.
The DWF was determined from a Kohlrausch fit to the long time tail
of the low pressure data with fixed $\beta=0.6$ and $\tau_Q(P)$.
The insensitivity of line shape parameter $\lambda$ to pressure
is supported by broad band light scattering measurements on the 
glass formers Cumen \cite{LiKO95} and Salol \cite{PrBB00} where in 
isothermal and isobaric experiments identical line shapes were found.
\begin{figure}[thb]
\begin{center}
\epsfxsize=120mm
{\epsffile[177 296 545 755]{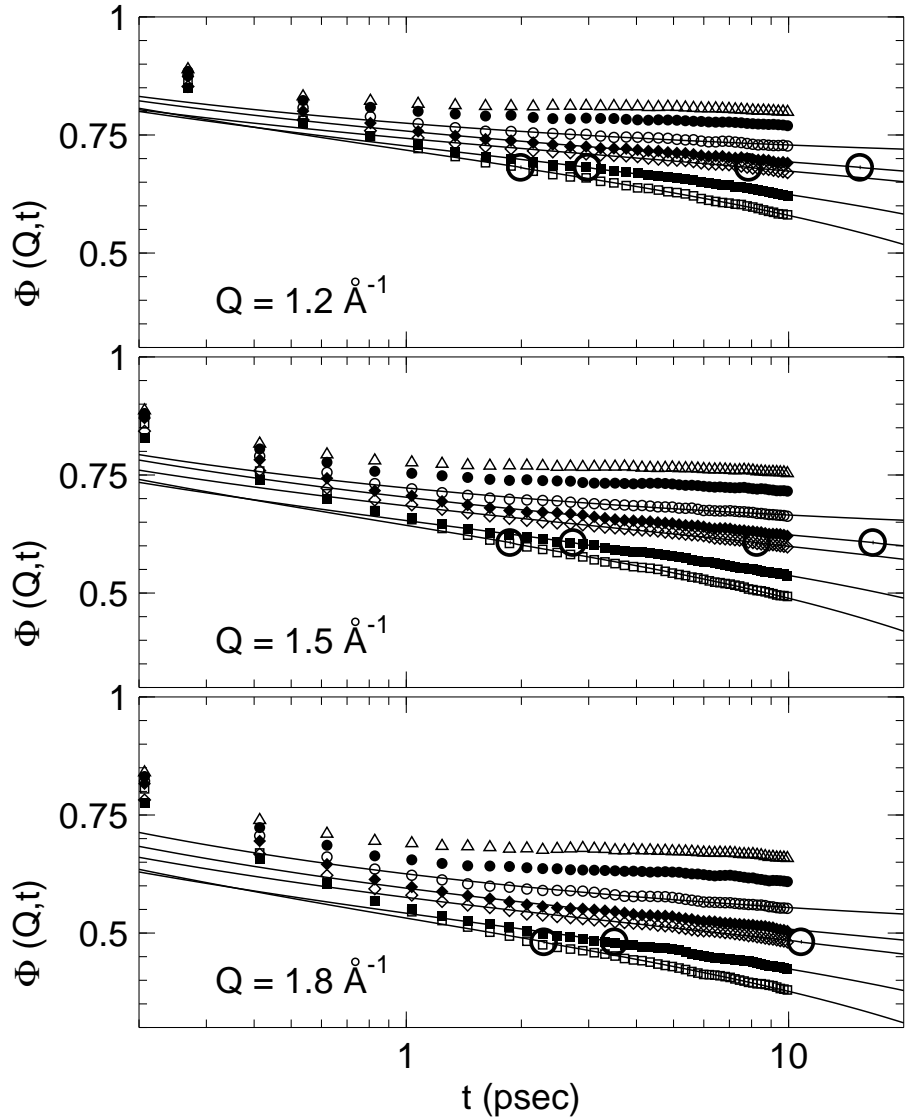}}
\end{center}
\caption{(*) Isothermal intermediate scattering function $S(Q,t)$ at 316\,K
obtained on IN6 at three different wave numbers $Q=1.2,1.5$ and
$1.8$\,\AA$^{-1}$ in the \brx\ regime.
Lines are fits with (\ref{beta-scaling-law}) with fixed line shape
parmeter $\lambda=0.77$ and fixed plateau values (indicated by the 
thick circles).
The pressures are from bottom to top: 0,1\,MPa, 15\,MPa, 45\,MPa,
60\,MPa, 90\,MPa, 150\,MPa and 200\,MPa.}
\label{Sqt-P}
\end{figure}

Figure~\ref{Sqt-P} shows isothermal pressure dependent tagged
particle correlation functions together with fits to (\ref{beta-scaling-law}).
The correlation functions at the highest pressures decay to a plateau
which is constant in time and higher than the value for the liquid state:
we are in the glassy state.
For the lower pressures one recognizes a two step decay as in the temperature
dependent case.
While the first decay on the sub--picosecond time scale due to the microscopic
dynamics is not covered by (\ref{beta-scaling-law}) it consistently
describes the bending of $\Phi(Q,t)$ into and out of the intermediate
plateau $f_Q$ over about one decade depending on the pressure.

\begin{figure}[thb]
\begin{center}
\epsfxsize=155mm
{\epsffile[78 530 559 746]{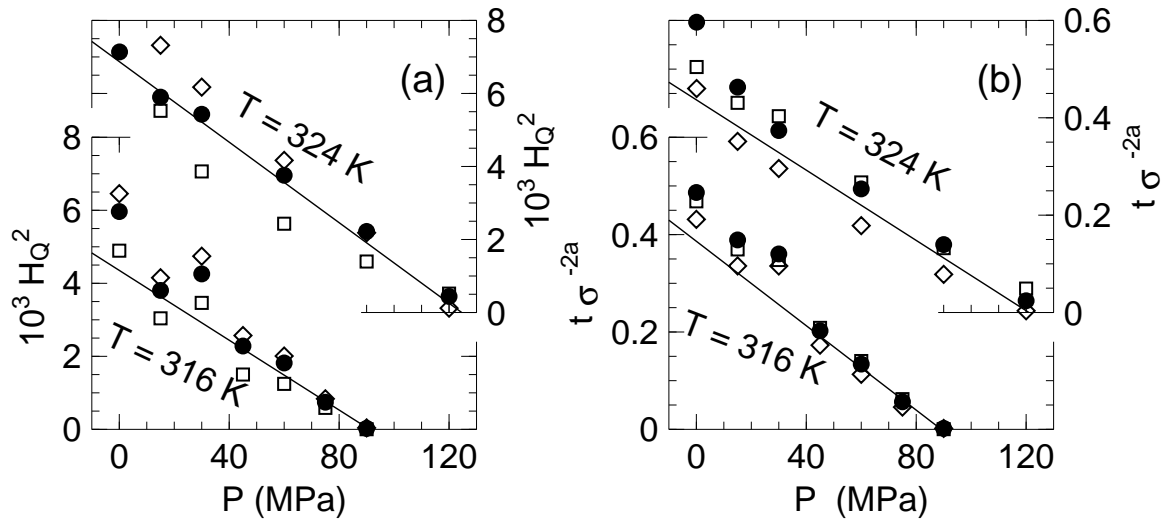}}
\end{center}
\caption{\brx\ amplitude $H_Q$ (a) and cross--over time $t_{\sigma}$ (b)
for two isotherms 316 and 324\,K and three wave numbers 
$Q=1.2 (\square), 1.5 (\bullet)$ and $1.8$\,\AA$^{-1}$ ($\lozenge$).
The data are linearized as $H_Q^2$ and $t_{\sigma}^{-2a}$ according to
(\ref{hq}); lines are fits to $Q$ averaged values.
Reproduced from \protect\cite{ToSW98}.}
\label{hq-ts-P}
\end{figure}

On approaching the cross--over from the liquid side, $f_Q$ varies only
weakly, whereas $H_Q$ and \t_s\ should become singular.
In figure~\ref{hq-ts-P} the power laws (\ref{hq}) are tested for two
isotherms by plotting $H_Q^2$ and $t_{\sigma}^{-2a}$ vs.\ $P$,
with $a=0.295$ determined by $\lambda$.

Asymptotically, a linear behavior is observed allowing for a consistent
determination of the mode--coupling crossover line $P_{\rm c}(T)$ or
$T_{\rm c}(P)$.

\begin{table}
\caption{Critical pressure $P_c$ and corresponding density $n_c$
as function of temperature.
The product $\Gamma_c= n \sigma^3 {(\epsilon/T)}^{1/4}$ 
is found to be constant (its errors are estimated  
from $\Delta T_c=\pm3$\,K and $\Delta P_c=\pm5$\,MPa).
The choice of $\sigma=7.6$\,\AA\ and $\epsilon=600$\,K
is explained in the text.
The values in the lower part stem from temperature dependent 
elastic scans where the critical temperatures can only be estimated.
Nevertheless, the $(T,P)$--combinations are compatible with the
phase boundery.}
\label{Gamma}
\begin{indented}
\item[]\begin{tabular}[th]{cccc}
\br
$T\;$(K) & $P_c\;$(MPa) & $n_c\;$(nm$^{-3}$) & 
$\Gamma_c$\\
\mr
290 & ~~0.1  & 2.840 & 1.495(4) \\ 
301 & ~42(3) & 2.875 & 1.499(6) \\ 
316 & ~90(7) & 2.906 & 1.497(5) \\ 
324 & 123(7) & 2.931 & 1.501(9) \\ 
\mr
306 & ~60(3) & 2.889 & 1.500(6)  \\
317 & 100(5) & 2.917 & 1.503(12) \\
\br
\end{tabular}
\end{indented}
\end{table}

\begin{figure}[thb]
\begin{center}
\epsfxsize=120mm
{\epsffile[178 514 554 791]{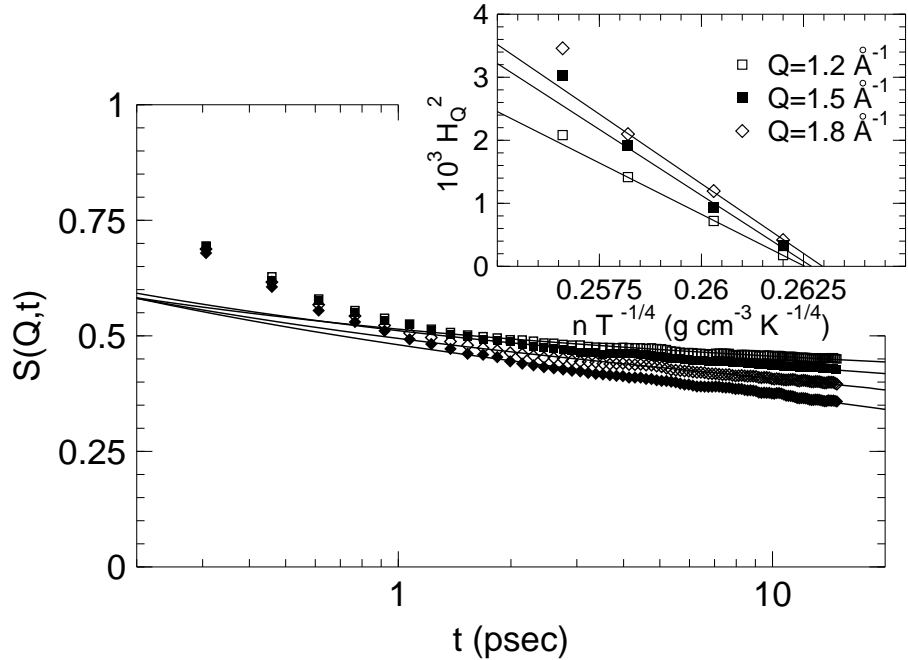}}
\end{center}
\caption{(*) Isochoric tagged particle correlation function $\Phi(Q,t)$ for 
$Q$=1.8 \AA$^{-1}$ at a density $n\simeq1.083$\,g/cm$^3$.
The $(T,P)$ combinations are from top to bottom:
(292,0.1); (300,12); (310,26.5); (319,39) in K and MPa.
Lines are fits with a fixed $t_{\sigma}(\Gamma)$.
The inset shows the results for the \brx\ amplitude $H_Q$ in linearised
form for three wave numbers all extrapolating to a consistent
single value of $nT^{-1/4} \propto \Gamma$.
For explanation of $\Gamma$ see text.}
\label{Sqt-isochorQ18}
\end{figure}

Together with $T_{\rm c}$ at ambient pressure and the $P_{\rm c}$ values
obtained from the elastic scans six points of the
dynamic phase boundary $P_c(T)$ are obtained as summarized in 
table~\ref{Gamma}.
The $(P_{\rm c},T_{\rm c})$ line has a slope 
${\rm d}T_c/{\rm d}P \simeq 0.26 -- 0.28$\,K/MPa
which is in remarkable accord with the slope determined from \arx,
viscosity and $T_{\rm g}(P)$ data:
the mode--coupling crossover line is parallel to lines of equal 
$\alpha$ response (see figure~\ref{Matraze}).
Lines of constant densities, however, have far steeper slopes of about 
0.7\,K/MPa.
It confirms that the temperature influences the dynamics 
not only via the free volume.
This finding is further illustrated by a direct comparison of isochoric 
correlation functions in figure~\ref{Sqt-isochorQ18}.
The loss of correlations is the faster the higher the temperature.


The above results imply that the dynamic crossover can be described by a
single separation parameter $\sigma(T,P)$ which, however, does not depend
on density alone.
Instead, it appears that density $n$ and temperature $T$ can be
condensed into an effective coupling $\Gamma \propto nT^{-1/4}$ which is
constant within experimental error as shown in table~\ref{Gamma}.
Furthermore, constraint fits to all individual $\Phi(Q,t;T,P)$
confirm that the temperature and pressure dependence of the amplitude and the
cross--over time can be described by the power law (\ref{hq}) in conjunction
with just one separation parameter $\sigma=(\Gamma-\Gamma_c)/\Gamma_c$
(figure~\ref{Sqt-isochorQ18}).
At least in a certain neigbourhood of $P_c(T)$ this parameter is linear in
the effective coupling $\Gamma\propto n T^{-3/k}$ which is known to
characterise equilibrium properties of a dense soft sphere fluid with
repulsive core potentials of the Lennard--Jones $r^{-k}$ type \cite{HaMD86}.
Note, that a plot $\log n_{\rm c}$ vs.\ $\log T_{\rm c}$ yields
a slope of 0.27 very close to the theoretically expected value 1/4 for $k=12$.

In a supercooled and compressed liquid, we are actually in a high--density
regime where the static structure factor $S(Q)$ is determined mainly by the 
repulsive part of the potential \cite{HaMD86,BoYi80,WeCA71}.
Since this repulsive part can be modeled well by a soft sphere system our 
results are thus evidence that for the glass transition
subtle changes in the static structure factor are responsible 
for the slowing down of the dynamics.
This view is also supported by computer simulations of 
binary soft sphere \cite{BeHH87,RoBH89,BaLa90} and Lennard--Jones
\cite{NaKo97} systems, in which the glass transition was found to occur at
constant~$\Gamma_c$.
It is astonishing that these results apply literally in a complex
molecular liquid like OTP.
In fact, the comparison with a Lennard--Jones liquid can be made almost
quantitative. 
To do so, we need the two parameters of the $6--12$ potential:
a van der Waals diameter $\sigma$ and a potential depth $\epsilon$. 
$\sigma=7.6$\,\AA\ is estimated from structural information \cite{Bon64}.
This is in good accord with the prepeak in the static structure factor at 
$Q \simeq 0.85\ldots0.9$\,\AA$^{-1}$ \cite{ToSW97b,Mas98}
and the scaling factor needed to describe the $Q$ dependence of the critical
Debye--Waller factors $f_Q^c$ and the amplitudes $h_Q$ in figure~\ref{fq-hq}.
A potential depth $\epsilon=600$\,K has been determined by calibrating a
three--site OTP model to experimental density and diffusivity \cite{LeWa94}.
Using this input, we obtain $\Gamma_c=1.50$ (table~\ref{Gamma}) which is 
closer to the Lennard--Jones result $\Gamma^{\rm LJ}_c\simeq 1.25$
\cite{Ben86a,NaKo97} than one could reasonably expect.

Bengtzelius \cite{Ben86a} calculated $T_{\rm c}(n)$ for different
densities for a Lennard--Jones system.
In order to compare our $T_{\rm c}(n)$ with his result we rescale his
axis such that they coincide at $P_0$.
Our slope  (5.02 10$^{-4}$\, gcm$^{-3}$) is in good accord
with the theoretical one (6.22 10$^{-4}$\, gcm$^{-3}$).
From the rescaling we get the two Lennard--Jones parameters:
the potential depth $\epsilon/k_{\rm B}\simeq 608$\,K which
astonishingly is the value found in the simulations \cite{LeWa94} and
the diameter $\sigma\simeq 7$\,\AA\ being in good accord with 
the \vdW\ radius.

\begin{figure}[thb]
\begin{center}
\epsfxsize=120mm
{\epsffile[104 532 435 746]{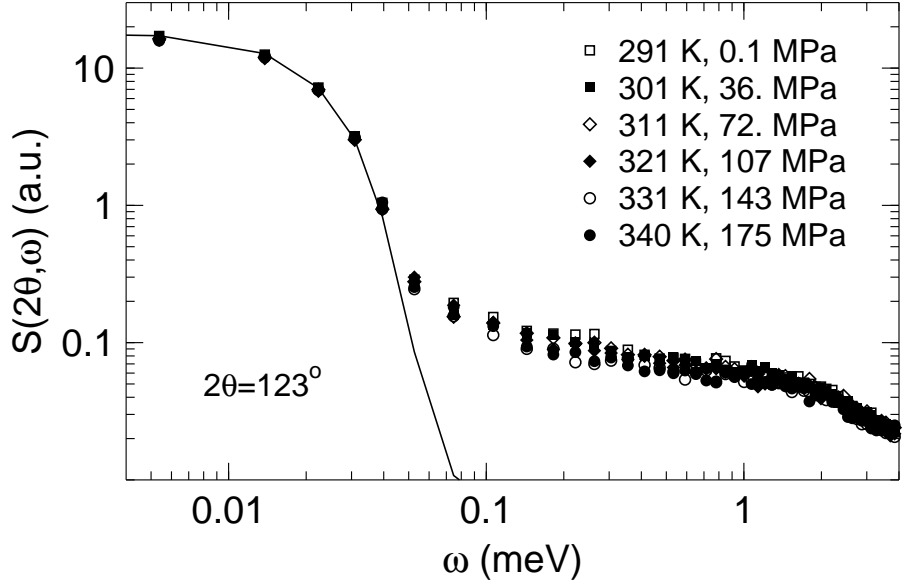}}
\end{center}
\caption{(*) Incoherent dynamic structure factor $S(2\theta,\omega)$
at a fixed scattering angle $2\theta=123^{\circ}$
for several combinations of temperature and pressure leading to the same 
value for $\Gamma$.
Note, the spectra are identical within experimental error.}
\label{S2thw-isochron}
\end{figure}

We have found in the pressure and temperature dependent measurements 
of the static structure factor that along an isochronous line 
($\tau_{\alpha}=const.$) $S(Q)$
does not change significantly (figure~\ref{Sofq}).
According to the theory the spectra $S(Q,\omega)$ should all be
identical (apart from the possibility of a trivial shift in the overall 
time scale $t_0$).
In order to test this behaviour directly we measured
the dynamics for several combinations of temperature and pressure leading
to the same $\Gamma$ 
(and not as usual along isothermal, isochoric or isobaric paths).
A direct comparision of the dynamic structure factors in 
figure~\ref{S2thw-isochron}
shows indeed that the sprectra are identical within statistical accuracy.
This implies that the plateau height $f_Q$, the $\beta$--relaxation
amplitude $h_Q$ and the cross--over time $t_{\sigma}$ are identical
along a constant $\Gamma$, {\it i.\ e.} along a constant 
$\tau_{\alpha}$--line.
Note, that for our purposes it is sufficient to compare $S(2\theta,\omega)$
after subtracting the empty can spectra.


\begin{table}[tbh]
\caption{Connection between expansivity $\alpha$, compressibility $\beta$ and
the repulsive part of the potential expressed via \eref{gamma-thermo} 
along the dynamic phase boundery {\it i.~e.}.}
\label{thermo}
\begin{indented}
\item[]\begin{tabular}[th]{ccccc}
\br
$T\;$(K) & $P_c\;$(MPa) & $\alpha 10^{4}/\;$K$^{-1}$ 
                             & $\beta 10^{4}/\;$MPa$^{-1}$ 
             & $T_{\rm c}(\frac{\beta}{{\rm d}T/{\rm d}P|_\tau}-\alpha)$ \\
\mr
290 & ~~0.1  & 7.2 & 5.0 & 0.31 \\
301 & ~42(3) & 6.4 & 4.4 & 0.28 \\
306 & ~60(3) & 6.1 & 4.1 & 0.26 \\
316 & ~90(7) & 5.7 & 3.8 & 0.25 \\
317 & 100(5) & 5.6 & 3.7 & 0.24 \\
324 & 123(7) & 5.3 & 3.6 & 0.24 \\
\br
\end{tabular}
\end{indented}
\end{table}
  

Interestingly, a relation between the effective coupling $\Gamma$ or, 
more exactly, the repulsive part of the potential and
some macroscopic quantities can be established.
To do so we assume the specific volume $v(T,P)$ to be linear dependent on 
temperature and pressure via the thermal expansivity $\alpha$
and the isothermal compressibility $\beta$:
\begin{equation}
v(T,P)=v(T_{\rm c},0)(1+\alpha (T-T_{\rm c})-\beta P_{\rm c}(T))
\end{equation}
with $T_{\rm c}$ the critical temperature and 
$P_{\rm c}(T)$ the critical pressure.
At the phase boundery $\Gamma^{-1}=vT^{1/4}=const.$, thus
we can derive the following
equation using $T=T_{\rm c}(1+(T-T_{\rm c})/T_{\rm c})$ and 
developping in a Taylor series up to first order
\begin{equation} \label{gamma-thermo}
T_{\rm c}(\frac{\beta}{{\rm d}T/{\rm d}P|_\tau}-\alpha) \simeq
\frac{1}{4} 
\end{equation}
where we have used the fact that the phase boundery follows
\begin{equation}
P_{\rm c}(T)=\frac{(T-T_{\rm c})}{{\rm d}T/{\rm d}P|_\tau}.
\end{equation}
Using our ($T_{\rm c},P_{\rm c}$) combinations, the values for 
the expansivity and compressibility at these \cite{NaKo89} and
${\rm d}T/{\rm d}P|_\tau \simeq 0.28$\,K/MPa gives us the result in 
table~\ref{thermo}.
Astonishingly, equation (\ref{gamma-thermo}) holds quite good
indicating a deeper link between 
the microscopic repulsive part of the potential 
and the macroscopic quantities expansivity and compressibility.

It would indeed be interesting to check on other fragile glass froming 
liquids, whether the critical mode--coupling phase boundery ist parallel 
to lines of constant $\alpha$--relaxation time and whether an effective 
coupling and a link to the expansivity and the compressibility can be 
established in a similar manner as we found for OTP.

\subsection{$\alpha$--Relaxation} \label{p-arx}
On the BS instrument IN16 we measured the slowing down of
the structural \arx\ as a function of pressure.
In a first survey, we could show that the mean relaxation time follows
the pressure dependence of the viscosity.
The shape of the \arx\ spectra was then determined in some long--running
scans.
\begin{figure}[!hb]
\begin{center}
\epsfxsize=120mm
\epsfysize=115mm
{\epsffile[92 206 576 746]{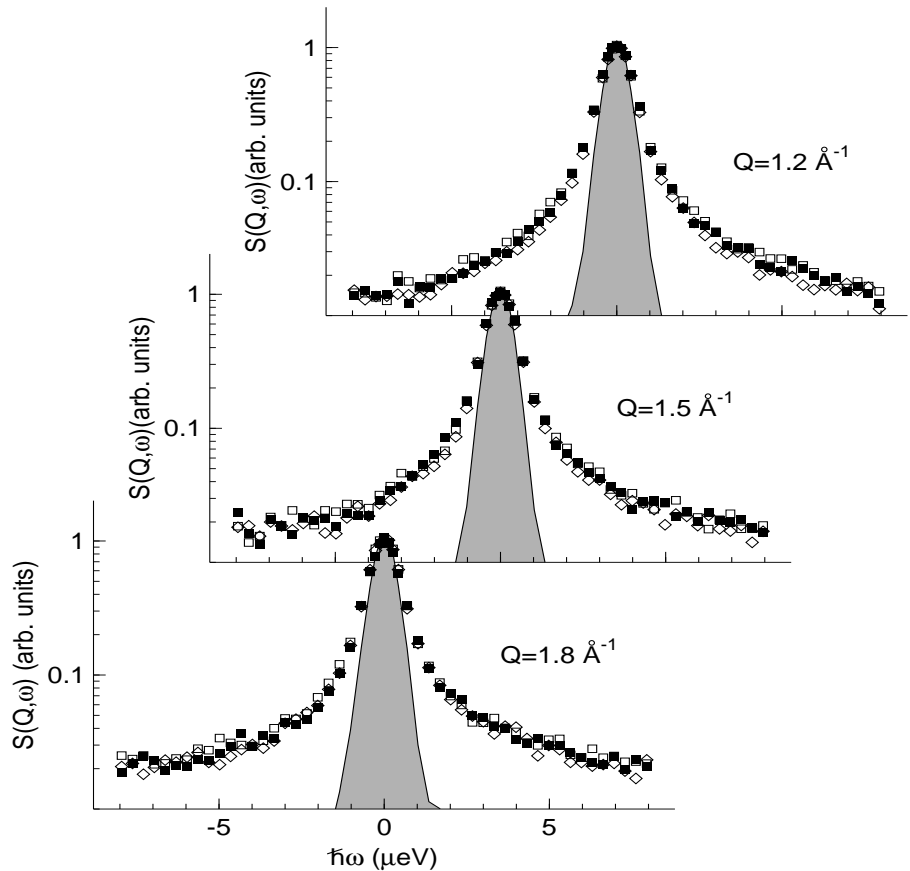}}
\end{center}
\caption{(*) \arx\ spectra from IN16 at $Q=1.2,1.5$ and 1.8\,\AA$^{-1}$
for three combinations of temperature and pressure leading to approximately
the same viscosity.
The spectra are scaled to their value at $\omega=0$.
The hached area represents the measured resolution function.
The $(T,P)$ combinations are: 306\,K, 0.1\,MPa ($\square$);
320\,K, 50\,MPa ($\blacksquare$) and 335\,K, 100\,MPa ($\lozenge$).
}
\label{Sqw-isoviscous}
\end{figure}

At three different points ($T,P$) of equal viscosity, the quasielastic
broadening of all spectra coincide as shown in
figure~\ref{Sqw-isoviscous} for three values of $Q$.
The microscopic relaxation time is proportional to
the macroscopic viscosity $\eta$ and the spectral
distribution is quite independent of pressure.
This is confirmed from photon
correlation \cite{FyDW83} and dielectric measurements \cite{NaEM87}.

By construction of a master curve for isothermal spectra at 320\,K
we can further illustrate the invariance
of the line shape and the viscosity scaling.
Motivated by the result of figure~\ref{Sqw-isoviscous} we use viscosity data 
\cite{SchKB98} to rescale the times $t\to{\hat t}=t \eta(P_0)/{\eta(P)}$ 
with $P_0=0.1$ MPa similar to the procedure described in section~\ref{t-arx}.
\begin{figure}[thb]
\begin{center}
\epsfxsize=120mm
{\epsffile[46 471 467 789]{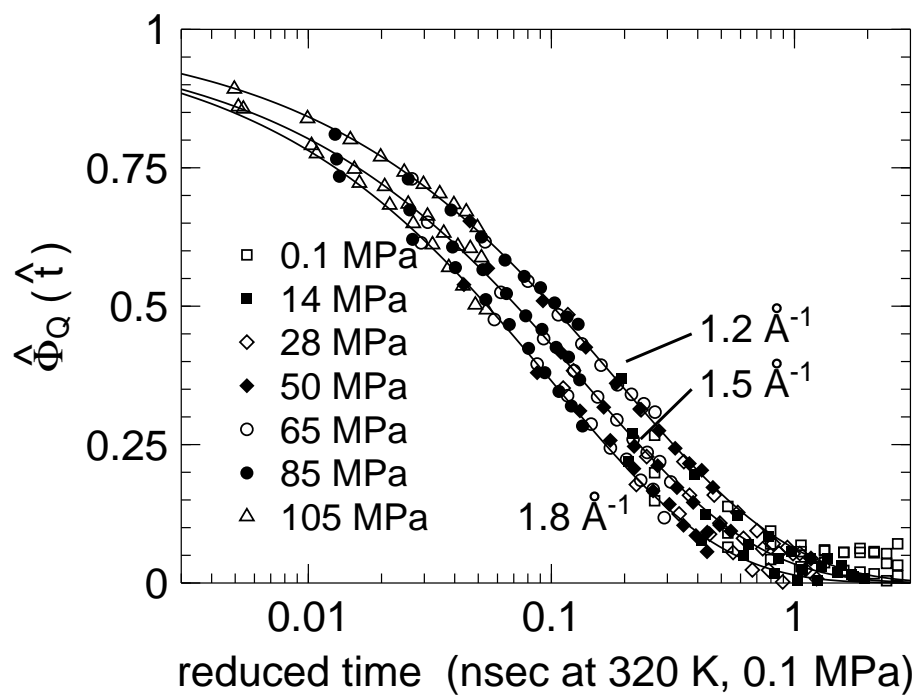}}
\end{center}
\caption{(*) Master curves $\hat \Phi_Q(\hat t)/A_Q(P)$ for the isothermal
$\alpha$--relaxation of the tagged particle fluctuations at 320\,K and
different pressures, obtained by rescaling times with viscosity $\eta(P)$.}
\label{Mqt-alpha-inc-P}
\end{figure}

\begin{figure}[thb]
\begin{center}
\epsfxsize=155mm
{\epsffile[51 465 568 810]{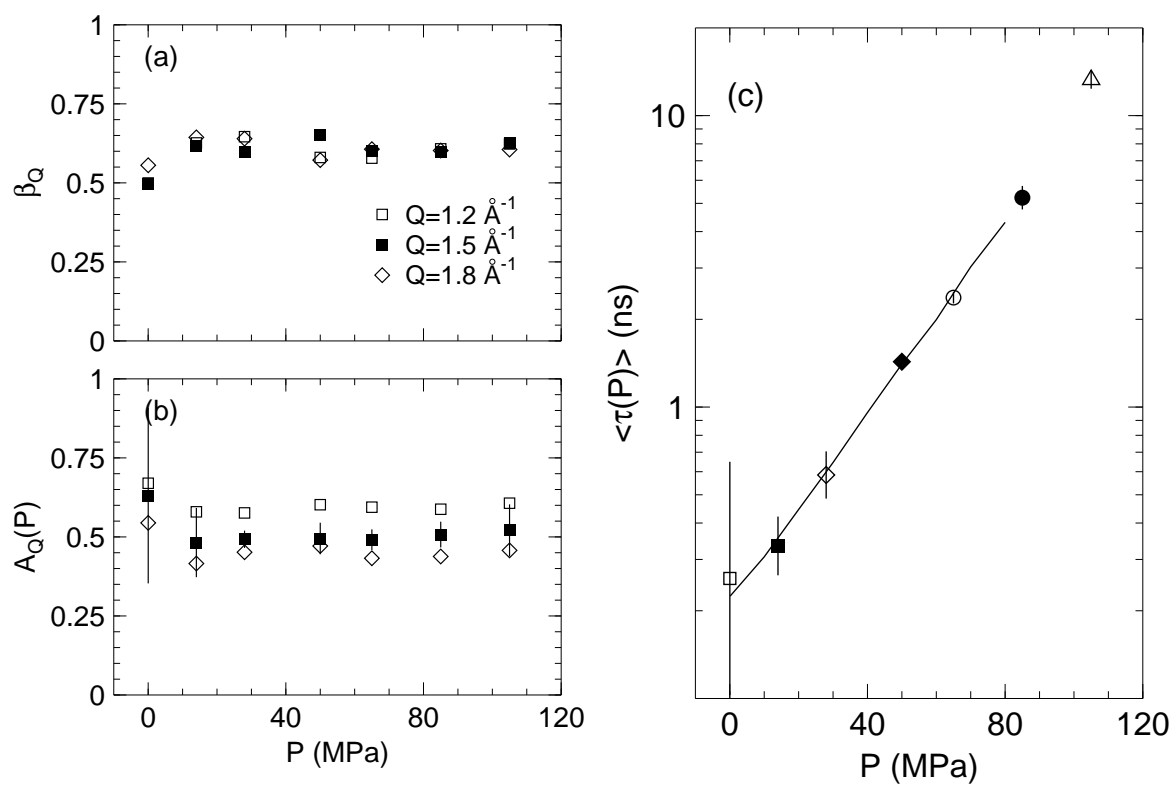}}
\end{center}
\caption{(*) Kohlrausch parameters $\beta_Q$ (a), $A_Q$ (b) and $\tm$ (c)
as obtained from the scaling analysis as a function of pressure.
In each fit, two parameters were fixed at the values obtained from the master
curve analysis.
Amplitude $A_Q$ and stretching exponent $\beta_Q$ are roughly independent
of pressure in the liquid state.
The mean relaxation time $\tm$ scales with visosity (line).
}
\label{Ab-Q-P}
\end{figure}

The rescaled $\hat\Phi(Q,\hat t)$ coincide over two decades as shown in
figure~\ref{Mqt-alpha-inc-P} for three $Q$ values.
The scaling is well satisfied. 
The \arx\ amplitude and the line shape parameter $\beta$ 
as derived from the iterative procedure are shown in figure~\ref{Ab-Q-P}.
Both parameters are nearly independent of pressure, which in turn justifies
the analysis of the \brx\ presented in the previous section.
The strechting parameter $\beta$ is similar to that derived 
from previous investigations at atmospheric pressure \cite{BaFK89}:
The distribution of relaxation times is pretty much 
uneffected by the pressure and temperature changes.
The mean relaxation times $\langle \tau(P) \rangle$ as obtained from 
Kohlrausch fits with fixed ${\langle A_Q \rangle}_P$ and $\beta$ and after 
elimination of the $Q$ dependence are also shown in figure~\ref{Ab-Q-P}.
It follows the same pressure dependence as the
viscosity, in agreement with the isoviscous measurements.


Thus, a time--temperature--pressure superposition principle holds.
It is well established by PCS and dielectric spectroscopy in OTP as well
as in some other fragile liquids \cite{NaMa83,PaZR96}.
Our measurements extend the validity of this scaling principle down to
the picosecond time range.

\begin{figure}[thb]
\begin{center}
\epsfxsize=120mm
{\epsffile[63 381 523 779]{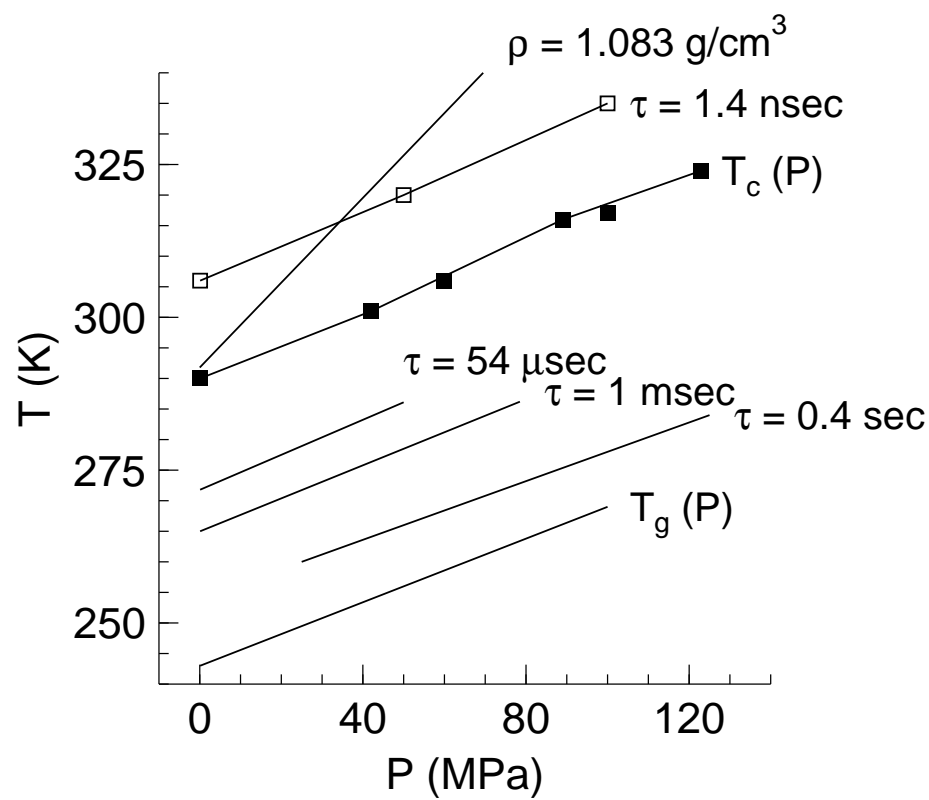}}
\end{center}
\caption{(*) Pressure and temperature dependence of isochronous \arx\ times.
Note that from $T_{\rm g}$ corresponding to relaxation time of the 
order of 100\,sec
down to nsec all slopes are identical. 
In particular the dynamic phase boundary $T_{\rm c}(P)$ is parallel to
lines of equal $\alpha$ response.
Constructed using values from \protect\cite{AtAn79,FyDW83,NaEM87}.}
\label{Matraze}
\end{figure}

The pressure dependence of isochronous relaxation times is compiled
in figure~\ref{Matraze}.

Lines of isochronous \arx\ times are connected by the same slope
of about $({\rm d}T/{\rm d}P)_{\tau} \simeq (0.28 \pm 0.01)$\,K/MPa
over about twelve orders of magnitude 
regardless of the time scale.
At long times the pressure dependence of the glass 
transition temperature $({\rm d}T_{\rm g}/{\rm d}P)$ \cite{AtAn79},
is the same as in the second to microsecond time range covered
by photon--correlation spectroscopy \cite{FyDW83} 
and dielectric spectroscopy \cite{NaEM87}
down to the sub--nanosecond regime covered here 
by neutron scattering.

\section{Vibrational Dynamics [36]} 
 \label{Vibrational_dynamics}
When the dynamic anomalies of the glass transition are discussed
it is particularly important for molecular glasses, which have a large
number of internal degrees of freedom, to study at least 
qualitatively the vibrational mode structure.
This seems to be even more important in view of the fact that low--frequency
vibrational features -- for obscure reasons called Boson peak -- 
persist well into the supercooled phase.

This can be done by comparison of the absolute frequency scales 
of the vibrational density of states between the
ordered and disordered state.
From single crystal dispersion measurements -- if available --
information about low--energy internal modes and hybridisation of
external and internal modes can be obtained.

\subsection{Single Crystal: Phonon Dispersions}
\begin{figure}[thb]
\begin{center}
\epsfxsize=155mm
{\epsffile[61 556 584 753]{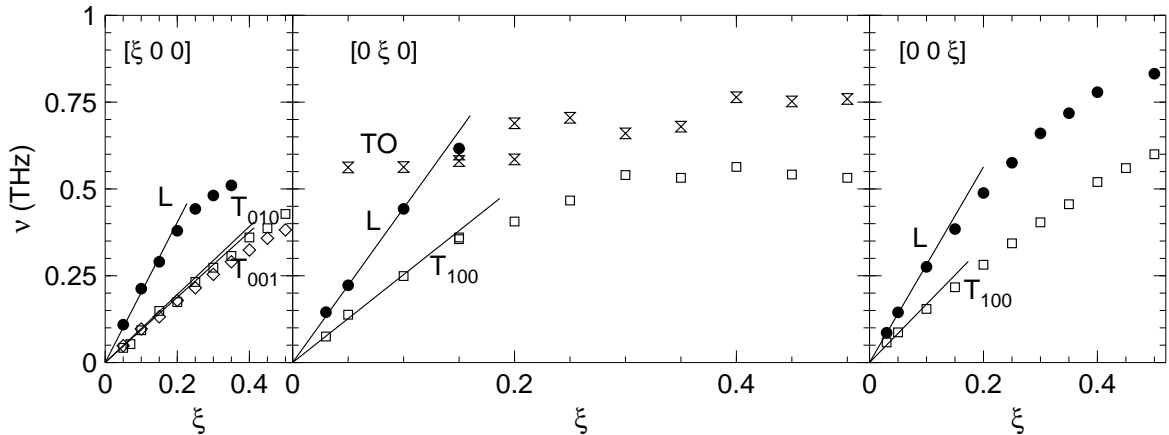}}
\end{center}
\caption{Phonon dispersion relations of crystalline OTP as
measured on the triple axis spectrometer IN12 at 200\,K.
The wave vectors are given in reciprocal lattice units, excitation 
energies in THz (1\,THz = 33.36\,cm$^{-1}$ = 4.1357\,meV).
Taken from \protect\cite{ToZF00}.}
\label{Dis-relat}
\end{figure}

The phonon dispersions of single crystal OTP at 200\,K
along the three main symmetry directions [100], [010], and [001]
are presented in figure~\ref{Dis-relat}.
Some optic--like phonons are detected as well, 
as an example, the lowest optic branch in [010] direction is 
included in figure~\ref{Dis-relat}.
However, for a detailed study of them thermal neutrons instead of 
cold ones would be needed.

In the small $Q$ limit the sound velocities of acoustic excitations
obtained from the initial slope are summarized in table~\ref{soundns}.
In all lattice directions,
longitudinal sound modes are almost twice as fast as transverse modes.
The single crystal data may also be compared with
sound velocities in the glass.
Results from Brillouin light scattering 
are summarized in table~\ref{TsoundG}.
Taking the simple arithmetic average over the three crystal axes,
the mean longitudinal and transverse sound velocities 
$\langle v_{\rm L,T} \rangle$
exceed those of the glass by about 20\,\% and 27\,\%, respectively.

\begin{table}
\caption{Sound velocities (in km/s)
for three phonon branches and different lattice directions, obtained from
linear fits to the low--$Q$ limit of the measured phonon dispersions.}
\label{soundns}  
\begin{indented}
\item[]\begin{tabular}{llllll}
\br
 &$v_{[100]}$ & $v_{[010]}$ & $v_{[001]}$ &
   $\langle v\rangle$ &${\langle v^{-3}\rangle}^{-1/3}$ \\
\noalign{\smallskip}\mr\noalign{\smallskip}
$v_{L}  $& 3.71  & 2.68  & 3.30 & 3.23 & 3.11 \\*[0.8em]
$v_{T_1}$& 1.82  & 1.52  & 1.97 && \\*[-.4em]
         &       &       &      & 1.75 &1.71 \\*[-.4em]
$v_{T_2}$& 1.67  & --    & --   &&\\
\noalign{\smallskip}
\br
\end{tabular}
\end{indented}
\end{table}

\begin{table}
\caption{Sound velocities in the glass as measured by Brillouin scattering,
using visible light or X--rays.}
\label{TsoundG}
\begin{indented}
\item[]\begin{tabular}{lllll}
\br\noalign{\smallskip}
  $T$ (K)&$v_{\rm L}$ (km/s)&$v_{\rm T}$ (km/s)&method&reference\\
\noalign{\smallskip}\mr\noalign{\smallskip}
220&2.94&1.37&light&\protect\cite{HiWa81}\\
223&2.63&&light&\protect\cite{MoCF98b}\\
200&2.70&&X-rays&\protect\cite{MoMR98}\\
\noalign{\smallskip}
\br
\end{tabular}
\end{indented}
\end{table}


\subsection{Glass versus Polycrystal: Density of Vibrational States} 
Without crystalline order, it is no longer possible
to measure selected phonon modes with well--defined polarization and
propagation vector.
For polycrystalline or amorphous samples,
the distribution of vibrational modes can be conveyed only
in form of a spectral density of states (DOS).

\begin{figure}[thb]
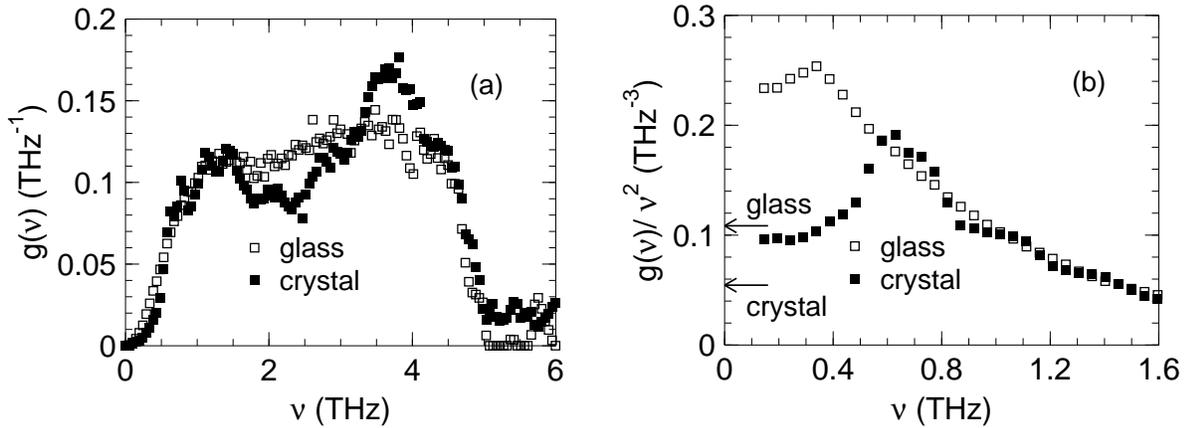

\begin{center}
\begin{minipage}[t]{75mm}
\epsfxsize=75mm
{\epsffile[16 584 239 753]{Bilder/Vibration/opp-02-dgc.ps}}
\end{minipage}
\hfill
\begin{minipage}[t]{75mm}
\epsfxsize=75mm
{\epsffile[23 584 244 753]{Bilder/Vibration/opp-03-dw2.ps}}
\end{minipage}
\end{center}
\caption{(a) Density of vibrational states of glassy and crystalline OTP 
at 100\,K.
(b) To emphasize the excess density of states of the glass over the crystal
$g(\nu)/\nu^2$ is shown.
The arrows mark the Debye limit $\nu \to 0$ calculated from sound
velocities and densities at 220\,K for the glass and at 200\,K for the
crystal.
Taken from \protect\cite{ToZF00}.}
\label{DOS_100K}
\end{figure}

The DOS of glassy and crystalline OTP at 100\,K are shown in
figure~\ref{DOS_100K}.
They were calculated directly from the spectra at the highest scattering 
angles $S(2\theta,\omega)$ without interpolation to constant $Q$.
Multiphonon contributions were calculated be repeated convolution of
$g(\omega)$ with itself and subtraction from $S(2\theta,\omega)$ in 
an interative procedure described in detail in \cite{WuKB93}.

Comparing our results with model calculations \cite{Bus82},
we assign all vibrations below $\nu_{\rm g}$ to
the 16 degrees of freedom needed to describe the crystal structure.
For 16 low--lying modes in a molecule with 32 atoms,
we expect an integrated DOS
\begin{equation}
    \int_0^{\nu_{\rm g}}\!{\rm d}\nu\, g(\nu)={16\over32}=0.5\,.
\end{equation}

Rather broad distributions are found for both phases showing a pronounced 
gap above $\nu_{\rm g}\simeq5$\,THz 
(1\,THz = 33.36\,cm$^{-1}$ = 4.1357\,meV).
In the glass a first shoulder around 1.5\,THz is followed by a second at 
3.5\,THz in accordance with results from Raman studies \cite{CrBA94,KiVP99}.
As expected, the crystal DOS is more structured, in particular in the 
low energy region. 
Distinct peaks at 0.6, 0.8, 1.1 and 1.5\,THz become apparent.
They are due to strong contributions from zone--boundary modes
(figure~\ref{Dis-relat}).
Compared to the glass, significant density is missing in the low 
energy region and in the range from 1.5 to 3\,THz, 
and reappears at higher frequencies around 3.5\,THz.
The gap for $\nu > 5$\,THz common for both states and indicates
a separation from high energy intramolecular modes.

The DOS of the polycrystalline sample
is in accord with the measured dispersion of the single crystal:
A small shoulder at 0.4\,THz can be attributed 
to the transverse acoustic zone--boundary phonons in $[100]$ direction,
and the main peak at 0.6\,THz corresponds 
to the transverse acoustic zone--boundary phonons in the other two lattice 
directions.
The peak at 0.8\,THz reflects the longitudinal acoustic zone--boundary 
phonon in $[001]$ direction and the transverse optic in $[010]$ direction.

In order to show the low energy modes on an enhanced scale,
we plot in figure~\ref{DOS_100K} $g(\nu)/\nu^2$ 
which in the one--phonon approximation is proportional to $S(Q,\nu)$ itself.
In this representation,
the excess of the glass over the crystal becomes evident.
A well defined frequency peak appears around 0.35\,THz which is down shifted 
with respect to the first peak of the crystal at 0.6\,THz
and superposed to a long tail which is similar for both, glass and crystal. 
Note that the maximum of the boson peak at 0.35\,THz is located 
below the lowest acoustic zone--boundary phonons in the crystal.

Another dynamic observable which can be obtained from neutron scattering
is the atomic mean square displacement $\langle r^2(T) \rangle$.
Roughly speaking, $\langle r^2(T) \rangle^{3/2}$ measures the volume
to which an atom remains confined in the limit $t\rightarrow\infty$.
For a large class of model situations 
(harmonic solid, Markovian diffusion, \ldots) 
it can be obtained directly
from the Gaussian $Q$ dependence of the elastic scattering intensity
\begin{equation} \label{Sq0}
         S(Q,\nu\!=\!0)=\exp(-Q^2 \langle r(T)^2 \rangle)\,.
\end{equation}
For a harmonic solid, 
\begin{equation} \label{msd}
      \langle r(T)^2\rangle = 
         {\hbar^2 \over 6 M k_{\rm B} T } \int_0^\infty\! {\rm d}\nu\,
         \frac{g(\nu)}{\beta} {\rm coth}(\frac{\beta}{2}) 
\end{equation}
(with $\beta=h\nu/k_{\rm B}T$)
crosses over from zero--point oscillations 
$\langle r^2(0) \rangle$
to a linear regime
$\langle r^2(T) \rangle \propto T$. 
Note, for a meaningful application of (\protect\ref{msd}), we have to
  assume the rigid--body limit (Sect.~3.1, Ref.~\protect\cite{WuKB93}).
  Therefore, $M$ has to be taken as the average atomic mass which in the case
  of OTP is 7.2.
In any real experiment,
which integrates over the elastic line with a resolution $\Delta\nu$,
one actually measures atomic displacements within finite times 
$t_\Delta\simeq 2\pi/\Delta\nu$.

\begin{figure}[thb]
\begin{center}
\epsfxsize=120mm
{\epsffile[68 525 349 746]{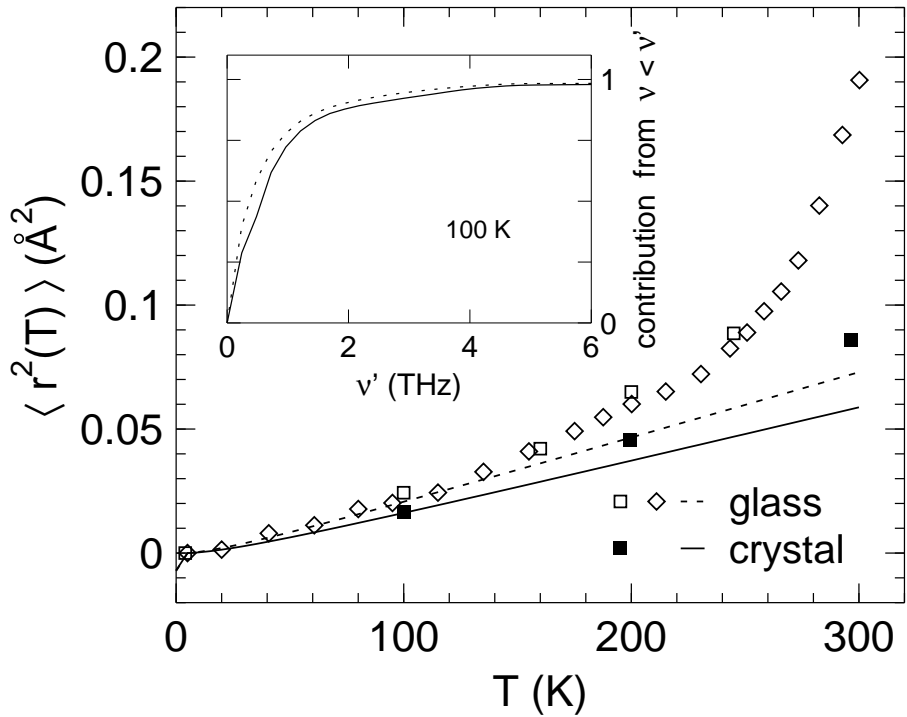}}
\end{center}
\caption{Mean square displacement $\langle r^2(T) \rangle$ of glassy OTP 
from elastic back-scattering (IN13, $\lozenge$)
and from time-of-flight spectra (IN6, plateau value of~$S(Q,t)$, $\square$).
For comparison, IN6 data of polycrystalline OTP are also shown 
($\blacksquare$).
The lines are calculated from the densities of states at 100\,K 
of the glass (dashed) and the polycrystal (solid line).
The inset shows the relative value of $d=\langle r^2\rangle ^{1/2}$ 
at 100\,K for the crystal and the glass 
when the integral (\protect\ref{msd}) is restricted to modes with 
$0<\nu<\nu'$.
Taken from \protect\cite{ToZF00}.}
\label{MSD-MSD_w}
\end{figure}

Figure~\ref{MSD-MSD_w} shows mean square displacements of OTP,
determined according to (\ref{Sq0}) 
from elastic back--scattering 
(fixed window scans on IN13, with $t_\Delta\simeq100-200\,$psec)
and from Fourier--deconvoluted time-of-flight spectra
(taking the plateau $S(Q,t_\Delta)$ with $t_\Delta\simeq5-10\,$psec
from Fourier transformed IN6 data).
For the glassy sample, a direct comparison can be made and shows
good agreement between IN6 and IN13.


The lines in figure~\ref{MSD-MSD_w}
are calculated through (\ref{msd}) from the DOS at 100\,K.
For low temperatures, the $\langle r^2(T) \rangle$ 
are in full accord with the values determined through (\ref{Sq0}).
This comparison can be seen as a cross--check between
the analysis of elastic and inelastic neutron scattering data.

Equation (\ref{msd}) not only gives the absolute value 
of~$\langle r^2(T) \rangle$,
but enables us also to read off which modes contribute most
to the atomic displacement.
To this end, we restrict the integration (\ref{msd}) to 
modes with $0<\nu<\nu'$.
The inset in figure~\ref{MSD-MSD_w} 
shows the relative value 
$\langle r^2(\nu';T) \rangle^{1/2} / \langle r^2(\nu;T) \rangle^{1/2}$ 
for $T=100\,$K as function of~$\nu'$.
Modes below 0.6\,THz in the crystal, or 0.4\,THz in the glass
contribute about 55\% to the total displacement;
90\% are reached only at about 2\,THz.
This means that the modes which are responsible for
the mean square displacement and the Debye--Waller factor
are not rigid body motions alone, 
but contain a significant contribution from 
other modes.

Anyway, one immediately recognizes that for all temperatures the
mean square displacement in the glass is larger than in the crystal.
For higher temperatures the measured mean square displacements
are clearly larger than expected from harmonic theory.
These anharmonic contributions to $\langle r^2(T) \rangle$ start
around 140--150\,K.
Around the same temperature deviations from the proportionality 
$\langle r^2(T) \rangle \propto T$ can be observed in 
coherent elastic scans and in the \Mob\ experiment \cite{VaFl79}
(figure~\ref{DWF-VaFl79} in section~\ref{t-squarerootsingularity}).
In addition, around 140--150\,K the temperature dependence of the 
experimental specific heat $c_{\rm P}(T)$ \cite{ChBe72} shows a 
slight change for both glass and crystal.

Anharmonicities can directly be observed via the temperature dependence 
of the density of states.
With increasing temperature the first moment, which provides a measure for 
the center of gravity, clearly shifts to lower frequencies
for the glass and even more for the crystal.
Note however, the strong increase above 240\,K in the glass is connected
with the glass transition.

\begin{figure}[thb]
\begin{center}
\epsfxsize=120mm
{\epsffile[79 525 360 746]{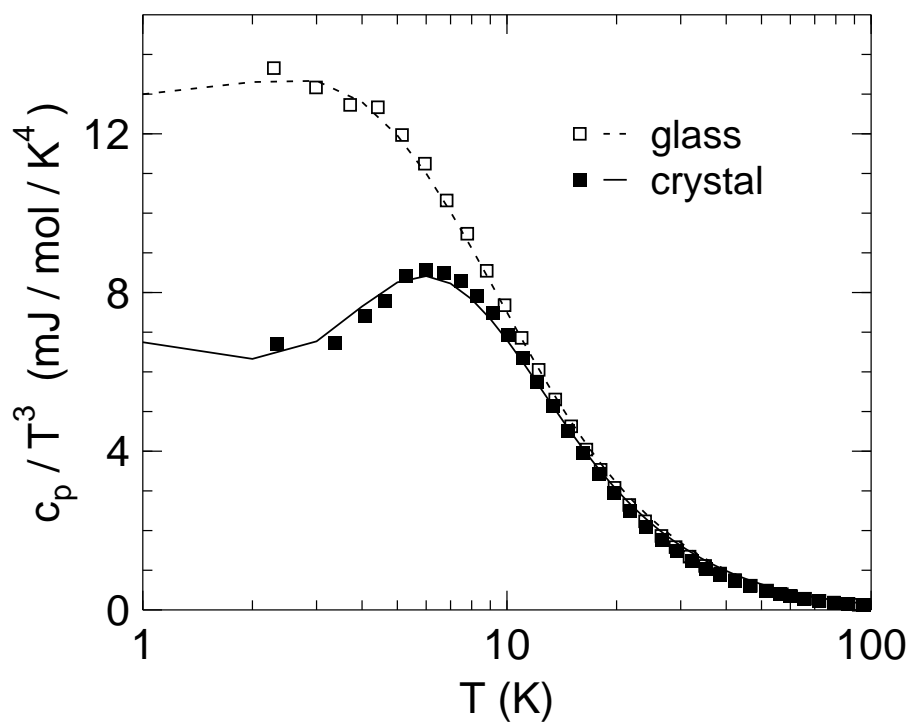}}
\end{center}
\caption{Experimental heat capacities $c_{\rm P}/T^3$ of
glassy ($\square$) and crystalline ($\blacksquare$) OTP \protect\cite{ChBe72},
compared to $c_{\rm P}/T^3$ calculated from the density of vibrational states 
$g(\nu)$ at 100\,K of the glass (dashed line) and the polycrystal 
(solid line).
Taken from \protect\cite{ToZF00}.}
\label{cpT}
\end{figure}

Our vibrational density of states can also be checked againts
the heat capacity, which for a harmonic solid is given by the integral
\begin{equation}\label{cv}
   c_{\rm P}(T) \simeq c_{\rm V}(T) = N_{\rm at}R\int_0^\infty\!{\rm d}\nu\, g(\nu)
             \frac{(\beta/2)^2}{\sinh^2(\beta/2)}\,.
\end{equation} 
With a Debye DOS, this yields the well--known $c_{\rm P}\propto T^3$.
Therefore, in figure~\ref{cpT} experimental data \cite{ChBe72} 
for the specific heat of glassy and polycrystalline OTP 
are plotted as $c_{\rm P}/T^3$.
In this representation,
a boson peak at 0.35\,THz is expected to lead to a maximum at about 4\,K.


The lines are calculated through (\ref{cv}) from the neutron DOS.
They agree  with the measured data for both glass and crystal in 
absolute units and over a broad temperature range.
Similar accord has been reported for a number of other systems
\cite{CrBA94,BeCA96,TaRV98,WuPC95,BuPN86}.

At higher temperatures,
the heat capacities of crystalline and glassy OTP differ only little
({\it e.g.} at 200\,K: $c_{\rm P}^{\rm cryst}= 182.8$\,J\,mol$^{-1}$\,K$^{-1}$ 
and $c_{\rm P}^{\rm glass}= 186.1$\,J\,mol$^{-1}$\,K$^{-1}$ \cite{ChBe72}).
Towards higher frequencies, the DOS becomes less sensitive
to the presence or absence of crystalline order 
and remaining differences (clearly visible in figure~\ref{DOS_100K})
are largely averaged out by the integral (\ref{cv}).
 
Altogether, these observations suggest that the mode structure in the glass
and the crystal is quite similar except for an overall softening in the 
glass.
Concerning the above presented analysis of quasielastic
scattering in the supercooled liquid we can state that
above 0.6\,THz a very large portion of inelastic 
intensity in $S(Q,\nu)$ of glassy OTP must arise from optic modes.
Hybridisation and coupling between inter- and intramolecular modes
plays an important role for frequencies higher than 0.6\,THz 
\cite{CrBG93,CrBA94}.
As a consequence, the quasielastic scattering, which is confined to  
below 0.25\,THz (1\,meV), is clearly dominated by rigid--body motions 
and the analysis in terms of relaxations remains reasonable.

\section{Conclusions}
As the reader has seen in the preceeding sections,
neutron scattering experiments on the fragile van der Waals liquid 
Orthoterphenyl provide clear evidence for a dynamic cross--over 
from a liquid--like to a solid--like motional mechanism occuring around the 
critical temperature $T_{\rm c}$.

Mode--coupling theory gives a quantitative consistent description of the 
dynamics in the supercooled state. 
Time-temperature-pressure superposition, known from Hz--MHz spectroscopy, 
is valid down to the picosecond time range, where macroscopic properties 
of the material evolve from their microscopic dynamics. 
Spectral line shapes observed by coherent and incoherent neutron 
scattering experiments are invariant under temperature and pressure 
variations. 
The mean relaxation time is proportional to the viscosity and states of 
equal kinetics are connected by the same slope d$T$/d$P$, regardless of the 
time scale.

Temperature and pressure dependent Debye--Waller factors show an anomalous 
decrease that can be described by a square root law. 
Whether or not the square root law results in a sharp singularity is 
however sometimes not compelling, in particular for small wave numbers 
$Q$ or at the structure factor maximum, {\it i.\ e.\ } where the effect is 
expected to be very small anyway. 
The lost scattering intensity reappears as an quasielastic process and is 
identified with the $\beta$--relaxation whose time correlations follow the 
predicted scaling behaviour. 
More importantly, all correlation functions can be described by the same MCT 
$\beta$ correlator (\ref{beta-scaling-law}) with 
{\it one consistent set} of exponents.
The fit parameters show the predicted $Q$ dependence and 
consistently converge to the same critical $T_{\rm c}(P)$.
They allow the unambigous determination of a dynamic crossover 
line $T_{\rm c}(P)$. 
This line is parallel to the lines of equal viscosity: 
$\alpha$-- and $\beta$--relaxation are driven by the same parameter. 
At least in a certain neighbourhood of $T_{\rm c}(P)$ this parameter is 
linear in the effective coupling $\Gamma  \propto nT^{-1/4}$ which in 
turn  ultimately depends on the short range interaction of the molecules. 
Our experiments have also allowed for to separate temperature from 
density effects and to relate at least qualitatively dynamic and 
structural changes.

Furthermore, we have excluded any intramolecular origin of the observed 
anomalies by measurement on partially deuterated samples and 
by comparison of frequency scales of the glass and the crystal. 
Any structural reason is excluded by the smooth temperature and pressure 
evolution of the static structure factor $S(Q)$. 
Also in that sense OTP represented a good model system for testing the 
predictions of MCT.

However, mode--coupling theory does not describes some other important 
aspects like slow processes in the vicinity of T$_{\rm g}$, slow secondary 
Johari--Goldstein $\beta$--relaxation \cite{JoGo70}, cooperativity, 
dynamic heterogeneity \cite{BoHD96} and the vibrational dynamics. 
We should also mention that in a molecular liquid the first critical 
exponent $a$ has never been observed purely and that the appearance of 
the so--called "knee" below the critical temperature is still an open 
question.

Experimentally, the identification of the dynamic transition is far from 
being trivial and the comparison of neutron scattering and MCT predictions 
still remains qualitative in many aspects. 
Complicated spectral shapes have to be measured with high precision, a good 
$Q$ resolution and over a large time or frequency window. 
We presented two different approaches that partly satisfy these demands: 
the combination of different spectrometers and the analysis using mastercurves.

On a fundamental level, the presence of a de Gennes type narrowing in 
OTP-d$_{14}$ reinforces our very basic assumption that incoherent and 
coherent neutron scattering probe self and density correlations of OTP 
molecules, respectively. 
Of course, we measure correlations of single H, D and C atoms which, 
in addition to a translational component, also contain contributions 
from rotational motion relative to the center--of--mass. 
Whether such contributions are relevant and more sophisticated MCT 
extentions have to be used can only be decided by joint use of different 
experimental techniques including molecular dynamics simulations with 
appropriate models.

For the time being we believe that in our study we have exploited the very 
limits of neutron scattering in its present state of the art. 
We have also demonstrated that some very interesting experiments like the 
separation of coherent and incoherent scattering are feasible, but better 
and more accurate results have to await new high--flux neutron sources. 
The total efficiency has to be increased by at least one order of magnitude
by increasing the incomming flux and / or by a much higher spectrometer 
efficiency.
Upgrading and development of all types of spectrometers is currently under 
way.

Other methods, {\it e.\ g.\ } synchrotron radiation, optic Kerr effect, 
impulsive stimulated light scattering or high frequency dielectric 
spectroscopy, among others, seem also to be very promising for the 
investigation of glassy dynamics in particular as they couple to different 
degrees of freedom. 
They may open the way to a comparison of experimental data with more 
sophisticated versions of the theory.

\ack
The author is particularly grateful to his colleagues
F.\ Fujara, W.\ Petry, H.\ Schober and J.\ Wuttke
for this fruitful and long--standing collaboration.
Much of the deuterated OTP and the single crystals have been prepared
by H.\ Zimmermann.
I owe special thanks to C.\ Alba--Simionesco, W.\ G\"otze, W.\ Kob,
F.\ Sciortino and H.\ Sillescu for many discussions and suggestions.
I am also very grateful to J.\ Seelig for the hospitality in his group
during the completion of the manuscript.
Financial support by the Bundesministerium f\"ur Bildung und Forschung
(BMBF) under project numbers 03{\sc si2mai}, 03{\sc fu4dor4} and 
03{\sc fu5dor1} and by the Institute Laue Langevin, Grenoble is gratefully 
acknowledged.

\end{document}